\documentclass[12pt,twoside,polish,english]{mbook}
\usepackage{graphicx}
\usepackage{amssymb}
\usepackage{amsmath}
\usepackage[T1]{fontenc}
\usepackage[cp 1250]{inputenc}
\usepackage{textcomp}
\usepackage{amsfonts}
\usepackage{epsfig}
\usepackage{here}
\usepackage{cite}
\usepackage{esint}
\usepackage{fancyhdr}
\usepackage{color}
\begin{document}

\pagestyle{empty}

\begin{center}
\vspace{3.75cm}
\end{center}
\begin{center}\end{center}
\begin{center}
{\LARGE UNIWERSYTET JAGIELLO{\'N}SKI}\\
{\Large WYDZIA{\L} FIZYKI, ASTRONOMII  }\\
{\Large I INFORMATYKI STOSOWANEJ}\\
{\Large INSTYTUT FIZYKI im. MARIANA SMOLUCHOWSKIEGO }\\

\vspace{2.75cm}

{\bf{\Huge Studia por{\'o}wnawcze oddzia{\l}ywania w niskoenergetycznych
uk{\l}adach \\ 
$pp\eta$ i $pp\eta^{\prime}$  \\}}

\vspace{1.5cm}

{\Large Pawe{\l} Klaja} 

\vspace{4.0cm}

\end{center}

\begin{flushright}
{\small praca przygotowana w Zak{\l}adzie Fizyki J\k{a}drowej \\
Instytutu Fizyki im. Mariana Smoluchowskiego, \\
Wydzia{\l}u Fizyki, Astronomii i Informatyki Stosowanej Uniwersytetu Jagiello{\'n}skiego, oraz \\
w Instytucie Fizyki J\k{a}drowej w Centrum Badawczym J{\"u}lich,\\
pod kierunkiem Prof. Paw{\l}a Moskala.  \\}
\end{flushright}
\vspace{0.5cm}
\begin{center}
    Krak{\'o}w 2009 
\end{center} 
  
   \cleardoublepage
   
\pagestyle{empty}

\begin{center}
\vspace{3.75cm}
\end{center}
\begin{center}\end{center}
\begin{center}
{\LARGE JAGIELLONIAN UNIVERSITY}\\
{\Large FACULTY OF PHYSICS, ASTRONOMY  }\\
{\Large AND APPLIED COMPUTER SCIENCE}\\
{\Large MARIAN SMOLUCHOWSKI INSTITUTE OF PHYSICS  }\\

\vspace{2.75cm}

{\bf{\Huge Comparative studies of the interaction in the low energy \\
$pp\eta$ and $pp\eta^{\prime}$ systems \\}}

\vspace{1.5cm}

{\Large Pawe{\l} Klaja} 

\vspace{4.0cm}

\end{center}

\begin{flushright}
{\small dissertation prepared at the Nuclear Physics Department \\
of the Marian Smoluchowski Institute of Physics, 
at Faculty of Physics, Astronomy and Applied Computer Science of the Jagiellonian University\\
and in Institute of Nuclear Physics at Research Centre J{\"u}lich,\\
guided by Prof. Pawe{\l} Moskal.  \\}
\end{flushright}
\vspace{0.5cm}
\begin{center}
    Cracow 2009 
\end{center}

   \cleardoublepage

\pagestyle{empty}
\begin{center}

\end{center}

\vspace{15.cm}

\begin{flushright}
\textbf{{\it "To s\k{a} kulki prosz\k{e} Pa{\'n}stwa, na 100\%."}\\
P.~M. }
\end{flushright}

   \cleardoublepage
   \pagestyle{empty}

\begin{center}
{\bf Abstract\\}
\end{center}

\vspace{1.0cm}

The COSY-11 collaboration measured the $pp~\to~pp\eta$ and
$pp~\to~pp\eta^{\prime}$ reactions in order to perform comparative studies of the 
interactions within the proton-proton-meson system.\\  
This thesis presents in detail the analysis of the $pp \to pp\eta^{\prime}$ reaction 
which was measured at the proton beam momentum of 3.260 GeV/c.\\
The elaboration results in differential distributions of squared invariant 
proton-proton ($s_{pp}$) and proton-$\eta^{\prime}$ ($s_{p\eta^{\prime}}$) masses, as well as in
angular distributions and the total cross section at an excess energy of 16.4 MeV.\\
The differential distributions $s_{pp}$ and $s_{p\eta^{\prime}}$ are compared to theoretical 
predictions and to the analogous spectra determined for the $pp \to pp\eta$ reaction.\\
The comparison of the results for the $\eta$ and $\eta^{\prime}$ meson production
rather
excludes the hypothesis that the enhancement observed in the invariant mass distributions
is due to the meson-proton interaction. \\
Further, the shapes of the distributions do not favour any of the 
postulated theoretical models.\\  

\vspace{1.0cm}

\begin{center}
{\bf Streszczenie\\}
\end{center}

\vspace{1.0cm}
W ramach grupy badawczej COSY-11 wykonano pomiary reakcji $pp~\to~pp\eta$ oraz
$pp~\to~pp\eta^{\prime}$ w celu przeprowadzenia studi{\'o}w por{\'o}wnawczych 
oddzia{\l}ywania w uk{\l}adzie proton-proton-mezon.\\ 
W rozprawie doktorskiej zaprezentowano analiz\k{e} danych z pomiaru reakcji $pp \to pp\eta^{\prime}$ 
wykonanego z wykorzystaniem wi\k{a}zki protonowej o p\k{e}dzie 3.260 GeV/c.\\ 
Wynikiem analizy przedstawionej w niniejszej pracy s\k{a} r{\'o}{\.z}niczkowe przekroje czynne  
w funkcji mas niezmienniczych proton-proton ($s_{pp}$) i proton-$\eta^{\prime}$ ($s_{p\eta^{\prime}}$), 
rozk{\l}ady k\k{a}towe oraz ca{\l}kowity przekr{\'o}j czynny dla energii wzbudzenia Q = 16.4 MeV.\\
Rozk{\l}ady r{\'o}{\.z}niczkowe $s_{pp}$ oraz $s_{p\eta^{\prime}}$ zosta{\l}y por{\'o}wnane 
z przewidywaniami teoretycznymi oraz analogicznymi widmami otrzymanymi dla reakcji $pp \to pp\eta$.\\
Por{\'o}wnanie rezultat{\'o}w otrzymanych dla produkcji mezon{\'o}w $\eta$ i $\eta^{\prime}$ 
pozwala na wykluczenie hipotezy, {\.z}e wzmocnienie obserwowane w widmach mas niezmienniczych 
jest powodowane oddzia{\l}ywaniem mezonu z protonem.\\
Co wi\k{e}cej, na podstawie kszta{\l}tu otrzymanych dystrybucji nie da si\k{e} 
rozstrzygn\k{a}{\'c} pomi\k{e}dzy poprawno{\'s}ci\k{a} za{\l}o{\.z}e{\'n} w 
postulowanych modelach teoretycznych.\\

   \tableofcontents

\chapter{Introduction}
\label{Introduction}
\pagestyle{myheadings}
  \markboth{\bf Chapter 1.}{\bf Introduction}

The knowledge of the meson-nucleon interaction at the hadronic level is
one of the major goals in nuclear physics, nowadays. Also, studies of meson structure
and production mechanisms constitute a huge interest of nuclear and particle physicists.
During the last decades many measurements produced interesting results \cite{habil},
but various questions are still open. \\
In the SU(3)-flavour scheme the $\eta$ and $\eta^{\prime}$ mesons belong to the
nonet meson family of pseudoscalar mesons.
The $\eta$ and $\eta^{\prime}$ mesons constitute a mixture of
states: singlet-$\eta_{1}$ and octet-$\eta_{8}$.\\
It is important to stress that the strength of the $p\eta$ and $p\eta^{\prime}$ 
interaction depends on the structure of the $\eta$ and $\eta^{\prime}$ mesons and 
is directly related to the singlet-$\eta_{1}$ and octet-$\eta_{8}$ contributions 
in the wave functions of these mesons \cite{bass, bass1, bass2}.\\
Taking into account the mixing angle the resulting
contribution of the various quark flavours in the $\eta$ and $\eta^{\prime}$ wave function
is almost the same.
However, in spite of the postulated similar quark structure the $\eta$ and $\eta^{\prime}$ mesons owe unexpectedly different features.\\
The most drastic ones are:
\begin{itemize}
\item
$\eta$ and $\eta^{\prime}$ mesons possess different masses, the $\eta$(547) meson is almost two times
lighter than the $\eta^{\prime}$(958) meson \cite{pdg}.
\item
The branching ratios of B and D$_{s}$ mesons for the decays into the $\eta^{\prime}$ meson are higher than
for the decays into the $\eta$ meson, and deviate strongly from model predictions \cite{jossop, branden}.
\item
There is no experimental evidence for baryonic resonance which decays by the emission of an $\eta^{\prime}$ meson
\cite{pdg, hagiwara} while e.g. the resonances N(1535) and N(1650) decay via the emission of
the $\eta$ meson with a significant probability \cite{pdg}.
\end{itemize}
These different features could imply a possible difference in the interaction of
$\eta$ and $\eta^{\prime}$ mesons with elementary particles, and indicate that also the
production mechanism of both mesons in elementary particle collisions might vary.\\
Due to the short life-time of the flavour neutral pseudoscalar mesons, experiments with meson beams or
targets are difficult or hardly feasible.
Therefore, the nucleon-meson interaction can be studied only via its influence
on the cross sections of the reactions during which they are produced (e.g. $NN \to NN\ Meson$) \cite{habil}.\\
Quantitative information about the interaction
can be gained from the shape of the excitation functions for the $pp \to pp\eta$ and $pp \to pp\eta^{\prime}$
reactions as well as from a comparison of those to the $pp\pi^{0}$ system.\\

Up to now, only the proton interaction with pions and the $\eta$-meson~\cite{habil} was studied more exhaustively due to much more
higher total cross sections in comparison to the cross section for the $pp \to pp\eta^{\prime}$ reaction.\\
Besides the excitation function also differential distributions of invariant proton-proton and proton-meson masses
constitute a sensitive tool for studies of the interaction within the meson-nucleon system.
The distribution of the proton-$\eta$ invariant mass
showed a clear enhancement in the region of small proton-$\eta$ relative momenta \cite{prc69}.
The observed effect could be explained by the non negligible role of the
proton-$\eta$ interaction in the final state \cite{fix,fix2}, the admixture of higher partial waves
during the $\eta$ production \cite{naka}, or the
energy dependence of the production amplitude \cite{deloff}.
Using only the $pp~\to~pp\eta$ data, it is however not possible to justify or falsify any of
the above mentioned hypothesis.

The endeavour to explain the observed enhancement
motivated the experiment and the analysis of the $pp \to pp\eta^{\prime}$ reaction which is presented in this thesis.
This experiment was performed using the COSY-11 detector setup \cite{brauksiepe, klajac11, strona},
installed at the cooler synchrotron COSY \cite{ring1} at the Research Centre J{\"u}lich in Germany.
It was conducted in an energy range close to the kinematical threshold for the $\eta^{\prime}$ meson production,
where the relative velocities of the produced particles are small.\\
The analysis and results of the
$pp \to pp\eta^{\prime}$ reaction measurement, conducted in order to determine the distribution
of events over the phase space for an excess energy range equal to the one
measured before for the $pp \to pp\eta$ reaction are presented in this thesis. 
The measurement was performed at the nominal beam momentum of 3.257 GeV/c corresponding 
to the nominal excess energy for the $pp\eta^{\prime}$ reaction equal to 15.5 MeV. 
The analysis of the $pp\eta^{\prime}$ data was performed in a similar way as it has been done for the
$pp\eta$ system.\\
The comparison of the differential distributions for the proton-proton and for the proton-meson invariant masses
in the $\eta$ and $\eta^{\prime}$ production could help to judge about the validity of postulated theories
concerning the observed enhancement and allows for a quantitative estimation of the
relative strength of the proton-$\eta$ and the
proton-$\eta^{\prime}$ interactions, provided that the effect is caused by the proton-meson interaction.\\

In the following chapter the current status of the $pp \to pp\eta$ and  $pp \to pp\eta^{\prime}$
total cross section measurements is presented.
Furthermore, the possibility of proton-$\eta$ and proton-$\eta^{\prime}$ interaction studies is discussed and
the description of the observables used in the analysis described in this thesis is given.
The results achieved for the $pp \to pp\eta$ measurement as well as the available theoretical descriptions
of the results are presented.\\
Chapter \ref{Experimental facility} is devoted to the presentation of the experimental facility
used to perform the $pp \to pp\eta^{\prime}$ measurement.
In the next chapter, the methods used for the calibration of the COSY-11 detectors
and their relative geometrical settings are presented. The time-space
relation of the drift chambers and the procedure for the time-of-flight calibration is described.
Also, the procedure used for the monitoring of the relative beam-target setting is discussed.\\
In chapter \ref{Identification of the reaction} the method of the $pp \to pp\eta^{\prime}$ reaction identification will
be depicted and the method of identifying measured and also unobserved particles will be given.\\
Further on, the procedure of the
luminosity ($L$) determination will be presented in chapter \ref{Luminosity determination}, and
in the consecutive chapters it will be shown how the absolute value and the spread of the beam momentum was
extracted, and the method for the determination of the position of the
drift chambers relative to the COSY-11 dipole will be explained.\\
Chapter \ref{Evaluation of the differential distributions} comprises the evaluation procedure
of the differential distributions. First, the kinematical fit procedure and then
the background subtraction due to the multi-pion production will be discussed. \\
The final results concerning the total cross section and
differential distributions are presented  in chapter \ref{Cross sections}. The acceptance corrections are discussed
and the achieved experimental results are compared to theoretical predictions.\\
The conclusions are presented in chapter \ref{Summary}.\\
In the appendices at the end of the dissertation some issues discussed in the thesis are explained in more detail.
In the first one the structure of the pseudoscalar meson nonet, meson masses and quark structure are presented.
The second one is devoted to the description of the parameterization of the on-shell proton-proton interaction.
General remarks about the combined analysis of the $\eta^{\prime}$ meson formalism in
photo- and hadro-production are presented in the third addendum,
and the linear energy dependence of the production amplitude is discussed in the last one.\\

\chapter{Low energy interaction within a proton-proton-meson system}
\label{Low energy interaction within a proton-proton-meson system}
\pagestyle{myheadings}

\markboth{\bf Chapter 2.}
         {\bf Low energy interaction within a proton-proton-meson system}
	
The interaction of hadrons is the reflection of the strong force
between the quarks, and provides information about the hadron structure and the
strong interaction itself \cite{habil}. In the framework of the optical model, the interaction between hadrons can be
expressed in terms of phase-shifts, which in the zero energy limit are described by the
scattering length and effective range parameters \cite{habil}.
These variables are quite well established for the (low-energy) nucleon-nucleon interaction \cite{noyes,naisse},
but they are poorly known for the nucleon-meson or meson-meson interactions.
The estimated real part of the scattering length of the $\eta$-proton potential,
depending on the method of the analysis and studied region, is 3 to 10 times \cite{habil}
larger than for the $\pi^{0}$-proton scattering ($a_{p\pi^{0}}$ = 0.13~fm) \cite{sigg1, sigg2},
while for the $\eta^{\prime}$ meson only an upper limit is known of 0.8 fm \cite{swave}.\\
The interaction of mesons (e.g. pseudoscalar mesons\footnote{The features of the pseudoscalar meson nonet are
described in appendix \ref{Pseudoscalar mesons}.}: $\pi, K, \eta, \eta^{\prime}$) with nucleons
could be deduced from the experiments realized by means of meson beams,
but such experiments are not feasible in case of the flavour neutral mesons
due to their short lifetime~\cite{habil, pdg}.
However, the study of their
interaction with hadrons is certainly accessible via their influence
on the cross section of reactions like $NN \to NN~Meson$ in which they are produced~\cite{habil}.
In such a case, the interaction within the final meson--nucleon system
will modify the shape of the excitation function and of the differential
distributions of invariant masses of the nucleon-nucleon-meson systems.

\section{Excitation functions for $pp \to pp\eta$ and $pp \to pp\eta^{\prime}$ reactions}
\label{Excitation functions}
\begin{center}
\begin{figure}[H]
\includegraphics[width=0.5\textheight]{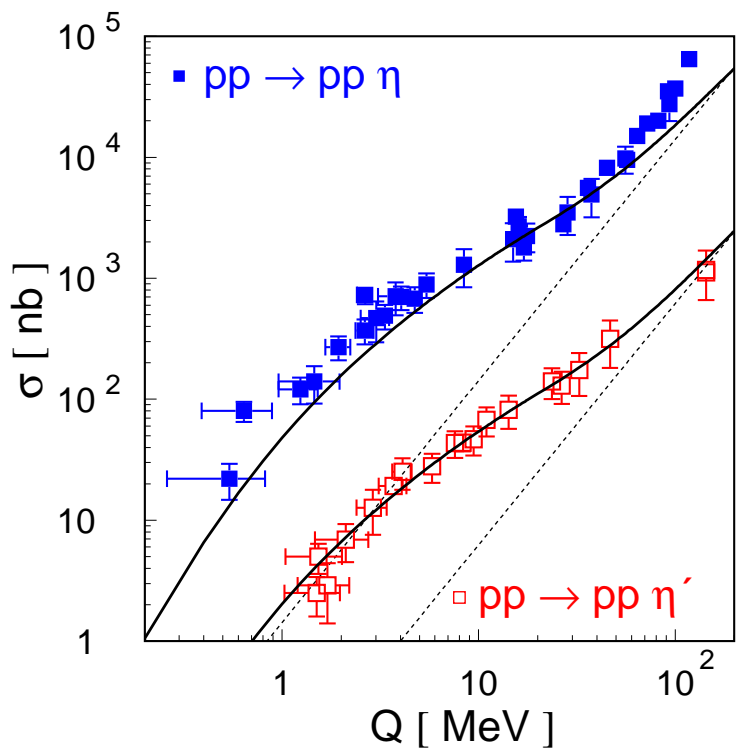}
\caption{The $pp \to pp\eta$ and $pp \to pp\eta^{\prime}$ excitation functions
     \cite{balestra, wurz, pm80, b474, khoukaz, hibou, chiav, calen2, berg, smyrski, calen}. The dashed lines indicate arbitrarily
     normalized functions obtained under the assumption of the
     homogeneous phase space occupation. Solid lines correspond to calculations of the phase space weighted
     by the proton-proton on-shell scattering amplitude \cite{swave}.}
\label{fig:cross}
\end{figure}
  \end{center}

Near the kinematical threshold measurements of nucleon-nucleon collisions allow to study
the particle production with a dominant contribution from one partial wave only.
In this energy range, the dependence of the total cross section as a function of the centre-of-mass excess energy
is predominantly determined by the available phase space and the interaction between the exit particles.
The excitation functions for the $pp \to pp\eta^{\prime}$ \cite{balestra, wurz, pm80, b474, khoukaz, hibou} and
$pp \to pp\eta$ \cite{hibou, chiav, calen2, berg, smyrski, calen} reactions are presented in figure \ref{fig:cross}. 
Comparing
the data to the arbitrarily normalized phase-space integral reveals that proton-proton
FSI enhances the total cross section by more than one order of magnitude for low energies. 
In case of the $\eta^{\prime}$ meson production one recognizes that the data are described
well assuming that the on-shell proton-proton amplitude exclusively determines the phase-space population.
This indicates that the proton-$\eta^{\prime}$ interaction is too small to  manifest itself in
the excitation function within the presently achieved statistical uncertainty.
However, for the $\eta$ meson production the enhancement is
by about a factor of two larger than in case of the $\eta^{\prime}$ meson
and cannot be described by the $pp$-FSI only.

\section{Comparison of $\eta$, $\eta^{\prime}$ and $\pi^{0}$ meson interaction with protons}
\label{Comparison interaction with protons}
The strength of the interaction deduced from the comparison of the
data and the lines in figure \ref{fig:cross} depends on the model of the proton-proton interaction
used in the calculations for the $pp\eta$ and $pp\eta^{\prime}$ systems \cite{swave}.
Therefore, in order to estimate a relative strength between the $p \eta$ and $p \eta^{\prime}$ interactions
in a model independent way one can compare the shape of the excitation function of the $pp \to pp\eta$
and $pp \to pp\eta^{\prime}$ reactions. Moreover, one can gain some quantitative information about these interactions
by a comparison of these shapes to the $pp\pi^{0}$ system \cite{habil},
since the $\pi^{0}$-proton scattering length is well known and amounts to  0.13~fm \cite{sigg1, sigg2}.\\
 \begin{figure}[H]
  \includegraphics[height=.33\textheight]{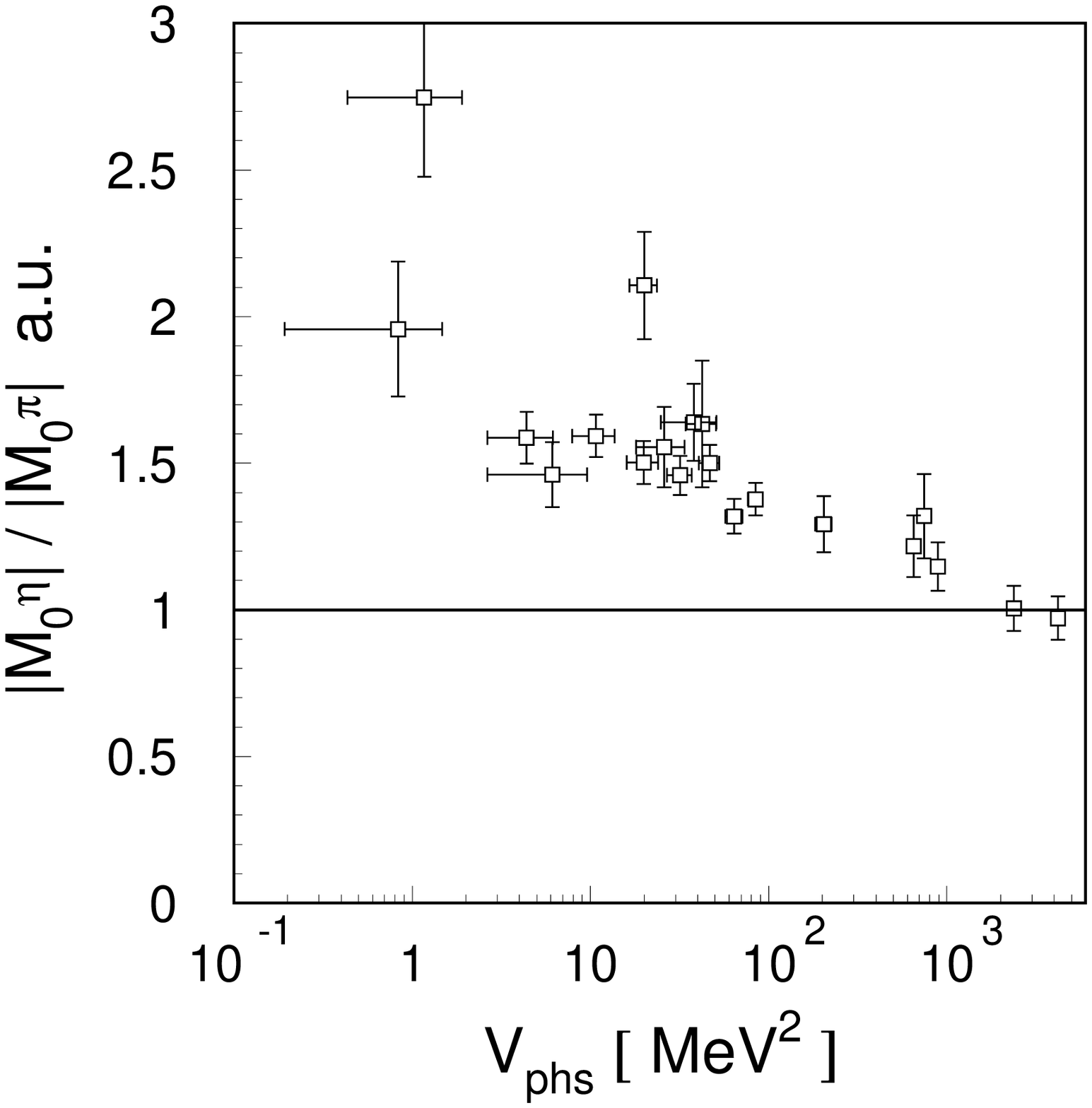}
  \includegraphics[height=.33\textheight]{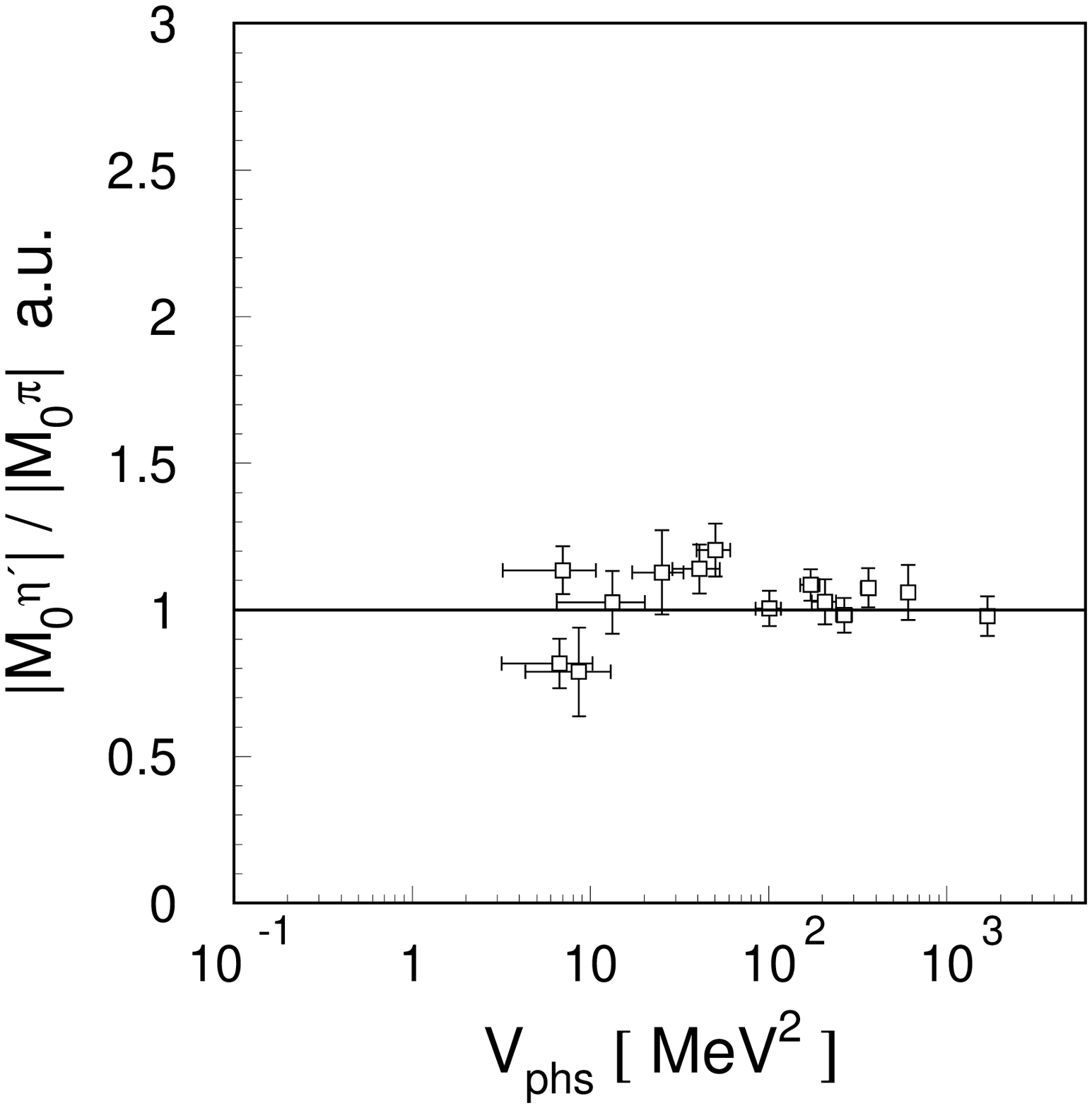}
  \caption{The ratios $|M_{0}^{\eta}|/|M_{0}^{\pi}|$ (left) and $|M_{0}^{\eta^{\prime}}|/|M_{0}^{\pi}|$ (right) 
  extracted from data calculating the $pp$-FSI according to the
  formulas from reference \cite{swave},
  and neglecting the proton-meson interaction \cite{habil}. The ratio is shown
  as a function of the phase space volume \cite{swave}.}
\label{fig:amplitude}
\end{figure}
For that purpose, one can compare only the dependence of the production amplitudes $|M_{0}|$
derived from the data taking into account the $pp$-FSI only.\\
The dependence of $|M_{0}^{\eta}|$ and $|M_{0}^{\eta^{\prime}}|$ as a function of the phase space volumes
normalized
to  $|M_{0}^{\pi}|$ are presented in figure \ref{fig:amplitude}. The $|M_{0}|$ for the
$\eta$, $\eta^{\prime}$ and $\pi^{0}$ mesons were extracted from data,
disregarding any proton-meson interaction.
When the neglected $\eta$($\eta^{\prime}$)-proton interaction would have been the same
as the one for proton-$\pi^{0}$, the points in the plots should have been consitent with unity
as can be seen
for the $pp \to pp\eta^{\prime}$ reaction, when really the interaction shows its weakness, independently of the prescription
used for the proton-proton final state interaction \cite{swave}. In the case of the $\eta^{\prime}$ meson production
its weak interaction with
nucleons at the low-energy range is expected due to the lack of any baryonic resonances
which could decay into a $N\eta^{\prime}$ system \cite{habil, hagiwara}\footnote{Recent calculations of K. Nakayama, 
Y. Oh and
H. Haberzettl predict also resonance state contributions
for the photo- and hadro-productions \cite{kanzo_c11}.}.\\

Statistical uncertainties allowed to get only a very conservative upper limit for the  real part
of the scattering length of the proton-$\eta^{\prime}$ potential resulting in:\\
$|Re~a_{p \eta^{\prime}}|~<~0.8~fm$~\cite{habil,b474}.\\
Thus, independent of the model used for the prescription of the $pp$-FSI,
from a comparison of the energy dependence of the production
amplitudes for the $pp \to pp\eta$, $pp \to pp\eta^{\prime}$ and $pp \to pp\pi^{0}$ reactions,
it was concluded that the interaction
within the proton-$\eta^{\prime}$ system is much weaker than the interaction between the proton and the $\eta$ meson.\\
Another possibility of learning about interactions within nucleon-nucleon-meson
systems is given by the differential distributions of the invariant masses.
This is why the present analysis of the $pp\eta^{\prime}$ system has been performed in a similar way
as it has been done earlier \cite{prc69} for the $pp\eta$ system. The determined $pp$ and $p$-meson
invariant mass distributions will
be used for a comparative study of the interaction within the proton-meson system.\\
In the next section the definitions of the studied observables which will be used in the further analysis are
presented.

\section{Definitions of observables}
\label{Definitions of observables}
\begin{center}
\begin{figure}[H]
\includegraphics[width=0.65\textheight]{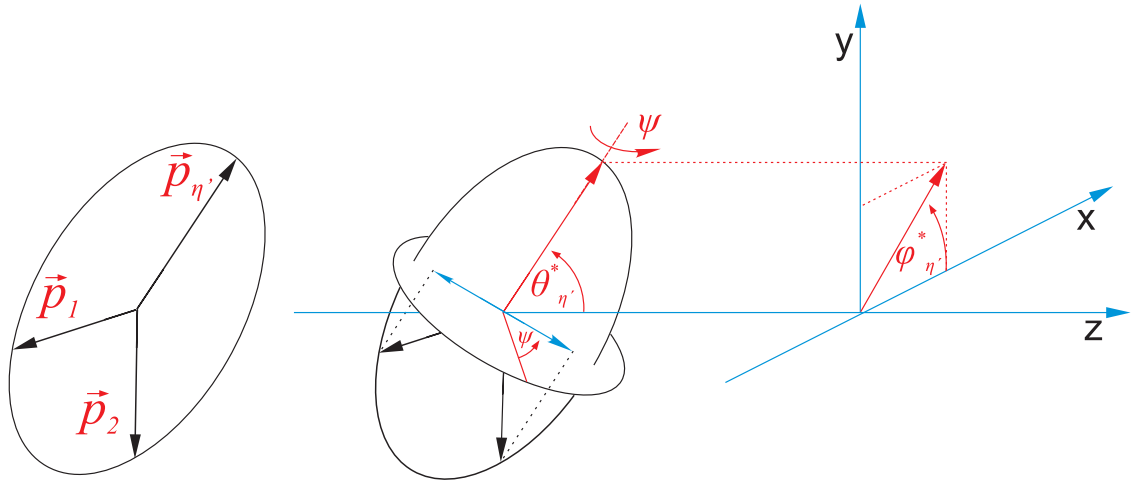}
\caption{Schematic definitions of the centre-of-mass kinematical variables used for the description of
the $pp\eta^{\prime}$ system. In the centre-of-mass frame the momentum vectors of the three outgoing particles are located within
the reaction plane. In this plane the relative motion of the ejectiles is fixed by the square of the
invariant masses $s_{pp}$ and $s_{p\eta^{\prime}}$. Three remaining variables the $\phi^{*}_{\eta^{\prime}}$, $\theta^{*}_{\eta^{\prime}}$
and $\psi$ angles are used to define the orientation of the emission plane in space.}
\label{fig:plane}
\end{figure}
  \end{center}

To describe the studied three particle ($pp\eta^{\prime}$) system one needs only five independent variables in the centre-of-mass system.
In this frame, due to energy and momentum conservation,
momentum vectors of protons and $\eta^{\prime}$ lie in one plane, called reaction plane.
In that plane (shown schematically in figure \ref{fig:plane}) the relative momenta of particles are described
by only two variables. These quantities may be chosen as square of the proton-proton invariant mass $s_{pp}$
and square of the proton-$\eta^{\prime}$ invariant mass $s_{p \eta^{\prime}}$.
Invariant masses depend on the relative velocity of the particles and
are therefore well suited for a description of the interactions between these particles.
Besides the relative movement of particles on the reaction plane three other variables have to be defined for
fixing the orientation of the reaction plane in the coordinate system.
In this thesis, by analogy to the evaluation of the $pp\eta$ system \cite{prc69}, the
azimuthal and polar angles of the $\eta^{\prime}$ meson momentum vector relative to the beam direction, denoted as
$\phi^{*}_{\eta^{\prime}}$ and $\theta^{*}_{\eta^{\prime}}$ are used, respectively, and the $\psi$ angle describing the rotation of the
reaction plane around the direction of the momentum vector of the $\eta^{\prime}$ meson. \\
The interaction between final state particles does not alter the orientation of the reaction plane \cite{prc69}.
Therefore, it will manifest itself only in the distribution of the invariant masses $s_{pp}$ or $s_{p \eta^{\prime}}$,
or generally in the population of the Dalitz plot ($s_{pp}$ vs. $s_{p \eta^{\prime}}$).\\
In the case of non-interacting particles in the final state these distributions
should correspond to a homogeneously populated phase space.
Therefore, their interaction should show up as a deviation from these expectation.

\section{$pp$ and $p-meson$ invariant mass distributions}
\label{invariant mass distributions}
Only two invariant masses of three subsystems are independent and therefore the whole accessible
information about the final state interaction can be shown in the Dalitz plot. One can
also use the projection of the phase-space distribution onto the invariant masses of
proton-proton or proton-meson subsystems \cite{prc69}.

\begin{figure}
  \includegraphics[height=.3\textheight]{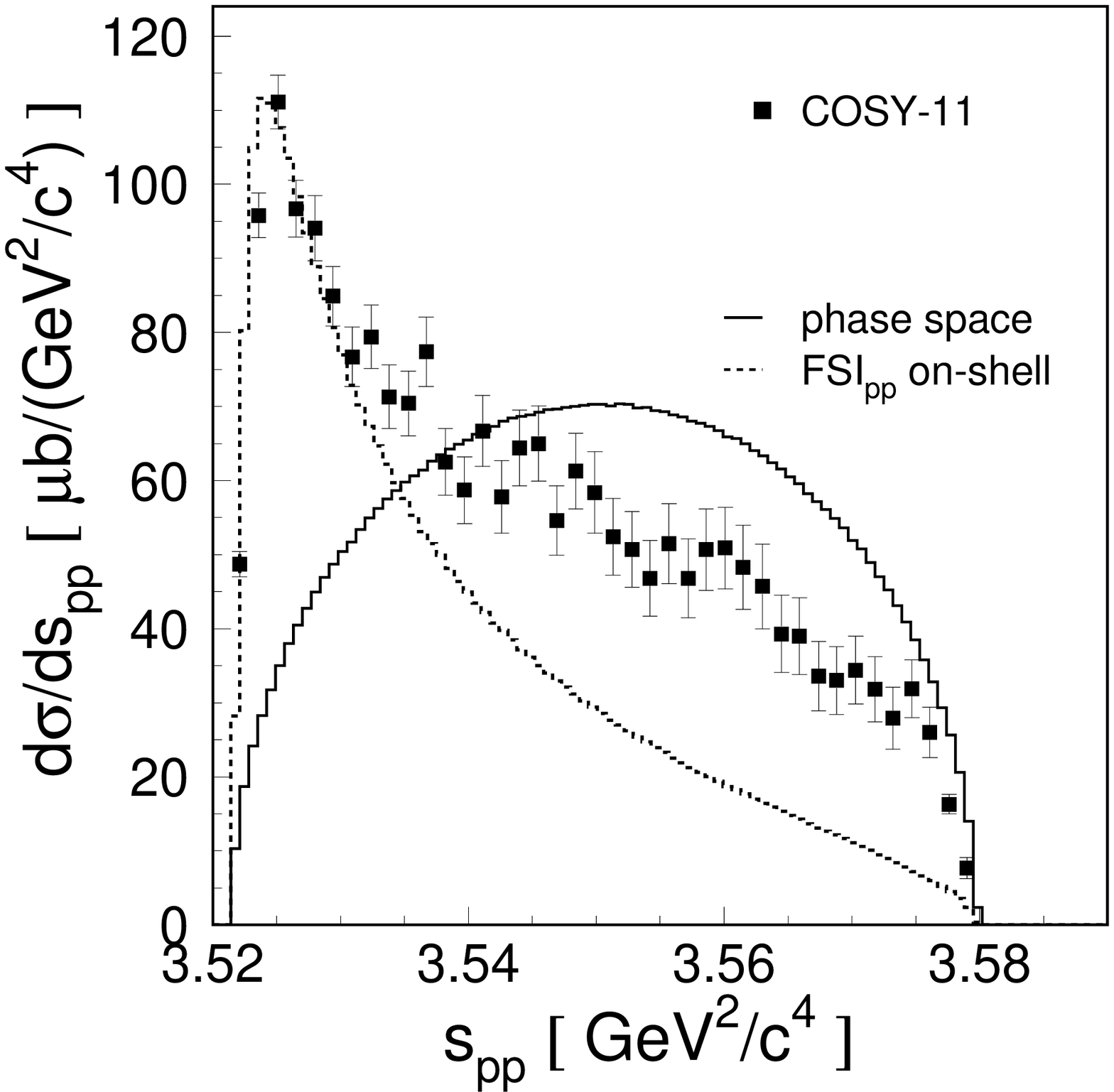}
  \includegraphics[height=.3\textheight]{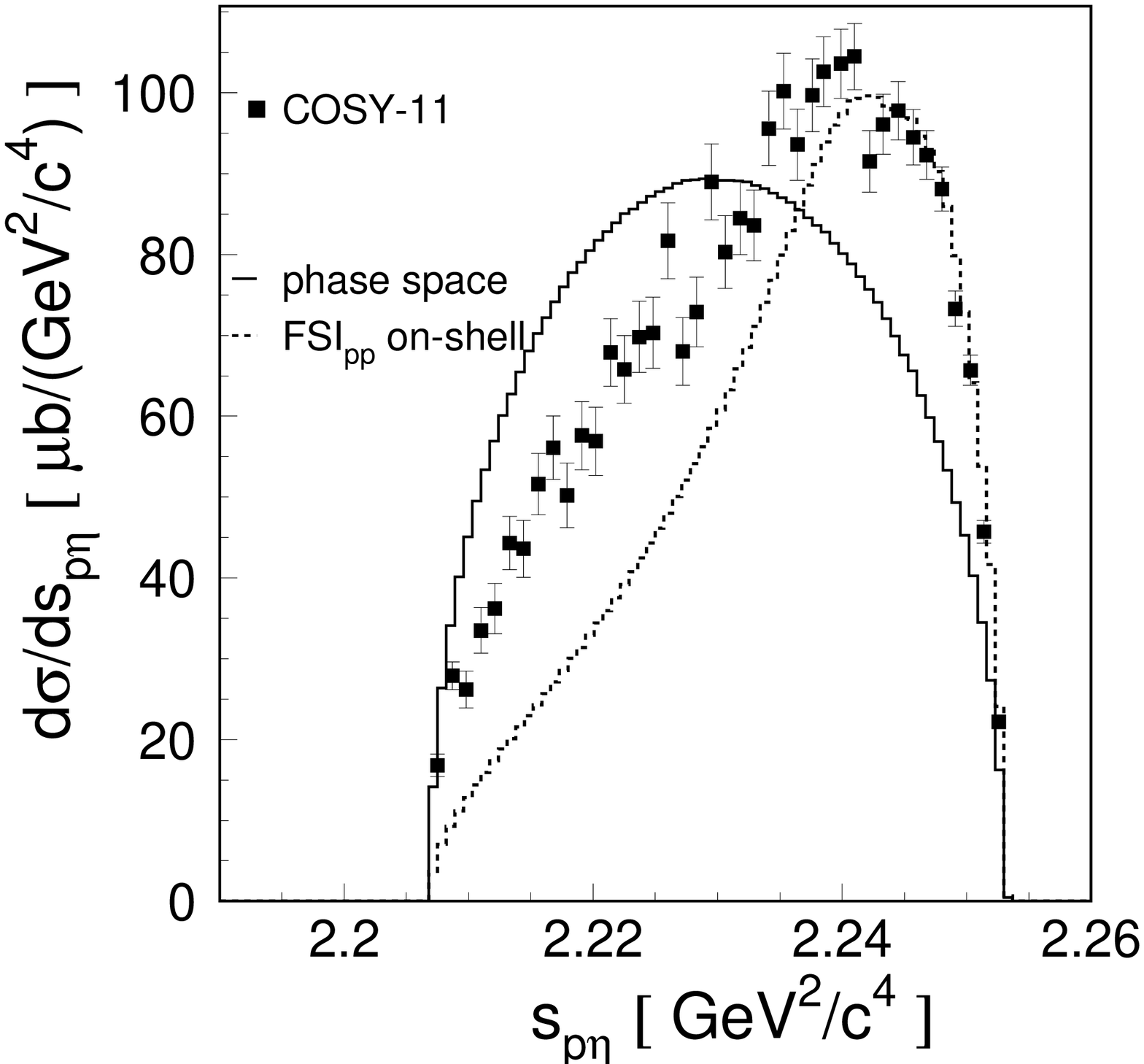}
  \caption{Distributions of the square of the proton-proton ($s_{pp}$) (left) and
  proton-$\eta$ ($s_{p \eta}$) (right) invariant masses determined experimentally for
  the $pp \to pp\eta$ reaction (closed squares). The integrals of the phase space weighted by the
  square of the proton-proton on-shell scattering amplitude
(dotted lines)-FSI$_{pp}$  have been normalized arbitrarily at small values of $s_{pp}$.
The expectations
under the assumption of a homogeneously populated phase space are shown as solid lines.}
\label{fig:phase}
\end{figure}

The qualitative phenomenological analysis of the determined
differential invariant proton-proton and proton-$\eta$ mass distributions revealed an
enhancement of the population density at the kinematical region corresponding to a small
proton-$\eta$ momentum.
The proton-proton and proton-$\eta$ invariant mass distributions determined for the
$pp \to pp\eta$ reaction at an excess energy of 15.5 MeV are presented
in figure \ref{fig:phase}\footnote{A similar enhancement has been observed for several measurements
performed at excess energies of 4.5 MeV \cite{prc69}, 10 MeV \cite{czyzyk, czyzyk1}, 15
and 41 MeV \cite{tof41}, and 72 MeV \cite{henrik}.}.
The dashed lines in both panels of the figure depict the results of
calculations where only the on-shell amplitude of the proton-proton interaction has been taken into account.

\begin{figure}
  \includegraphics[height=.3\textheight]{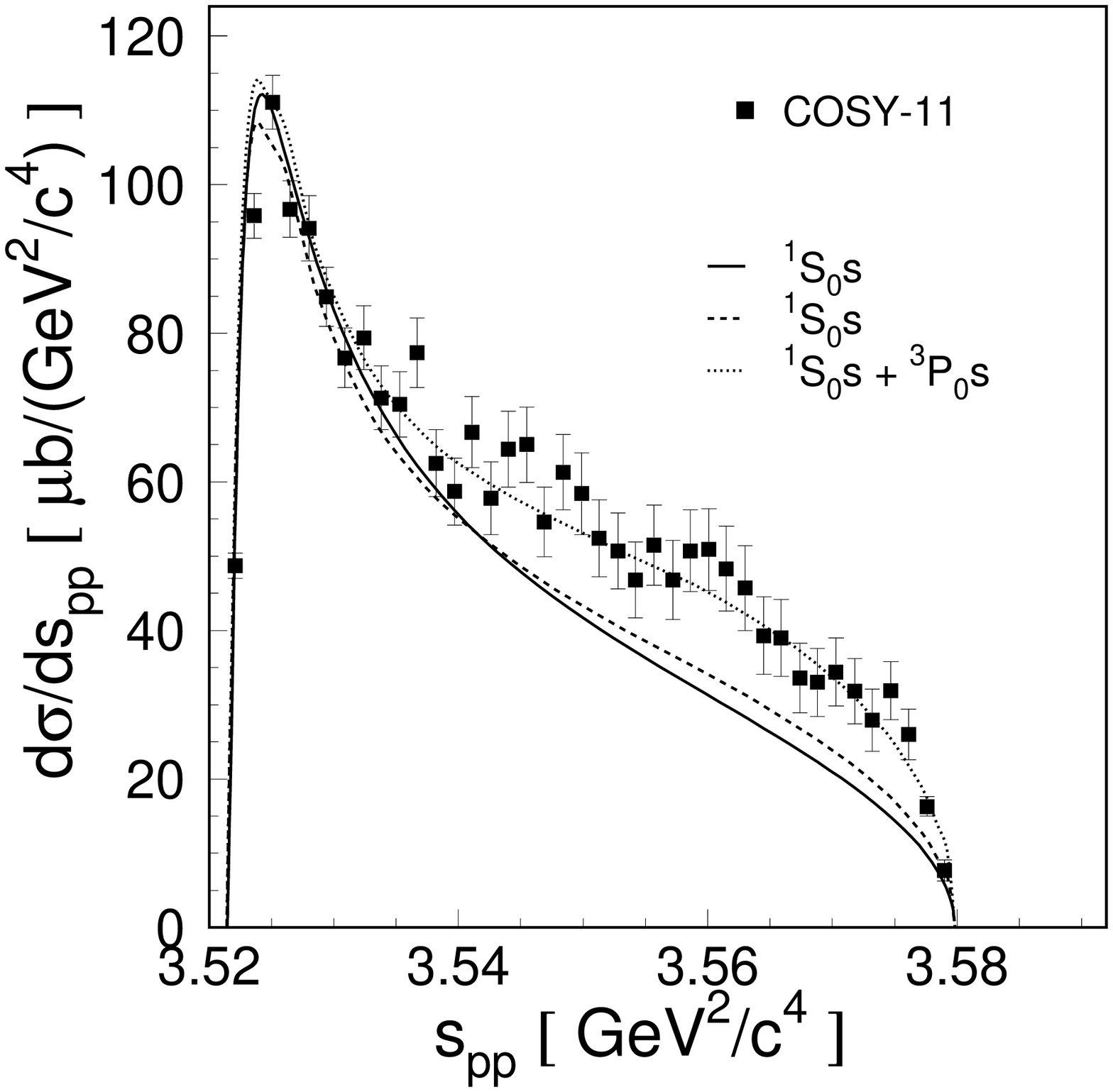}
  \includegraphics[height=.3\textheight]{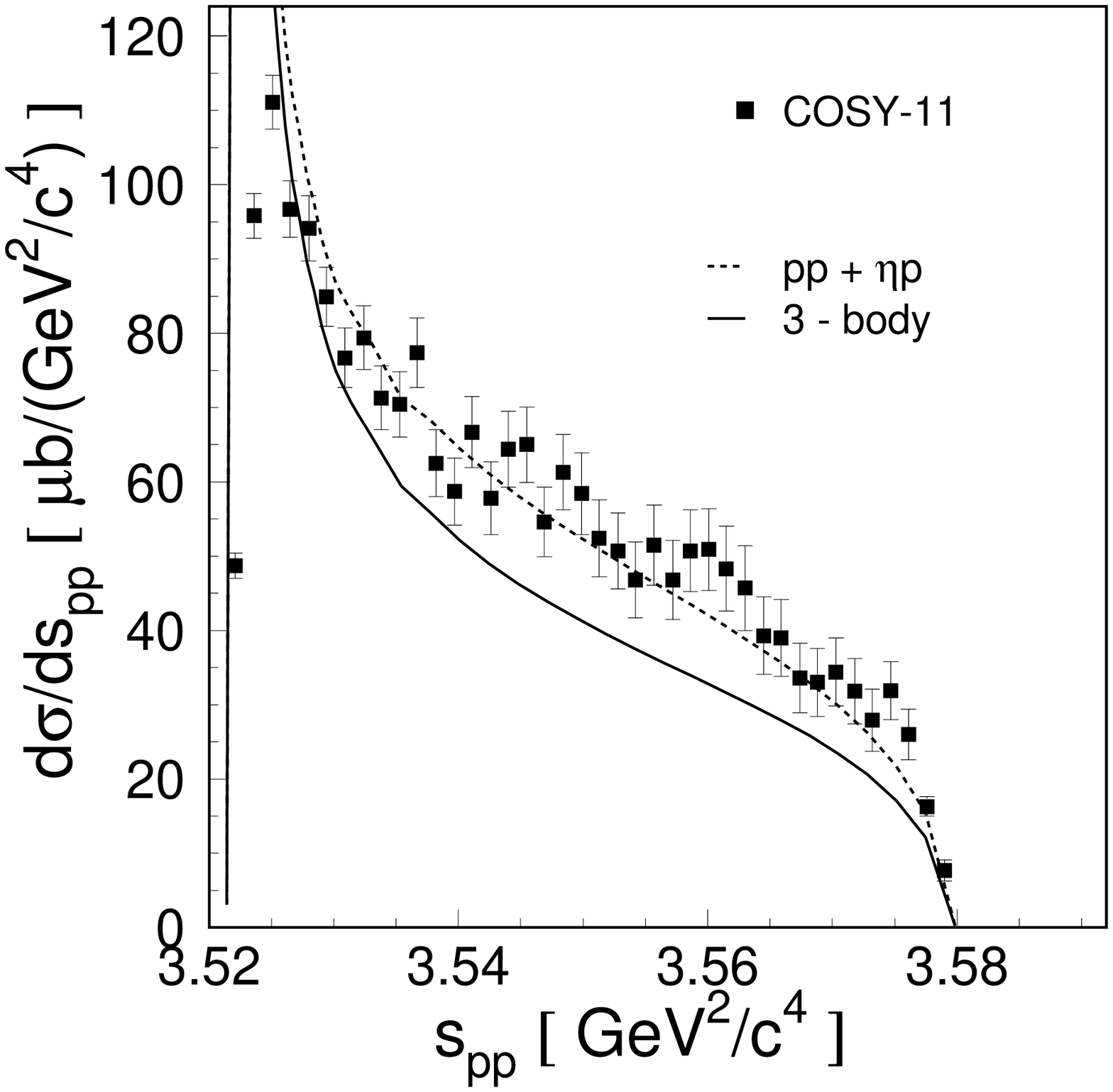}
  \caption{(Left panel) Distribution of the square of the proton-proton ($s_{pp}$) invariant mass
  determined experimentally for the $pp \to pp\eta$ reaction. Solid and dashed lines correspond to
  calculations under the assumption of a $^{3}P_{0} \to ^{1}\!\!S_{0}s$ transition according to the
  models described in \cite{baru} and \cite{naka}, respectively. The dotted line shows
  the result of calculations with inclusion of the $^{1}S_{0} \to ^{3}\!\!P_{0}s$ contribution
  as suggested in \cite{naka}.
  (Right panel) The same data as in the left panel but compared with three-body calculations \cite{fix, fix2}.
  The solid line was determined with a rigorous three-body approach. The dashed line depicts
  the situation if only pairwise interactions ($pp + p \eta$) are allowed.}
\label{fig:theory}
\end{figure}

In those calculations the enhancement factor has been estimated as the square of the on-shell
proton-proton scattering amplitude derived using the modified Cini-Fubini-Stanghellini formula
including the Wong-Noyes Coulomb corrections \cite{swave}.\\
 One can easily see that the mentioned effect is too large
to be described by the on-shell inclusion of the proton-proton FSI.\\

In fact a better description is achieved when contributions from higher partial waves or
off-shell effects of the proton-proton potential are taken into account. These calculations
compared to the experimentally determined differential proton-proton invariant mass distribution
are presented in figure \ref{fig:theory}.
In the left panel of this figure the experimentally determined differential
cross section as a function of the squared invariant
proton-proton mass is compared to the calculations of V. Baru and collaborators \cite{baru}
under the assumption of a $^{3}P_{0} \to ^{1}\!\!S_{0}s$
transition according to the models described in \cite{baru}, depicted as the solid line.\\
Dashed and dotted lines on the left panel of figure \ref{fig:theory} represent the calculations of
K.~Nakayama and his group \cite{naka}. The authors claim that the contribution of the S-wave alone is unable to explain the
observed enhancement in the squared proton-proton invariant mass distribution. Seeking for the better description
they postulate that the shape of the enhancement can be reproduced by folding the relative momentum of the
proton-proton subsystem with the available phase space \cite{naka} suggesting that the enhancement
could be the consequence of the $pp$ P-wave in the final state.\\
Calculations assuming a $^{3}P_{0} \to ^{1}\!\!S_{0}s$ transition
correspond to the dashed curve and the result of calculations with the inclusion of the $^{1}S_{0} \to ^{3}\!\!P_{0}s$ contribution
is depicted by the dotted line. Although the dotted line corresponding to calculations
based on the stronger P-wave contribution is in quite good agreement to the experimental determined differential
distribution of the proton-proton invariant mass, it underestimates the total cross section data taken for
the $pp \to pp\eta$ reaction near the kinematical threshold (Q~lower~than~30~MeV)~\cite{habil}.\\

\begin{figure}[H]
  \includegraphics[height=.3\textheight]{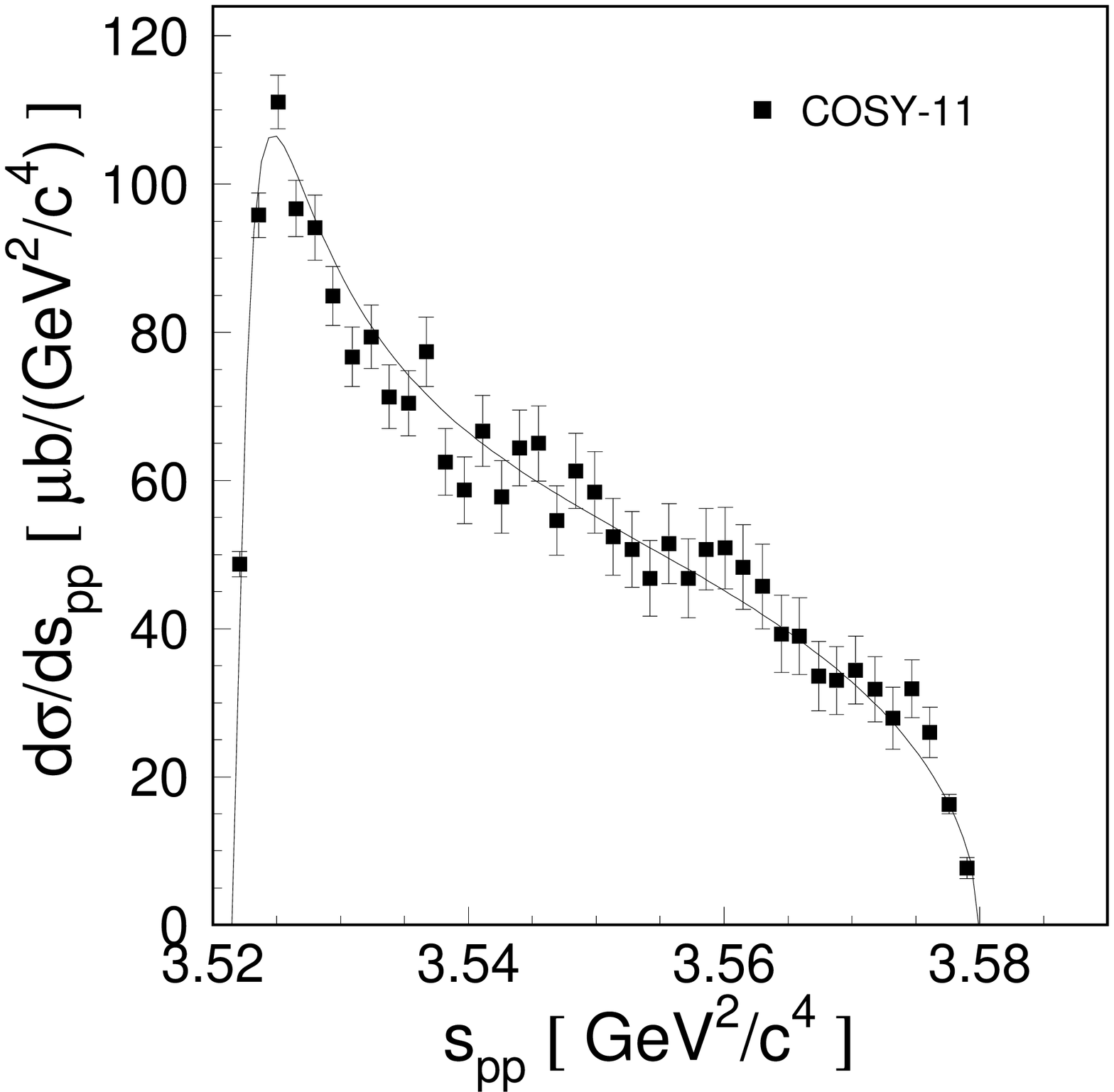}
  \includegraphics[height=.3\textheight]{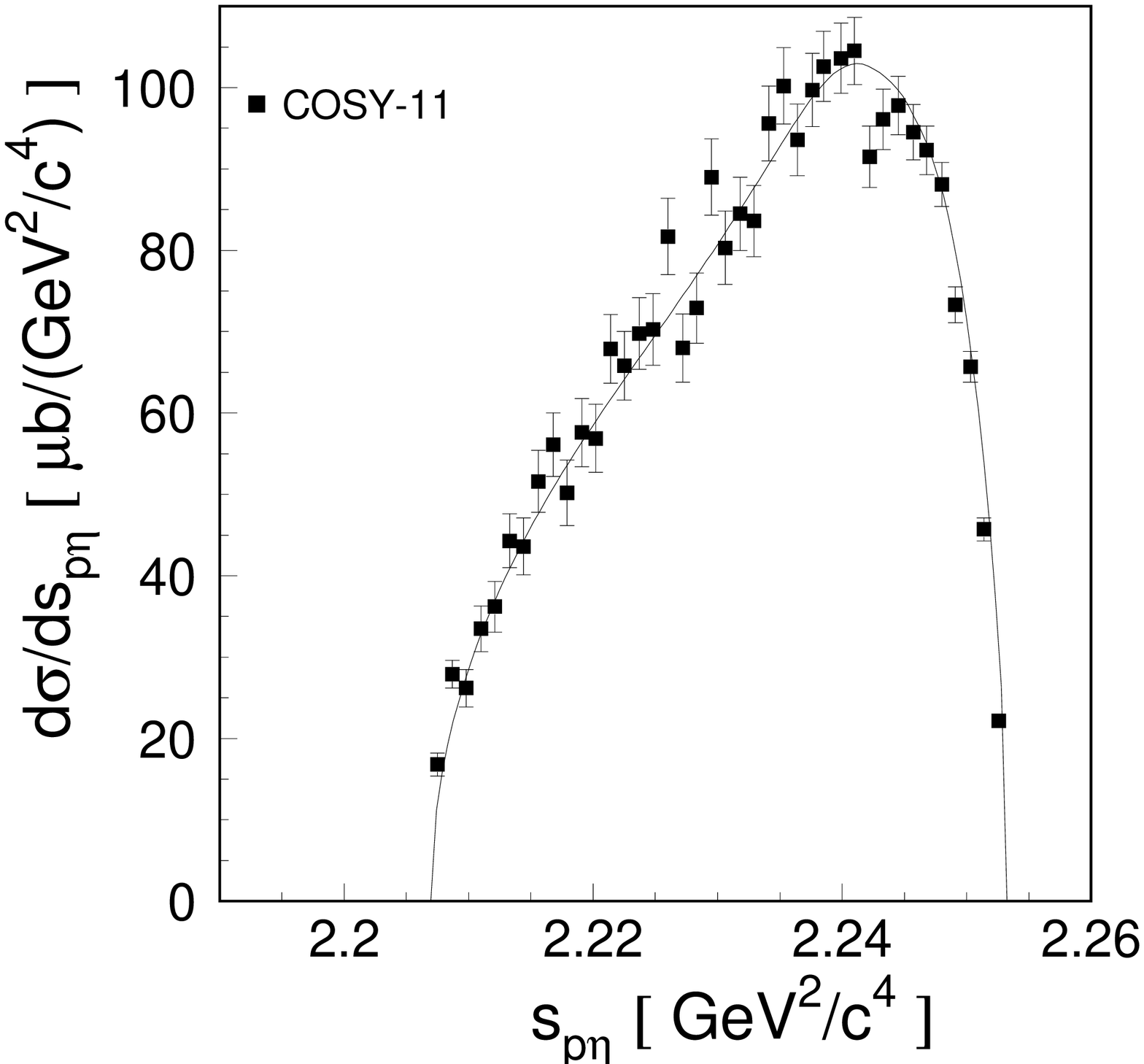}
  \caption{Distributions of the square of the proton-proton ($s_{pp}$) (left) and
  proton-$\eta$ ($s_{p \eta}$) (right) invariant masses determined experimentally for
  the $pp \to pp\eta$ reaction (closed squares). The experimental data are compared
  to calculations performed assuming the linear energy dependence of the production
  amplitude as proposed by A. Deloff \cite{deloff} depicted by solid lines.}
\label{fig:th_deloff}
\end{figure}

On the other hand, the discussed effect can
in principle be assigned to changes of the production amplitude, since in calculations
by V. Baru et al. \cite{baru}, and by K. Nakayama and collaborators \cite{naka} the production amplitude was
nearly constant.\\
An analysis guided by the assumption of a linear energy dependence of the production amplitude
was performed by A. Deloff \cite{deloff}. The squared invariant mass distributions $s_{pp}$ and $s_{p \eta}$
determined for the $pp \to pp\eta$ reaction measured at an
excess energy of Q~=~15.5~MeV are compared to calculations performed by A. Deloff in figure \ref{fig:th_deloff}.\\
The squared invariant mass distributions could be quite well reproduced by lifting the standard on-shell approximation in the
enhancement factor and allowing for a linear energy dependence in the leading
$^{3}P_{0} \to ^{1}\!\!S_{0}s$ partial wave amplitude \cite{deloff}. Those calculations are in contradiction to the
suggestion of Nakayama \cite{naka},
giving evidence that higher partial waves play only a marginal role.\\

At this point, the observed enhancement could be explained by three different hypotheses:\\
i) a significant role of {\bf proton-$\eta$ interaction} in the final state,\\
ii) an {\bf admixture of higher partial waves} or\\
iii) an {\bf energy dependence of the production amplitude}.\\
Based on the $pp \to pp\eta$ data only,
it is not possible to verify any of those models. These contingencies
motivated the work presented in this thesis which is an analysis of a high statistics
$pp \to pp\eta^{\prime}$ reaction measurement in order to determine the distribution
of events over the phase space for an excess energy of Q = 15.5 MeV the same one as
selected before for the $pp \to pp\eta$ reaction.
The comparison of the differential distributions for the proton-proton and proton-meson invariant masses
in the $\eta$ and $\eta^{\prime}$ production could help to judge between postulated explanations
of the observed effect and may allow for a quantitative estimation of the proton-$\eta$ and
proton-$\eta^{\prime}$ interaction.\\
The experimental facility, the method of the analysis, and achieved results for the $pp \to pp\eta^{\prime}$ reaction will be presented in the
following chapters.\\

\chapter{Experimental facility}
\label{Experimental facility}
\markboth{\bf Chapter 3.}
         {\bf Experimental facility}
	
The measurement of the $pp \to pp\eta^{\prime}$ reaction was conducted using the
cooler synchrotron COSY and the COSY-11 detector setup. Both facilities will be
described in this chapter.

\section{Cooler Synchrotron COSY}
\label{Cooler Synchrotron COSY}
\begin{center}
\begin{figure}[H]
\includegraphics[width=0.65\textheight]{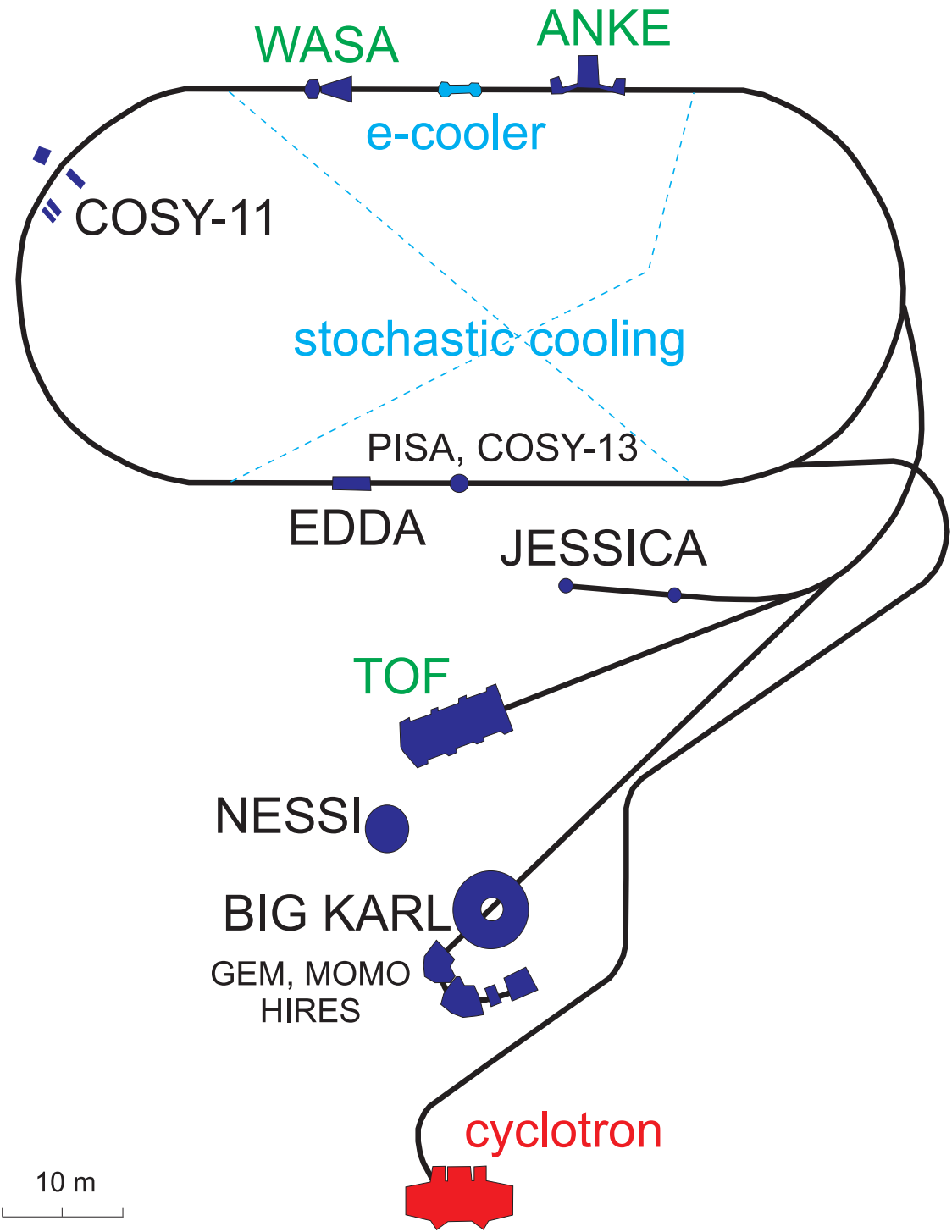}
\caption{
Schematic view of the floor plan of the COSY synchrotron.
Marked in violet are internal-beam \cite{brauksiepe, wasa, pisa, edda1, cosy13, anke} and external-beam \cite{tof,
jessica, nessi, gem, momo, hires} detector setups.
Facilities COSY-TOF, WASA and ANKE
 labelled in green are presently still in operation. The positions of electron and stochastic cooling devices are
 indicated and the stochastic cooling signal lines are
presented by the light-blue dashed lines. }
\label{fig:cosy}
\end{figure}
  \end{center}

The COoler SYnchrotron (COSY) \cite{ring1} is located at the Institute of
Nuclear Physics of the Research Centre J{\"u}lich in Germany.
The facility was designed to accelerate polarized and unpolarized proton and
deuteron beams in the momentum range from 0.3~GeV/c up to 3.7~GeV/c.
The sketch of the whole accelerator complex is presented in
figure~\ref{fig:cosy}.
The total length of the synchrotron ring is 184 meter. There are two straight
40 meter sections, and two bending sections with 24 dipole magnets.\\

The experimental installations at the synchrotron can be classified as two groups;\\
a) the detectors installed inside the COSY ring: WASA \cite{wasa,
wasastrona}, COSY-11 \cite{brauksiepe, klajac11, strona},
PISA~\cite{pisa}, EDDA \cite{edda1}, COSY-13 \cite{cosy13}, and ANKE \cite{anke}
and \\
b) outside of the COSY ring
at external beam lines: COSY-TOF \cite{tof}, JESSICA \cite{jessica}, NESSI \cite{nessi},
GEM \cite{gem}, MOMO \cite{momo}, and HIRES \cite{hires}.  \\
Some of those experiments are already completed and no longer in operation
(labelled in black) and the others are still in operation (labelled in green) in figure~\ref{fig:cosy}.\\

The COSY synchrotron is equipped with electron and stochastic cooling devices
which are used to decrease the momentum and spatial spread of the beam \cite{cooling}.\\
In the case of electron cooling, the electrons with velocities equal to the nominal proton beam
velocity are injected at a straight section of the synchrotron. This operation causes that
faster protons are decelerated and slower ones are accelerated.\\
Stochastic cooling uses an electromagnetic device, the so called pick-up unit, which measures the beam deviation from the
nominal position at one point of the accelerator and corrects it by transmitting a correction signal
through the shortest way to the kicker unit at the other side of the beam pipe. It
causes not only a shift to the nominal beam orbit, but also decreases the spread of the
transversal and longitudinal momentum components \cite{cooling1, cooling2}.

\section{COSY-11 detector setup}
\label{COSY-11 detector setup}
The measurement of the $pp \to pp\eta^{\prime}$ reaction is based on the registration
of the two outgoing protons and reconstruction of their momenta. The $\eta^{\prime}$ meson
is identified using the missing mass technique.

\begin{center}
\begin{figure}[H]
\includegraphics[width=0.63\textheight]{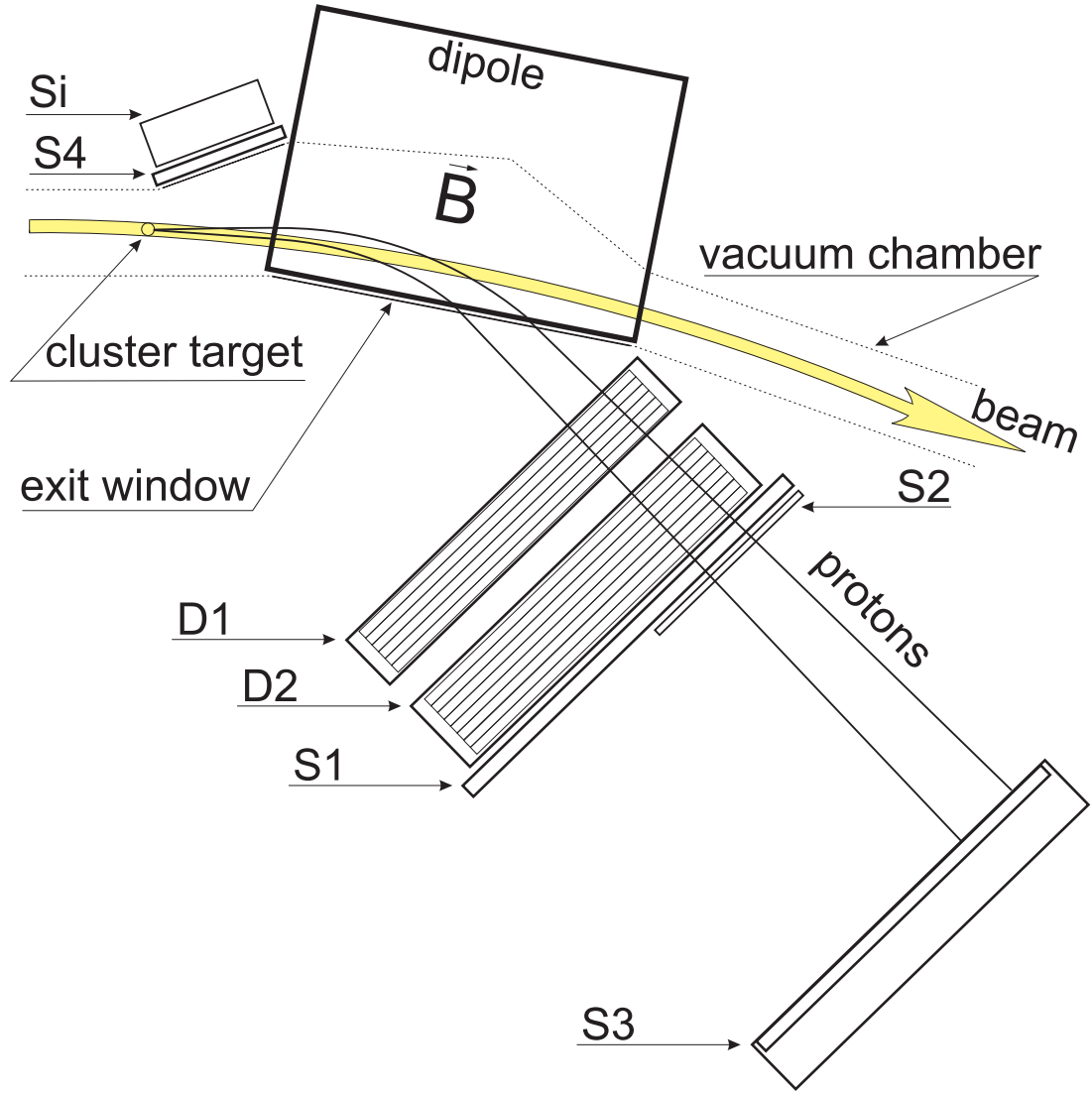}
\caption{
Schematic view of the COSY-11 detection facility \cite{brauksiepe}. Note, that
only those detectors which were used during the measurement of the
$pp \to ppX$ reaction are presented.
Protons originating from the $pp \to ppX$ reaction are bent in the dipole magnetic field,
and leave the vacuum chamber through the exit window. Afterwards they are
detected in the two drift chambers D1 and D2, in the scintillator hodoscopes S1 and S2, and
in the scintillator wall S3. The scintillation detector S4 and the silicon pad detector Si are
used in coincidence with the D1, D2 and S1
detectors for the registration of the elastically scattered protons.}
\label{fig:setup}
\end{figure}
  \end{center}
The COSY-11 facility is one of the internal detector setups installed inside
the COSY synchrotron tunnel at a bending section of the ring.
It is mounted next to one of the dipole magnets,
and benefits from the dipole magnetic field which is used for the particle separation from the beam.\\

A schematic view of the COSY-11 apparatus is presented in figure \ref{fig:setup}.
The figure illustrates also schematically the tracks of protons outgoing from
the $pp \to ppX$ reaction. Two outgoing
protons possessing smaller momenta than the beam momentum, are bent in the dipole magnetic field
towards the detector system.
They leave the vacuum chamber throughout the exit window made out of a 30 $\mu$m layer of aluminum
and 300 $\mu$m of a carbon fiber carrier material with an averaged density of
2.1 g/cm$^{3}$~\cite{brauksiepe} and
are detected using the drift chambers D1 and D2, the scintillator hodoscopes S1 and S2, and the scintillation 
wall S3\footnote{In the measurement of the meson production in the quasi-free $pd \to pnp_{spectator}X$
reaction dedicated neutron \cite{neut, neut1, neut2, neut3} and spectator \cite{spec, jpg, spec1}
detectors were installed in addition. They were however
not used for the measurement described in this thesis.}.\\

The target\footnote{The $H_{2}$ cluster target specifications are described in the references \cite{target, target1}.
The dimensions of the cluster target used during the measurement are described in details in chapter
\ref{Determination of the spread and the absolute value of the beam momentum}. }
used during the experiment,
was realized as a beam of $H_{2}$ molecules grouped inside clusters of up to about $10^{6}$~atoms.
The average density of the target was around $5 \cdot 10^{13}$ atoms/cm$^{2}$ \cite{target1}.
It was installed in front of the
dipole magnet as it can be seen schematically in figure \ref{fig:setup}.

The drift chambers D1 and D2 were used for the determination of the particles trajectories.
Those two planar drift chamber stacks are spaced by 70 cm \cite{brauksiepe, chambers}.
Their active area is 1680 mm wide and 433 mm high.
Drift chamber D1 (standing closer to the bending magnet) consists of six detection planes. The first two with vertical wires,
two with wires inclined by $+31^{o}$ and two inclined by $-31^{o}$. The D2 drift chamber is built in the same scheme, but
it is extended by two additional planes with vertical wires.

The wires in adjacent planes of
each pair are shifted by half of the cell width to resolve the left-right position ambiguity
with respect to the sense wire. The chosen configuration of the detection planes allows to
perform the measurement of the horizontal and vertical coordinates and enables a unique multi-hit event
identification \cite{chambers}.

A charged particle crossing
the drift chambers produces gas ionization inside the drift cells, filled with a gas
mixture of one to one argon and ethane at atmospheric pressure.
The electron drift time to the sense wire is a measure of the distance between the passing particle
track and the sense wire (see section \ref{Space-time relation for drift chambers}).
In the case of particle tracks oriented
perpendicular to the detection planes, the maximum drift time corresponding to
the maximum drift path of 20 mm equals to 400 ns.\\

Determined particle trajectories in the data analysis are traced through the magnetic field of the dipole back to the
target. Therefore, it is possible to reconstruct the momentum vectors of outgoing particles at the
reaction point. The reconstruction of the momentum vectors of the
registered particles combined with the information about the time-of-flight between the S1 and the S3
detectors allows for the calculation of the particle mass and by this the particle type identification. \\

The S1 scintillating hodoscope is built out of sixteen identical, vertically arranged modules, read out
from both sides (top and bottom) by photomultipliers. The modules with $45 \times 10 \times 0.4~ $cm$^{3}$
dimensions are arranged
with small vertical overlap ($1~$mm \cite{brauksiepe, S1}) in order to avoid "not covered" space in the
geometrical acceptance. The S1 detector is used as the "start" for the time-of-flight measurement.

The S2 scintillating hodoscope, similar as S1, consists of sixteen scintillation modules
with the dimension of  $45 \times 1.3 \times 0.2~ $cm$^{3}$ \cite{brauksiepe}.

The S3 scintillating detector delivers the "stop" information for the time-of-flight measurement.
It is built from one non-segmented scintillating wall with the dimension of $220 \times 100 \times 5~ $cm$^{3}$.
It is viewed by a matrix of 217 photomultipliers \cite{brauksiepe, S3, moskalphd}, occupying the edges of
equilateral triangles with the sides of 11.5~ cm.  \\

The S4 scintillation counter together with the silicon pad detector (depicted in figure~\ref{fig:setup} as Si)
are used for registration of the recoil protons from the proton-proton elastic scattering \cite{brauksiepe}.
The silicon pad detector \cite{brauksiepe} consists of 144 pads with dimensions of  $22.0 \times 4.5 \times 0.28~ $mm$^{3}$.
Each pad is read out separately.

\section{Trigger logic}
\label{Trigger logic}
In the experiment two independent trigger branches were used, in order to
detect the $pp~\to~pp\eta^{\prime}$ and $pp~\to~pp$ reactions.\\
The main trigger used for the $pp~\to~pp\eta^{\prime}$ reaction was based on the following conditions:
\begin{equation}
T_{pp \to pp\eta^{\prime}} = (S1_{\mu \ge 2}\lor S1_{\mu = 1, high}\lor S2_{\mu \ge 2}\lor S2_{\mu = 1, high}) \land S3_{\mu \ge 2},
\label{equ:trigg}
\end{equation}
where $\mu$ denotes the multiplicity of segments in the S1 and S2 scintillation hodoscopes, and the
number of photomultipliers which have fired in the S3 detector. The subscript $high$ stands for
a high amplitude signal in the S1 and S2 detectors which was implemented
for triggering events when two particles cross the same segment \cite{moskalphd}.
The hardware threshold for the high amplitude was
set high enough to reduce the
number of single particle events considerably, and low enough to accept
most events (almost~100~$\%$) with two protons passing through one segment of S1 \cite{moskalphd}.\\
The trigger used for the selection of elastic scattering reactions was based on the coincidence
between signals from the S1 and S4 detectors:
\begin{equation}
T_{pp \to pp} = S1_{\mu = 1}\land S4,
\label{equ:trigg_elas}
\end{equation}
where the S1 hodoscope was used for the registration of
forward scattered protons and the S4 scintillation detector
was registering recoil protons.\\
The detectors were positioned to cover a large part of the kinematics of the $pp \to pp$ elastic scattering. Due to the
high rate of the $pp \to pp$ reaction only every 128'th event was registered for the further
analysis. The number of the elastic scattering events was later used for the
luminosity determination (see chapter \ref{Luminosity determination}).

\chapter{Calibration of the detector setup}
\label{Calibration of the detector setup}
\markboth{\bf Chapter 4.}
         {\bf Calibration of the detector setup}

In this chapter the method used to calibrate the COSY-11 detectors 
  and their relative settings will be presented. In particular, the time-space 
  relation of the drift chambers and the procedure of time-of-flight calibration will be described. 
 In addition, the procedure of monitoring the relative beam-target setting will be discussed.

\section{Space-time relation for drift chambers}
\label{Space-time relation for drift chambers}
\begin{figure}[H]
  \includegraphics[height=.3\textheight]{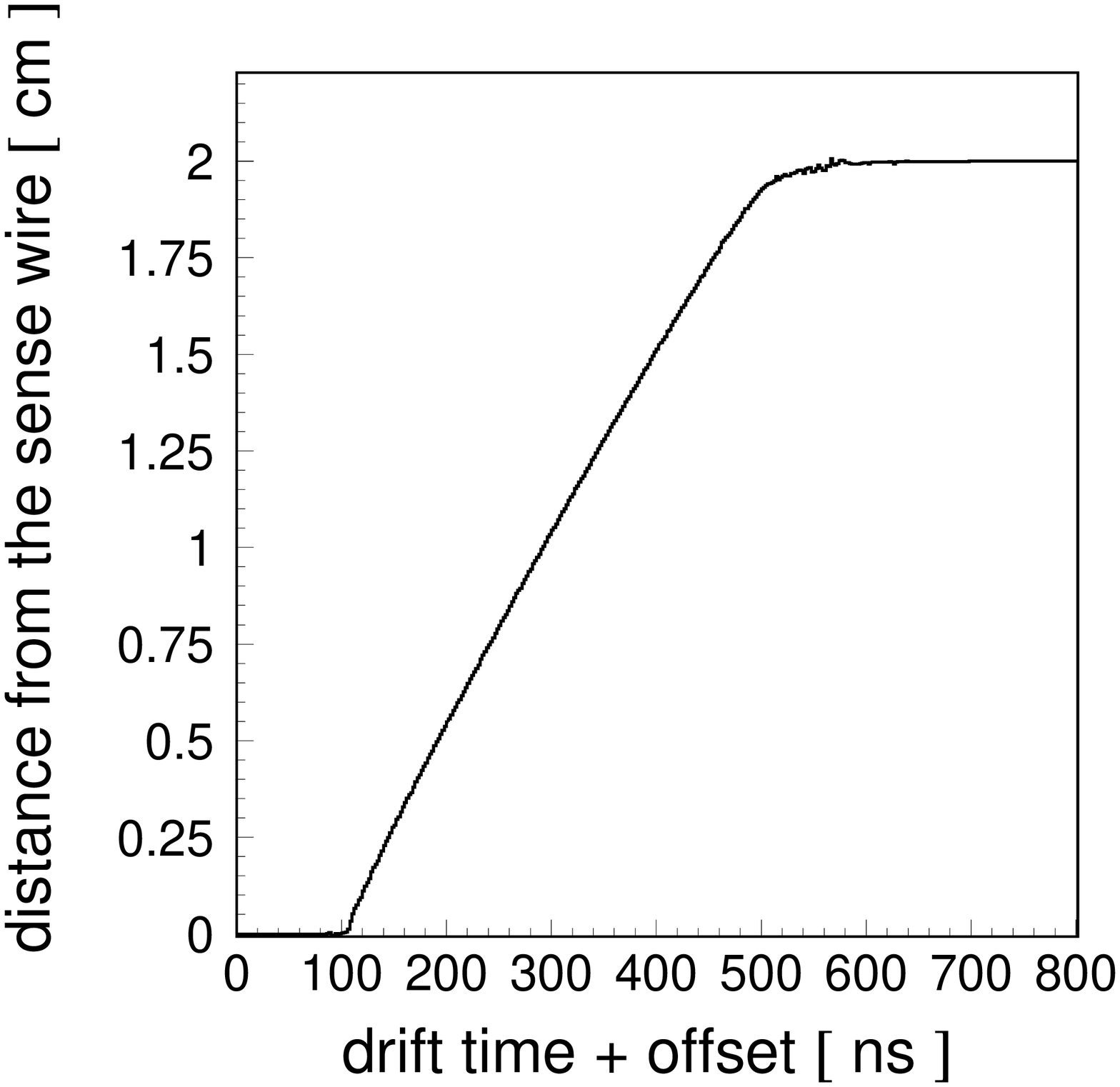} 
  \includegraphics[height=.3\textheight]{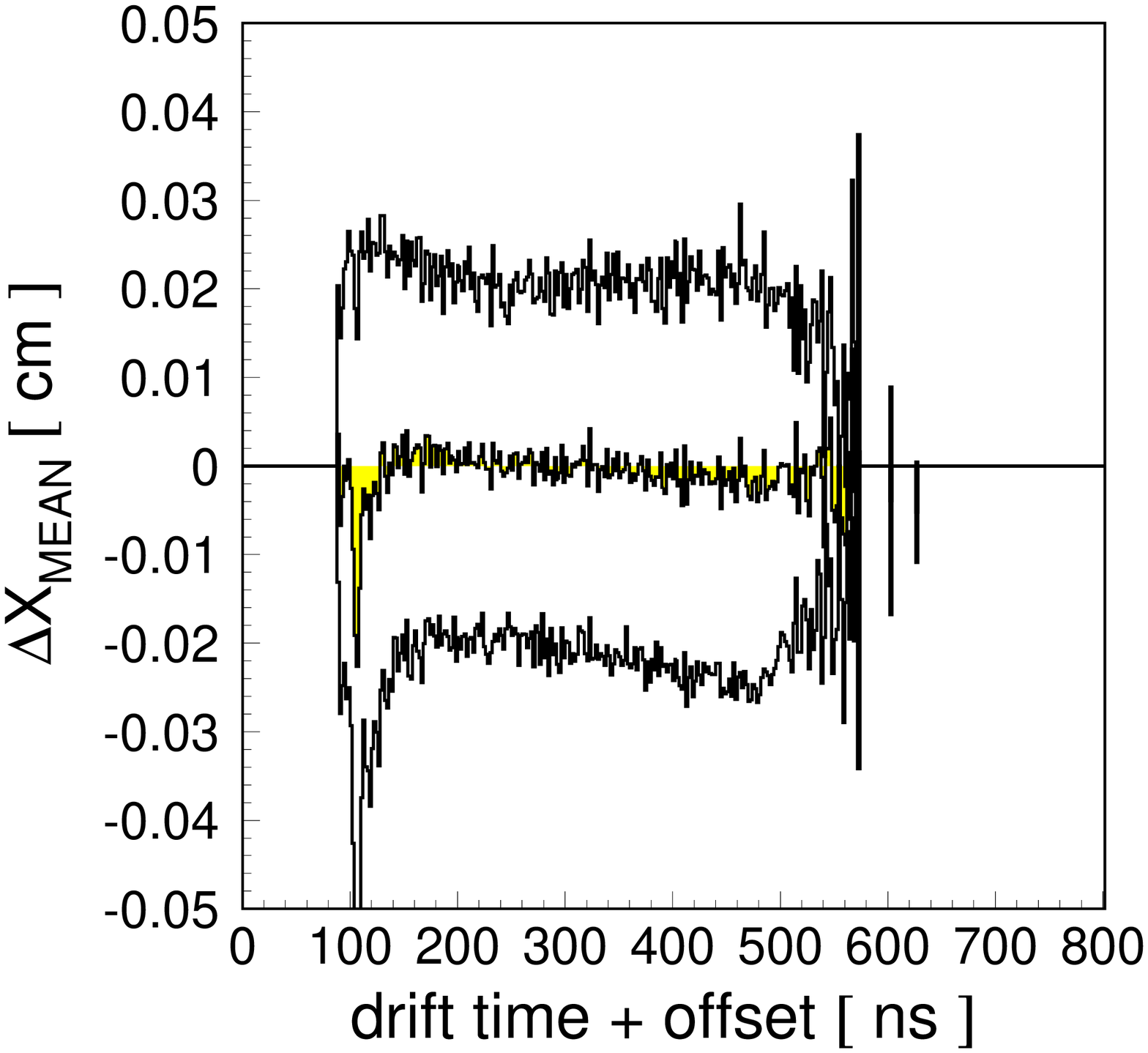} 
  \caption{(Left panel) Time-space calibration - 
  relation between the distance from the sense wire and the drift time. 
  (Right panel) Corrections of the time-space relation, where the middle histogram represents the 
  mean value of $\Delta$$X$, 
  the upper and lower histograms visualize one standard deviation ($\sigma$) 
  of the $\Delta$$X$ distribution. See the text for the explanation of $\Delta$$X$.}
\label{fig:t_s_cal}
\end{figure}

The drift chambers D1 and D2 consist of 6 and 8 planes of wires, respectively. 
They provide the information about the drift time of electrons (to the sense wires) produced along 
the trajectory of charged particles passing through the chambers. 
In order to reconstruct those trajectories one needs to establish a relation 
between drift time and distance between the particle track and the sense wire (Fig. \ref{fig:t_s_cal} (left)). 
Due to the sensitivity of the drift velocity to the atmospheric pressure, humidity and gas mixture changes \cite{thomas}, 
the data 
used for the calibration procedure were divided into time intervals of about 3-8 hours, with a similar number of collected events.  
The calibration function was determined for each interval applying the procedure of iterative improvements \cite{moskalphd}. 
Starting with an approximated function of the space-time relation, the distance X 
from the particle trajectory to the sense wire has been calculated. Then, a straight line to the obtained 
points was fitted and further on, assuming that 
it corresponds to the real particle track, the deviation $\Delta$$X$ , 
between the measured distance of the particle track from the sense wire and the one from the fitting procedure 
was calculated. 
Next, having a certain amount of data, one could determine a mean value of $\Delta$$X$ as a function 
of the drift time (presented in figure \ref{fig:t_s_cal} (right)). The $\Delta$$X_{MEAN}$ was subsequently used for a 
correction of the time-space relation. Next, the improved function has been used 
for the track reconstruction similarly as in the first step. The whole procedure was repeated until 
the corrections were smaller than the statistical uncertainty of the $\Delta$$X_{MEAN}$.\\
The averaged spatial resolution of the drift chambers achieved in the experiment discussed in this thesis 
amounted to  250 $\mu$m (rms).\\
  
\section{Time-of-flight calibration of scintillator detectors} 
\label{Time-of-flight calibration of scintillator detectors}  
\begin{center} 
\begin{figure}[H]
 \includegraphics[height=.35\textheight]{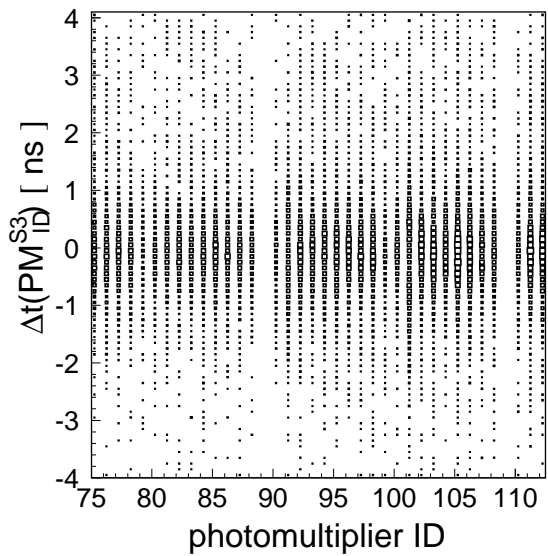}
  \caption{Distribution of the time difference between the time $tof_{rec}$ calculated from the reconstructed particle momentum 
  and the measured time $tof_{S1-S3}$ 
  between the S1 detector and a particular photomultiplier in the S3 detector, 
  as a function of 
  the photomultiplier ID in the S3 detector. As achieved after second iteration, the figure depicts only a fraction of PM's 
  of the S3 detector (75-112).}
\label{fig:s3_off}
\end{figure} 
\end{center}

The scintillator detectors S1 and S3 "start" and "stop", respectively, are 
used for the time-of-flight measurement. S1 consists of 16 scintillator plates 
with photomultiplier readout from both sides, and S3 is a scintillator wall
read out by a 217 
photomultipliers matrix. In order to obtain the proper information about the 
time-of-flight between both detectors, one needs to determine time "offsets"  
for all photomultipliers i.e. the relative differences in transition time of the signal  
from the photomultiplier to the TDC unit.\\
Let us denote $t_{S1}$ as the real time when a particle crosses the S1 detector and $t_{S3}$ when it 
crosses S3. 
Then, the time-of-flight can be calculated as follows: $tof_{S1-S3} = t_{S3}-t_{S1}$.\\ 
The measured TDC values for a single photomultiplier in S1 and S3 detectors read:
\begin{equation}
TDC_{S1}=t_{S1}+t_{y}+t^{S1}_{walk}(PM)+t^{S1}_{offset}(PM)-t_{trigger},
\label{equ:tdc_s1}
\end{equation}
\begin{equation}
TDC_{S3}=t_{S3}+t_{pos}+t^{S3}_{walk}(PM)+t^{S3}_{offset}(PM)-t_{trigger}.
\label{equ:tdc_s3}
\end{equation}
In both equations the time stamp $t_{trigger}$ (denoting the time of the trigger
signal) is identical.  
The index $y$ corresponds to the distance between 
hit position and the edge of the scintillator close to the given photomultiplier in the S1 detector 
and $pos$ stands for the distance between the 
hit position and the photomultiplier in the S3 detector. 
The abbreviation $t_{walk}$ denotes the corrections for the $time-walk$ effect, i.e. the signal time dependence on the 
signal amplitude \cite{leo}. Any dependence of $t_{y}$ is cancelled by taking the 
 average between the times measured by the upper and lower photomultipliers \cite{moskalphd} 
 and this can be calculated from the known trajectories. 
Thus, the only unknown variables are the time offsets $t_{offset}$ for both detectors. 
For a first approximation, the time difference in the S1 detector can be achieved by 
taking into account signals from the particles crossing the overlapping parts of the modules.
Next, for the S3 detector the time offset can be extracted from the comparison between 
$tof_{S1-S3} = t_{S3}-t_{S1}$ and the time-of-flight calculated from the particle momentum reconstructed 
via curvature in the magnetic field ($tof_{rec}$). 
Then, iteratively, using the obtained S3 offset one can determine 
the time offsets for the S1 detector. After two iterations, the time offsets 
for both detectors can be established.  
{\mbox{As~an~example,~the~distribution 
$\Delta t(PM^{S3}_{ID})$ determined as }}
\begin{center}
$\Delta t(PM^{S3}_{ID})=tof_{S1-S3}(PM^{S3}_{ID})-tof_{rec}$
\end{center}
for a group of 
photomultipliers of the S3 detector are presented in figure \ref{fig:s3_off}. 
The time offsets for the photomultipliers in the S3 detector are obtained on the basis of the 
time differences between $tof_{S1-S3}$ and $tof_{rec}$ presented in figure \ref{fig:s3_off}. 
They were adjusted such that this difference is equal to zero.\\
 
\section{Monitoring of relative beam-target settings}
\label{Monitoring of relative beam-target settings}  
\begin{center}  
\begin{figure}[H]
\includegraphics[width=.65\textheight]{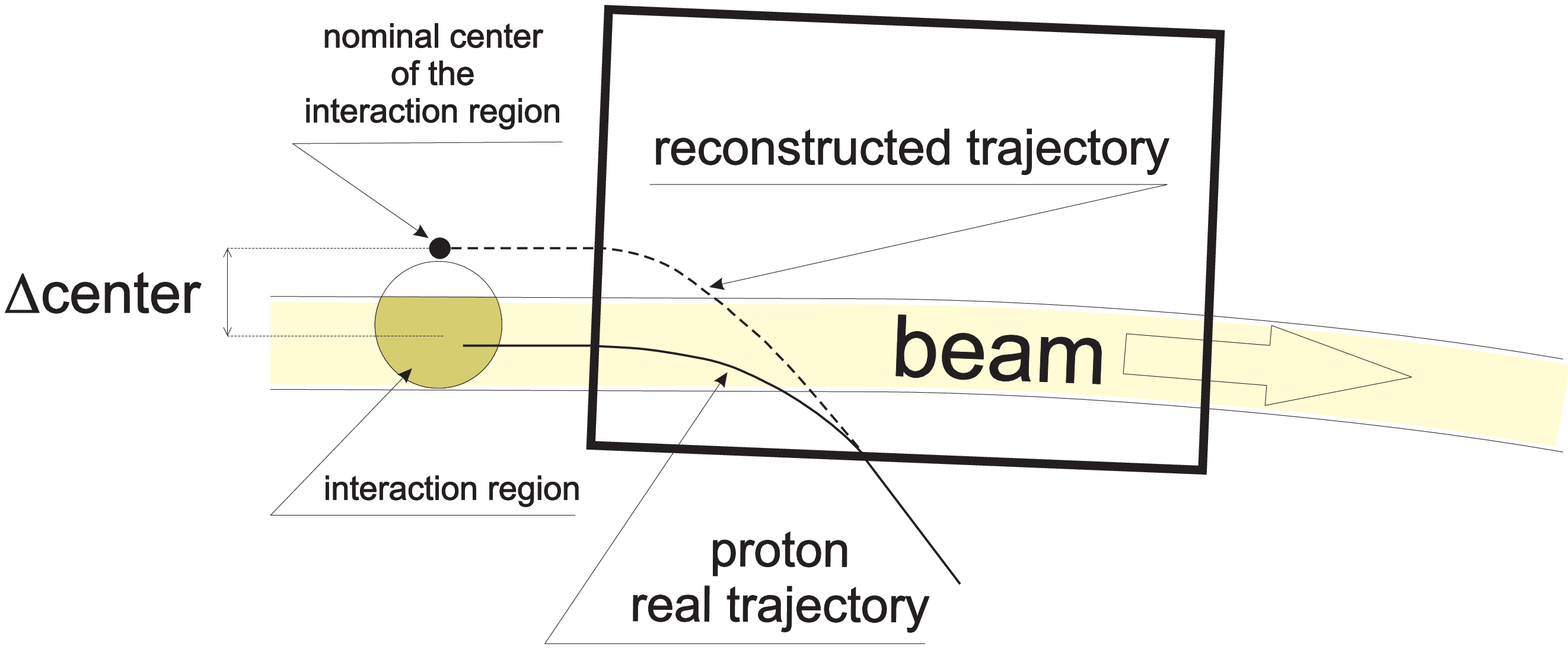}
\caption{Pictorial definition of the deviation of the center of the interaction region
from its nominal position ($\Delta center$).}
\label{fig:delta_center}
\end{figure}
\end{center}

\begin{center}
\begin{figure}[H]
\includegraphics[width=0.63\textheight]{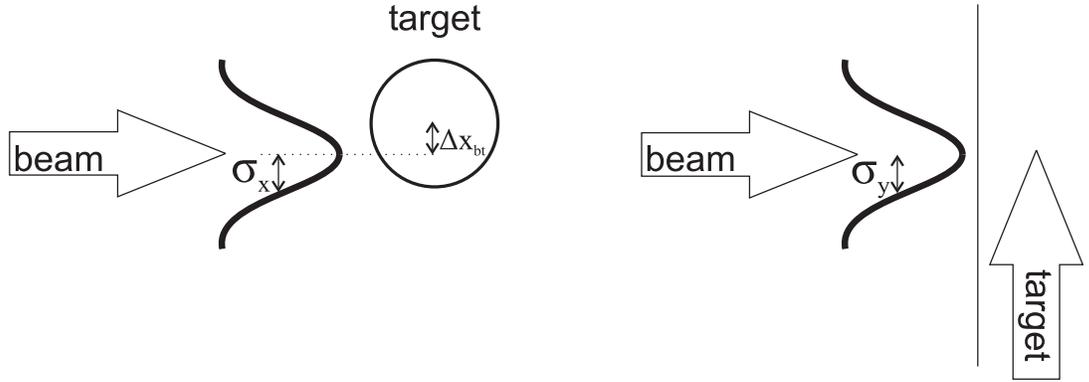}
\caption{Schematic view of the relative target and beam settings. Left panel depicts the view from above,
right presents a side view. $\sigma_{X}$ and $\sigma_{Y}$ denote the horizontal and vertical standard deviations of the assumed
Gaussian distributions of the proton beam density, respectively. $\Delta_{X_{bt}}$ denotes the distance
between the centers of the proton target and beam. The figure is adapted from \cite{monitoring}.}
\label{fig:b_t}
\end{figure}
\end{center}

Possible changes of the position where the beam crosses the target could have
significantly influenced the momentum reconstruction and as a consequence
could worsen the determination of the mass of the undetected particle. 
Therefore, it is important to monitor the position of the 
beam and target overlap.
The center of the beam-target overlap can be determined from the momentum distribution of the elastically scattered 
protons~\cite{monitoring}.
\begin{center}  
\begin{figure}[H]
\includegraphics[width=.35\textheight]{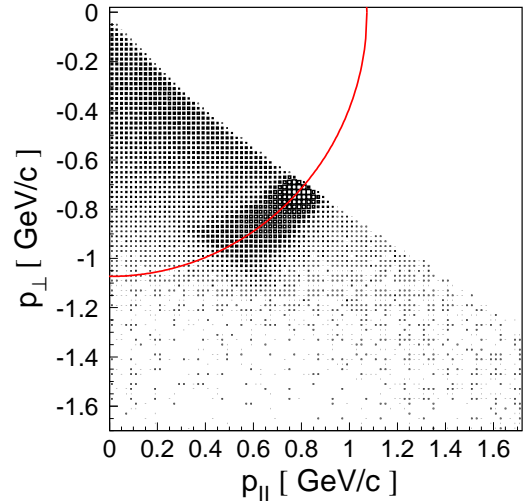}
\caption{The distribution of the perpendicular $p_{\perp}$ versus parallel
               $p_{\parallel}$ momentum components for $pp \to pp$ elastic
               scattering events at a beam momentum of 3.260 GeV/c. The
               solid line corresponds to the kinematical ellipse. 
	       Changes of the event density along the kinematical ellipse 
	       reflect the angular dependence of the cross section for the $pp \to pp$ reaction.}
\label{fig:p_p}
\end{figure}
\end{center}
The mean value of the distance between the expected kinematical ellipse
and the experimental points (shown in figure \ref{fig:p_p}) may be used
as a measure for the deviation of the center of the interaction region
from its nominal position ($\Delta center$). 
A pictorial definition of $\Delta center$ is presented in figure \ref{fig:delta_center} 
and the beam-target geometrical conditions are depicted in figure \ref{fig:b_t}. 
By assuming a wrong interaction center the reconstruction results 
in a wrong momentum determination and the $pp \to pp$ events are not centered around the expected kinematical ellipse. 

\begin{figure}[H]
\includegraphics[width=.33\textheight]{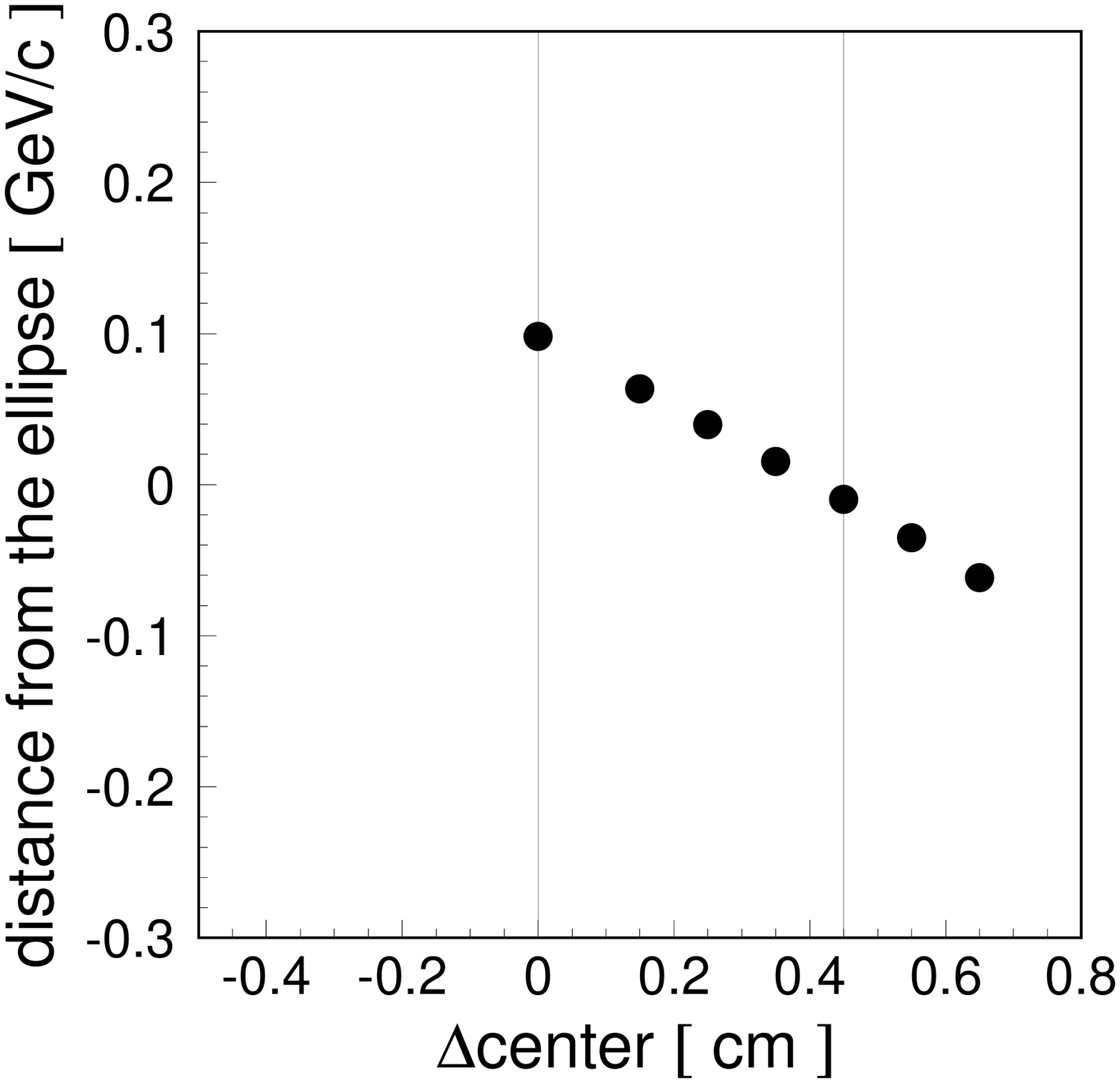}
\includegraphics[width=.33\textheight]{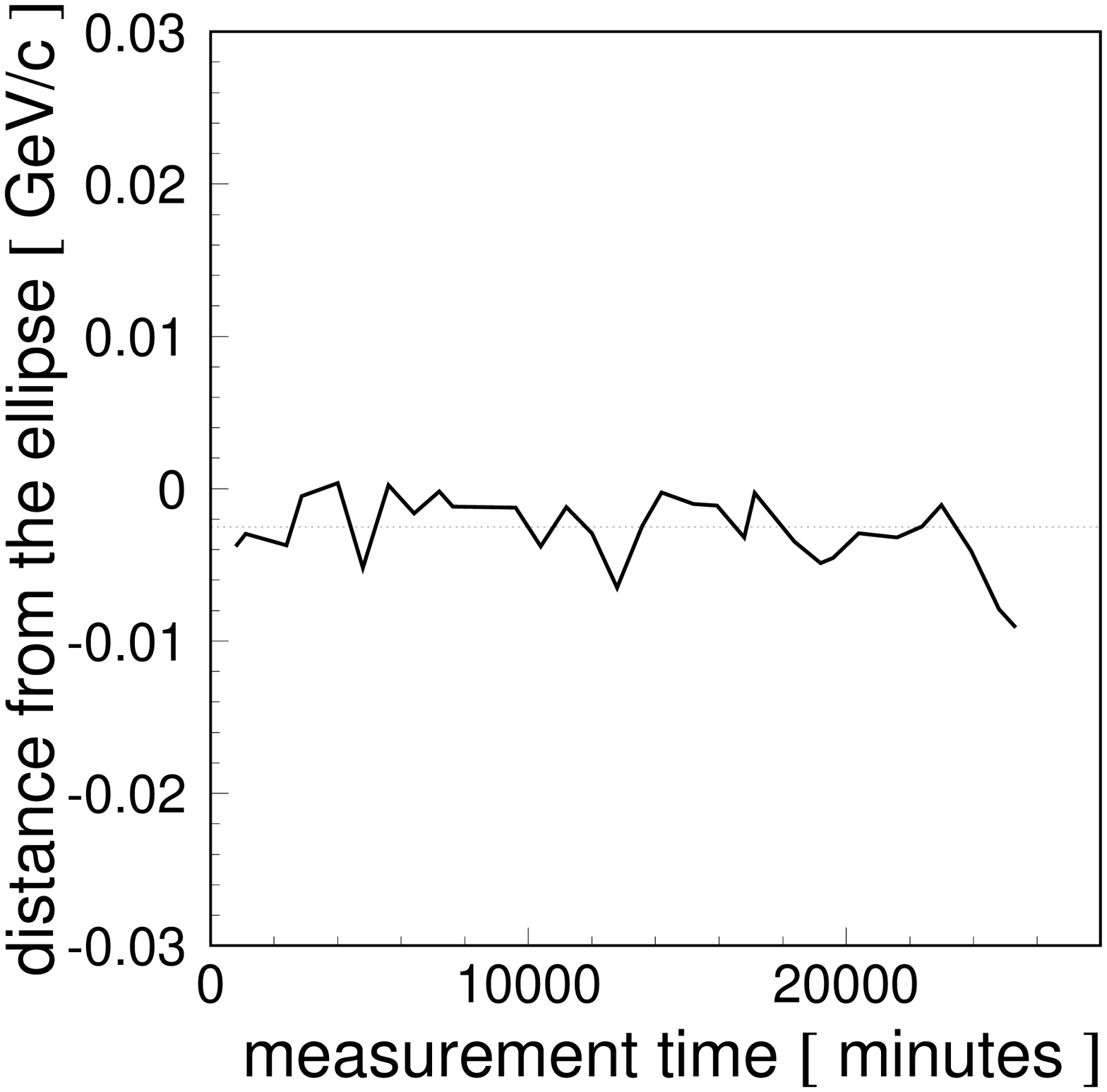}
\caption{(Left panel) The distance between the expected ellipse and the center of
         experimental distribution on the ($p_{\perp}$,$p_{\parallel}$) plot (figure~\ref{fig:p_p})
         versus the deviation of the center of the interaction region
from its nominal position ($\Delta center$). 
(Right panel) The deviation of the distance from the kinematical ellipse as a function
         of the time of the measurement. The mean value of the distance from the ellipse has been plotted
         for 13 hours intervals. In the analysis the value of $\Delta center$ was set to 0.45 cm.}
\label{fig:dist_beam}
\end{figure}

In the left panel of figure \ref{fig:dist_beam} the mean distance of experimental 
$pp \to pp$ events from the expected kinematical ellipse is shown as a function 
of $\Delta center$ assumed in the analysis. As can be seen, 
the center of the interaction region differs by 0.45 cm from the nominal one.     
In the right panel of figure \ref{fig:dist_beam} it is demonstrated that
the beam-target conditions were stable during the course of the experiment. Fluctuations
seen in the figure are within the statistical error in the determination of the
mean value of the distance from the ellipse. The variations are at a level of $10^{-3}$
and, as can be inferred from the plot presented in the left panel of figure \ref{fig:dist_beam},
correspond to shifts of the interaction 
center by less than 0.01 mm. Thus, the variations of the center of the interaction region 
can be safely neglected in the further analysis.

\chapter{Identification of the $pp \to pp\eta^{\prime}$ reaction}
\label{Identification of the reaction}
\markboth{\bf Chapter 5.}
         {\bf Identification of the $pp \to pp\eta^{\prime}$ reaction} 
 
In the following chapter the method of identifying the $pp \to pp\eta^{\prime}$ reaction will be described.

\section{Identification of protons}
\label{Identification of protons}
The measurement was based on the registration of two outgoing protons originating from the $pp \to ppX$ reaction. 
\begin{center}
\begin{figure}[H]
  \includegraphics[height=.45\textheight]{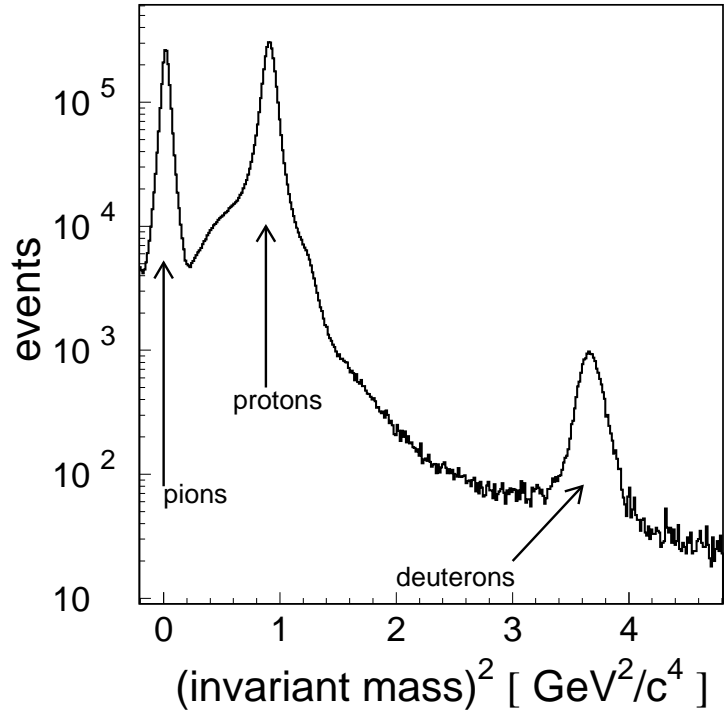}
  \caption{Distribution of the squared invariant mass of the registered particles. (Note the logarithmic scale on vertical axis.) 
  Signals from measured pions, protons, and deuterons are easily distinguished.}
\label{fig:invm}
\end{figure}
\end{center}
After choosing only two-track events, the protons were identified by the determination of their 
rest masses. The particle mass was calculated according to the following formula:
\begin{equation}
  m^{2}=\frac{\vec{p}\ ^{2}(1-\beta^{2})}{\beta^{2}},
\label{equ:invmass}
\end{equation}
where $\vec{p}$ and $\beta$ are denoting the momenta and velocities of particles, respectively, which were determined 
in an independent way  ($\vec{p}$ from the curvature of the trajectory in the dipole and 
$\beta$ from the time of flight between S1 and S3). The distribution of the squared masses of the particles   
is shown in figure \ref{fig:invm}. Clearly visible are signals from pions, protons and deuterons. \\
For the further analysis, particles with reconstructed masses in the range from 0.2 to 1.5~GeV$^{2}$/c$^{4}$ 
were assumed to be protons.\\

\section{Identification of the $\eta^{\prime}$ meson}
\label{Identification of the meson}
In the present experiment the decay products of the $\eta^{\prime}$ meson were not measured, therefore it was 
impossible to identify its production on an event-by-event basis. Even in experiments detecting 
all decay products an unambiguous identification of a single "$\eta^{\prime}$ production event" 
is not possible, but the background would be much smaller.
\begin{center}
\begin{figure}[H]
  \includegraphics[height=.45\textheight]{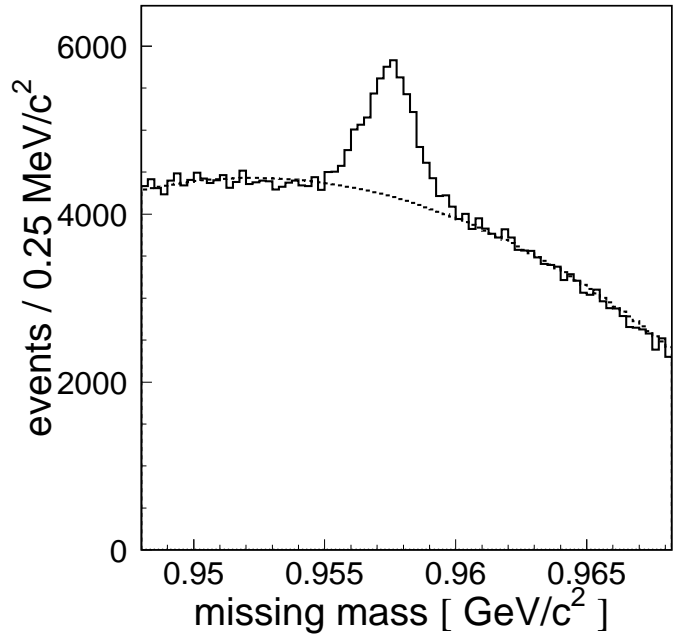}
  \caption{Missing mass spectrum for the $pp \to ppX$ reaction measured at 
  a beam momentum of $P_{B}=$ 3.260 GeV/c. The peak originating from the $pp \to pp\eta^{\prime}$ reaction 
  is clearly seen on top of a multi-pion production background. The dashed 
	line corresponds to a fit with a second order polynomial to the data outside the signal from the
	$\eta^{\prime}$ meson.}
\label{fig:mm}
\end{figure}
\end{center} 
The number of $pp \to pp\eta^{\prime}$ events was determined using the 
missing mass technique. This method is based on the knowledge 
of the protons four-momenta before and after the reaction. Denoting: 
$\mathbb{P}$$_{b}=(E_{b},\vec{p}_{b})$, $\mathbb{P}$$_{t}=(E_{t},0)$, $\mathbb{P}$$_{1}=(E_{1},\vec{p}_{1})$ and 
$\mathbb{P}$$_{2}=(E_{2},\vec{p}_{2})$ as the four-momenta of 
the proton beam, proton target, and first and second outgoing proton, respectively, 
one can use the following formula, in the case 
of the $pp \to ppX$ reaction, to calculate the mass $m_{X}$ of the unregistered particle:\\
\begin{center}
$m^{2}_{X}=E^{2}_{X}-\vec{p}\ ^{2}_{X}=( \mathbb{P}$$_{b}+  $$\mathbb{P}$$_{t}- $$\mathbb{P}$$_{1}- $$\mathbb{P}$$_{2})^{2}=$
\end{center}
\begin{equation}  
  =(E_{b}+E_{t}-E_{1}-E_{2})^{2}-(\vec{p}_{b}+\vec{p}_{t}-\vec{p}_{1}-\vec{p}_{2})^{2}.
\label{equ:missmass}
\end{equation}
In figure \ref{fig:mm} the missing mass spectrum determined experimentally for the $pp~\to~ppX$ reaction for the 
whole data sample is presented. The spectrum includes a broad distribution 
from multi-meson production and the well defined peak originating from the $\eta^{\prime}$ meson production.\\
The smooth behaviour of the experimental multi-pion production 
background, which could be verified by Monte Carlo simulations studies (see section \ref{Background subtraction}), 
allows for a simple polynomial fit. The knowledge of the smooth behaviour of the cross section~\cite{jpg, marcin},  
assures that in the range of the signal, the multi-pion background should be flat.\\
The dashed line in figure \ref{fig:mm} corresponds to a second order polynomial
fitted to the experimental background. Indeed, it can be seen that the fit reproduces 
the shape of the background satisfactory well.\\
\begin{center} 
 \begin{figure}[H] 
  \includegraphics[height=.45\textheight]{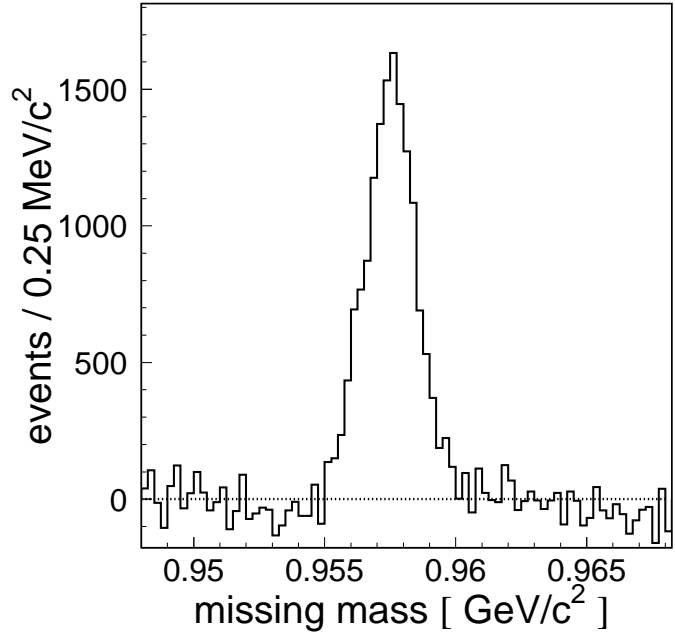}
        \caption{Result of the background subtraction using the second order polynomial fit presented in figure \ref{fig:mm} as a dashed line.}
\label{fig:mm_pol}		
\end{figure} 
\end{center} 
In figure \ref{fig:mm_pol} the experimental missing mass spectrum after the background subtraction is presented.  
The background was approximated by a second order polynomial fit as depicted in figure \ref{fig:mm} by a dashed line. 
The total number of registered and reconstructed $pp \to pp\eta^{\prime}$ reactions amounts to about 15000.\\
Here, the statistics achieved in the measurement is only illustrated and 
the possibility of the background determination is shown. A detailed discussion 
of the subtraction of the multi-pion production background  
for differential cross sections will be comprehensively described in section \ref{Background subtraction}.\\

\chapter{Luminosity determination}
\label{Luminosity determination}
\markboth{\bf Chapter 6.}
         {\bf Luminosity determination}

\begin{center}  
\begin{figure}[H]
\includegraphics[width=0.63\textheight]{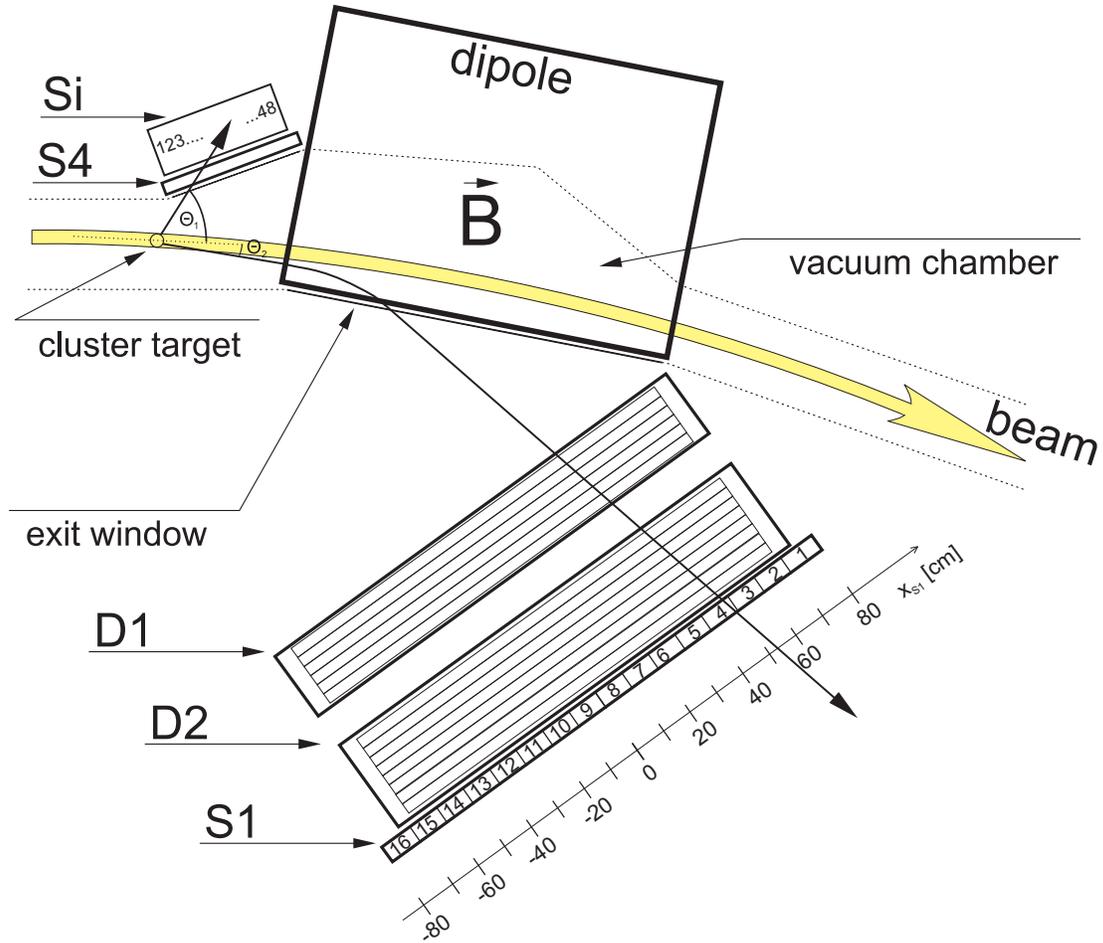}
\caption{Schematic view of the COSY-11 apparatus presenting the detectors used for the registration of the 
$pp \to pp$ elastic scattering. 
The superimposed lines show an example of trajectories from elastically scattered protons with the laboratory angles 
$\theta_{1}$ and $\theta_{2}$. 
One can compare this figure to figure \ref{fig:setup}. 
The $x_{S1}$ axis indicates the size of detector 
modules (for further description see text).}
\label{fig:setup_elas}
\end{figure}  
  \end{center}

In order to determine the absolute values of the differential cross sections, the 
luminosity ($L$) integrated over the measurement time has to be established.
For that purpose, the analysis of the $pp \to pp$ reaction, in 
order to establish the number of elastic scattering events was performed.\\
A schematic view of the COSY-11 detector setup with superimposed
tracks of elastically scattered protons is shown in figure \ref{fig:setup_elas}.
One can evaluate the luminosity ($L$) according to the formula:
\begin{equation}
\frac{\Delta N(\theta^{*}_{2})}{\Delta\Omega^{*}(\theta^{*}_{2})}~=~
\frac{d \sigma^{*}}{d \Omega^{*}}(\theta^{*}_{2})~\cdot~L,
\label{equ:cross}
\end{equation}
where $\frac{d \sigma^{*}}{d \Omega^{*}}$$(\theta^{*}_{2})$ denotes the known differential cross section \cite{edda} and 
$\Delta N$$(\theta^{*}_{2})$ indicates the number of elastically scattered protons at a solid 
angle $\Delta\Omega^{*}$ around the proton emission angle $\theta^{*}_{2}$ in the centre-of-mass system. 
In the further analysis, the available range of the $\theta^{*}_{2}$ angle ($44^{o}$~to~$66^{o}$) was divided  
into 11 bins with a width of $2^{o}$. The tracks of elastically scattered protons 
resulting in signals in the S1 detector with a coincident signal in S4 from the second proton covers the horizontal 
axis of the S1 detector, marked in figure \ref{fig:setup_elas} as $x_{S1}$, from 
   40 cm to 75 cm, which corresponds to a $\theta^{*}_{2}$ angle range 
   from $44^{o}$ to $66^{o}$.
 \begin{center} 
  \begin{figure}[H]
 \includegraphics[height=.45\textheight]{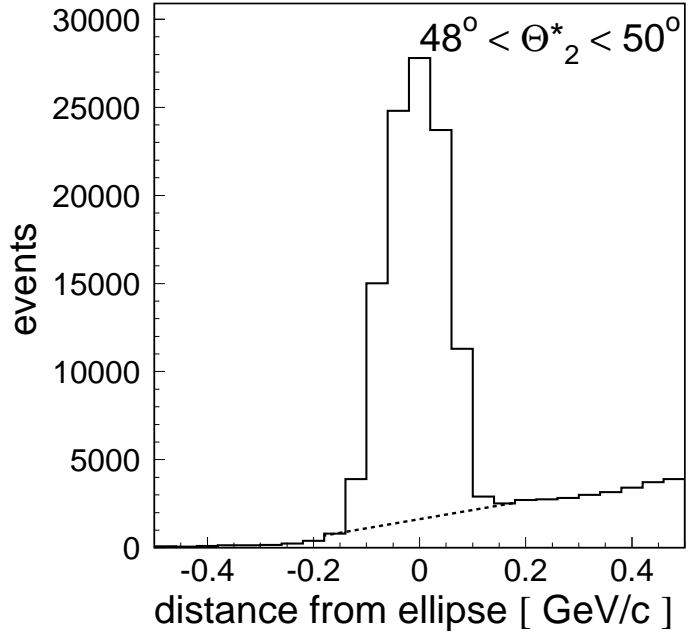}
        \caption{Projection of the event distribution along the kinematical ellipse for a 
	centre-of-mass proton scattering angle of $\theta^{*}_{2}$ in the range from $48^{o}$ to $50^{o}$, corresponding 
	to the range in the S1 detector from $x_{S1} = 67.0~cm$ to $x_{S1} = 70.0~cm$.}
\label{fig:49}
\end{figure}    
  \end{center}  
As an example, the distribution of elastically scattered protons at 
the $\theta^{*}_{2}$ angular range from $48^{o}$ to $50^{o}$ is presented in figure \ref{fig:49}. 
This distribution shows the projection of the experimental data along the 
kinematical ellipse. The number of events (reduced by the background indicated 
by the dashed line) is used for the calculation of the luminosity.\\
The signal from elastically scattered protons can be clearly separated from the flat multi-pion scattering background.
The solid angle $\Delta\Omega^{*}$ or the certain angular bin, is calculated using the Monte-Carlo method, as follows:
\begin{equation}
\Delta\Omega^{*}~=~\frac{4\pi~N_{accepted}}{2~N_{0}}~[sr],
\label{equ:omega}
\end{equation}
where $N_{0}$ stands for the number of proton-proton elastic scattering events in the corresponding angular range and 
$N_{accepted}$ constitutes the number of events in the considered bin of the $\theta^{*}_{2}$ angle, 
which could be registered and identified. 
In particular, the analysis in the following manner was done. First, $N_{0} = 2 \cdot 10^{7}$ events has been generated, 
calculating the response of the COSY-11 detectors, and then those 
events have been analysed using procedures applied for the experimental data evaluation in 
order to determine the number of $N_{accepted}$ events for each  $\theta^{*}_{2}$ angle interval.\\
\begin{center} 
  \begin{figure}[H]
 \includegraphics[height=.45\textheight]{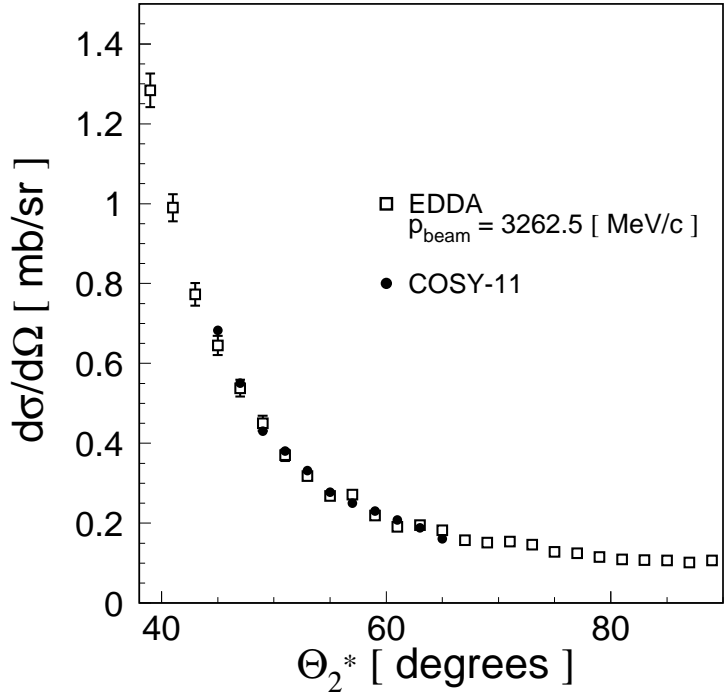}
        \caption{Differential cross section for the proton-proton elastic scattering.
	 The result of this thesis (closed circles) measured at 
  a beam momentum of $P_{B} = 3.260$ GeV/c
  was scaled 
  in amplitude to the cross section measured 
  by the EDDA collaboration shown by  open squares~\cite{edda}.}
\label{fig:mmass_lum}
\end{figure}
\end{center}
Figure \ref{fig:mmass_lum} indicates the angular distribution of the differential cross 
section for elastic proton-proton scattering obtained in the experiment (closed circles). 
The amplitude of that distribution was fitted to the data of the EDDA experiment 
including only one free parameter being the integrated luminosity (see eq. \ref{equ:cross}). The extracted 
integrated luminosity for the experiment described in this thesis amounts to ${\bf L~=~(5.859 \pm 0.055)~pb^{-1}}$.\\

The knowledge of the luminosity value will allow for the overall normalization of the 
derived differential cross section 
as a function of the $s_{pp}$ and $s_{p\eta^{\prime}}$ invariant masses, angular distributions and 
total cross section which will be discussed in chapter \ref{Cross sections}.

\chapter{Determination of the spread and the absolute value of the beam momentum}
\label{Determination of the spread and the absolute value of the beam momentum}
\markboth{\bf Chapter 7.}
         {\bf Determination of the spread and the absolute value of the beam momentum}

In order to perform realistic simulations
of the studied reactions, in particular to determine the acceptance and to calculate the covariance
matrix, it is mandatory to know the absolute value and the spread of the beam momentum.
The discussed measurement of the $pp \to pp\eta^{\prime}$ reaction was nominally performed at the same value of
excess energy Q as the $pp \to pp$ reaction measurement with Q = 15.5 MeV which
corresponds to a nominal proton beam momentum of $P_{B} = 3.257$ GeV/c.\\
The precision of the absolute beam momentum adjustment of the COSY synchrotron is about $10^{-3}$ \cite{cooling}
which in this case corresponds to $\sim$3 MeV/c.\\
The beam momentum dependence of the mean value of the missing mass distribution presented in figure
\ref{fig:mm_pol}, was studied 
in order to determine the actual value of the beam momentum more accurate.\\
The beam momentum was calculated using the formula:
\begin{equation}
m_{X} = \sqrt{s} - 2m_{p} = \Big( 2m^{2}_{p} + 2 m_{p} \sqrt{P^{2}_{B} + m^{2}_{p}} \Big) ^{1/2} - 2m_{p},
\label{equ:Q}
\end{equation}
where $\sqrt{s}$ denotes the total energy in the centre-of-mass frame, $P_{B}$ stands for the proton
beam momentum, and $m_{p}$ corresponds to the proton mass.\\
The beam momentum of $P_{B} = 3.260$ GeV/c
has been determined by adjusting the $P_{B}$ such that the mean value of the missing mass peak is equal to the $\eta^{\prime}$
meson mass. The determined value of the beam momentum differs by 0.003 GeV/c
from the nominal one.
This deviation is in agreement with results of analogous analysis performed in previous measurements \cite{moskalphd}.\\
The determined value of the excess energy amounts to {\bf (Q~=~16.39$~\pm~$0.01$~\pm~$0.4)~MeV},
where the errors indicate statistical and systematic uncertainty, respectively.
The dominating systematic uncertainty was established in \cite{prc69} and the statistic uncertainty of
the excess energy was determined using the following formula:
\begin{equation}
\Delta Q = \sqrt{\left( \frac{dQ}{dP_{B}} \right) ^{2}\cdot (\Delta P_{B})^{2}},
\label{equ:delta_Q}
\end{equation}
where $\Delta P_{B}$ was obtained from the linear relation to the missing mass
as described in~\cite{moskalphd}.\\

\begin{center}
\begin{figure}[H]
\includegraphics[width=.45\textheight]{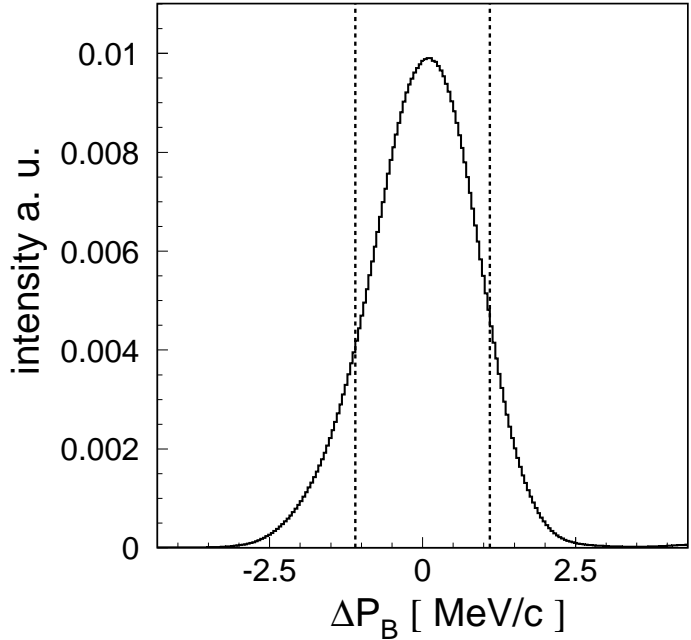}
\caption{Spectrum of the beam momentum distribution integrated over the whole measurement time.
The value of $\Delta P_{B} = 0$ corresponds to a beam momentum of 3.260 GeV/c.
The dashed lines mark the beam momentum dispersion for the extended 9 millimeter target
used in this experiment.}
\label{fig:dispersion}
\end{figure}
  \end{center}
After the determination of the real value of the absolute beam momentum, now its spread will be determined.
One can calculate the beam momentum spectrum from the frequency spectrum of the COSY beam
(Schotky spectrum measured during the experiment) using the below formula \cite{optics}:
\begin{equation}
\frac{\Delta f}{f}~=~\eta_{B} \cdot \frac{\Delta P_{B}}{P_{B}},
\label{equ:frequency}
\end{equation}
where $f$ and $P_{B}$ denote beam frequency and beam momentum, respectively, $\eta_{B}$ is a parameter which depends on
the beam optics, i.e. the electric and magnetic fields in the synchrotron. During the experiment,
the $\eta_{B}$ parameter was established to be $\eta_{B}$~=~0.12~\cite{eta}. The spectrum of the beam momentum obtained during
the experiment is shown in figure~\ref{fig:dispersion}.\\
The dispersion of the beam momentum depends on the magnetic field along the ring.
Protons on the outer routes have a "longer way" than those on the inner side of the beam, and the trajectories are
different from the nominal value.
When at a certain point the particle position deviates by $\Delta$$x$ from the nominal ($x_{0}$) position and
possesses the relative momentum deviation $\Delta P_{B} / P_{B}$, one can relate this value by means of the known dispersion ($D$) using
the following formula:
  \begin{equation}
\Delta x = D \cdot \frac{\Delta P_{B}}{P_{B}},
\label{equ:disp}
\end{equation}
where $D = \beta \cdot D_{mod}$ with $D_{mod} = 14~m$ \cite{eta, disp} and the particle velocity $\beta$.
Applying in the upper formulas the value of the beam momentum $P_{B}~=~3.260$~GeV/c, $\beta~=~0.961$,
and using a 9 millimeter wide target, 
one obtains $\Delta P_{B}~=~1.1$ MeV/c. This value is depicted in
figure \ref{fig:dispersion} as the area between the dashed lines. Thus,
the maximum beam momentum spread in the interaction point is $\Delta P_{B}~=~\pm~1.1$~MeV/c.

\chapter{Fine tunning of the relative dipole-chamber settings}
\label{Fine tunning of the relative dipole-chamber settings}
\markboth{\bf Chapter 8.}
         {\bf Fine tunning of the relative dipole-chamber settings}

In section \ref{Space-time relation for drift chambers} a space-time calibration of the drift chambers was described. 
In this chapter a procedure for the determination of their position relative to the COSY-11 dipole will be shown. 
For this purpose, the missing mass method for the $pp \to ppX$ 
and the kinematical ellipse for the $pp \to pp$ reaction, as described in the previous chapters, will be used.\\

\begin{figure}[H]  
 \includegraphics[height=.3\textheight]{p_vs_p.eps}
 \includegraphics[height=.3\textheight]{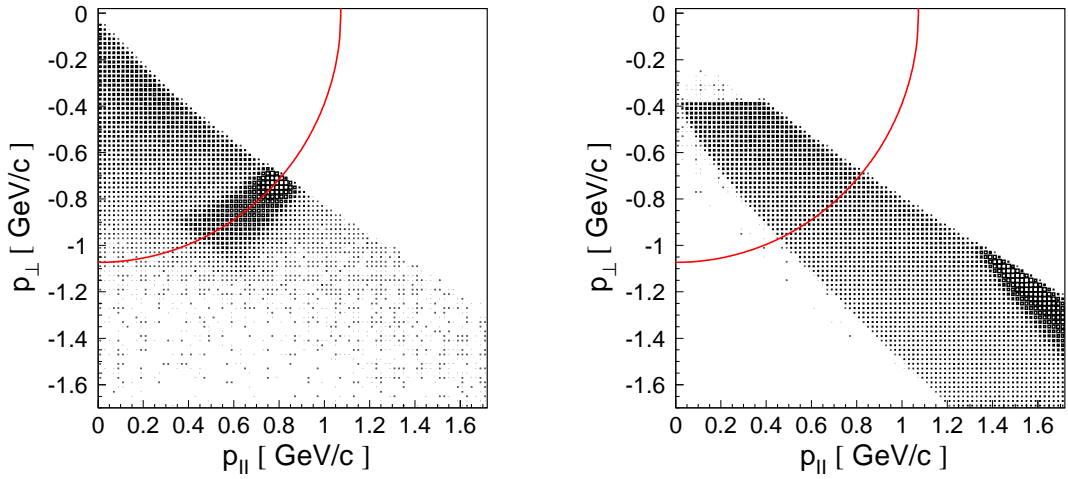}
        \caption{Distributions of the perpendicular $p_{\perp}$ versus the parallel
               $p_{\parallel}$ momentum components of measured protons for $pp \to pp$ elastic
               scattering events at a beam momentum of 3.260~GeV/c. The
               solid line corresponds to the kinematical ellipse where the elastically scattered events are 
	       expected. The distribution 
	       is shown for $\Delta\alpha~=~0.04^{o}$ (left), and for $\Delta\alpha~=~-0.9^{o}$ (right), 
	       respectively. }
\label{fig:p_pee}	       	
\end{figure}

The proper knowledge about the position of the drift chambers is needed in 
order to achieve the optimal resolution for the momentum reconstruction. 
In the horizontal plane the position of the drift chambers stack can be defined by two variables: the 
horizontal shift of the chambers $\Delta{x}$ and their inclination $\Delta\alpha$.
\begin{center}  
\begin{figure}[H]
\includegraphics[width=.6\textheight]{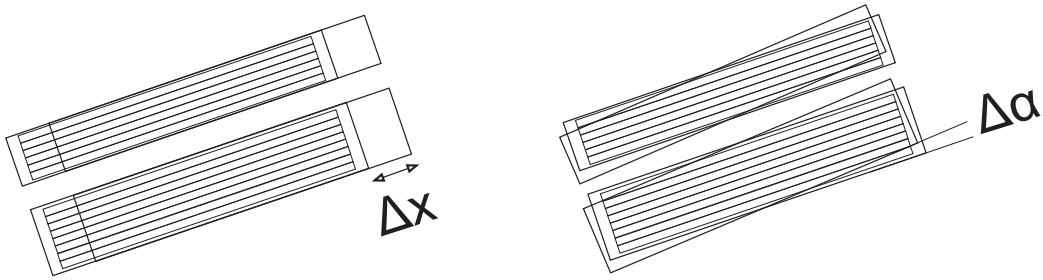}
\caption{Schematic view (from top) with the pictorial 
definitions of the horizontal shift of the chambers $\Delta{x}$ and their inclination $\Delta\alpha$. 
These parameters were used in the analysis to determine the proper arrangement of the D1 and D2 drift chambers 
relative to the COSY-11 dipole.}
\label{fig:chambers}
\end{figure}  
  \end{center}
The parameters $\Delta{x}$ and $\Delta\alpha$ influence the reconstruction of particle 
momenta and further more indirectly influence the missing mass 
resolution. In order to fix these parameters, an analysis of the proton elastic scattering 
events was performed, since the distribution of the perpendicular $p_{\perp}$ versus parallel 
$p_{\parallel}$ momentum components for the $pp \to pp$ reaction is sensitive 
to changes of the position of the drift chambers. 
The distribution of the $p_{\perp}$ versus $p_{\parallel}$ momentum components for the $pp \to pp$ elastic
scattering determined at a beam momentum of 3.260 GeV/c, presented earlier in 
chapter \ref{Monitoring of relative beam-target settings}, 
is shown in figure \ref{fig:p_pee}. In the left panel, the situation when 
the parameters are properly adjusted is presented. In the right panel the situation is shown 
when the inclination $\Delta\alpha$ was changed by $\sim 0.9^{o}$.  
Studies with a variation of $\Delta\alpha$ in the range from 
$-0.9^{o}$ to $0.9^{o}$ and of $\Delta{x}$ ranging from $0.0~$cm to $1.0~$cm were performed, 
where simultaneously the $p_{\perp}$ versus $p_{\parallel}$ spectrum and the changes of the width 
of the $\eta^{\prime}$ peak in the missing mass spectra were controlled.
The lowest width of the missing mass peak (see Fig. \ref{fig:mm}) and the 
best fit of the event distribution to the kinematical ellipse for the $pp \to pp$ elastic
               scattering distribution (see left panel of figure \ref{fig:p_pee}) has been found for the values of
	         $\Delta{x}~=~0.45~$cm and $\Delta\alpha~=~0.04^{o}$.\\
A comparison of the left and right histograms in figure \ref{fig:p_pee} shows the sensitivity of this distribution 
to the parameter $\Delta\alpha$, which could be determined within an accuracy of better than $0.01^{o}$.

\chapter{Evaluation of the differential distributions}
\label{Evaluation of the differential distributions}
\markboth{\bf Chapter 9.}
         {\bf Evaluation of the differential distributions}
	 	 
\section{Kinematical fit}
\label{Kinematical fit}
In order to search for small effects like proton-meson interaction on the population 
density of the phase-space, it is of importance to account for any possible changes of the measured distributions 
due to the finite resolution of the detector system, which may 
alter the shape of the spectrum especially close to the kinematical limit. 
Therefore, one must include the experimental resolution in the theoretical calculations 
and in order to improve the effect of the resolution a kinematical fitting of the data was performed. Both procedures require 
the knowledge of the covariance matrix. \\

\begin{center}  
\begin{figure}[H]
\includegraphics[width=.4\textheight]{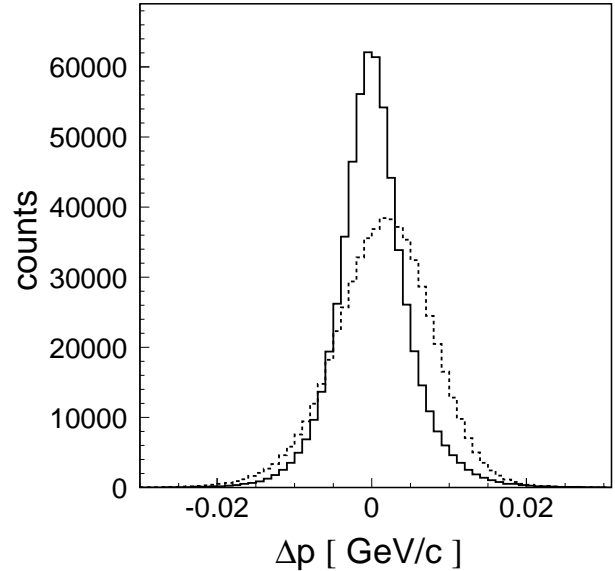}
\caption{Spectrum of the difference between the simulated and reconstructed proton momentum. The dashed 
                      line denotes the spectrum before kinematical fit and the solid line corresponds 
                      to the situation after the fitting procedure.}
\label{fig:delta_p}
\end{figure}  
  \end{center}

\begin{center}  
\begin{figure}[H]
\includegraphics[width=.45\textheight]{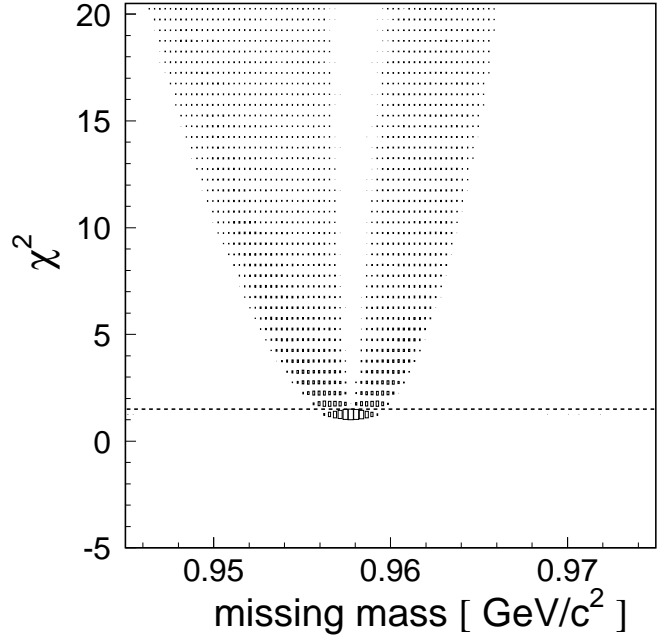}
\caption{The $\chi^{2}$ distribution as a function of the missing mass. 
	The dashed line at $\chi^{2}$ = 1.5 visualizes 
	the cut used in the analysis for the separation between kinematically fitted events ($P^{fit}$) 
	and events without the fit ($P^{exp}$).}
\label{fig:chi2}
\end{figure}  
  \end{center}

In order to calculate the covariances and variances between each combination of the registered proton momentum 
components,  
$2 \cdot 10^{7}$ reactions of the type $pp \to pp\eta^{\prime}$ have been generated including all arrangements and conditions from the experiment. 
Then, the whole simulated sample of events has been analysed with the same procedures as used for the experimental data. 
In this step for each simulated event the pair of the real (generated) protons momenta $\vec P_{1,gen}$, $\vec P_{2,gen}$  and 
the pair of momenta reconstructed from the simulated response of the detectors: $\vec P_{1,rec}$, $\vec P_{2,rec}$ are accessible. 
The available kinematical information about an event may be expressed in form of the six dimensional momentum vector:
$P = [p_{1x},p_{1y},p_{1z},p_{2x},p_{2y},p_{2z}]$ including the components of the 
reconstructed momenta of both protons  $P_{1,rec} = [p_{1x},p_{1y},p_{1z}]$ and $P_{2,rec} = [p_{2x},p_{2y},p_{2z}]$.
The covariance between the $i^{th}$ and $j^{th}$ components of $P$ was determined as the average 
of the product of deviations between the reconstructed and generated values.
The formula used to establish the covariance matrix elements, reads as follows:   
\begin{equation}
cov(i,j) = \frac{1}{N}\sum_{n=1}^{N} (P^{n}_{i,gen}-P^{n}_{i,recon})(P^{n}_{j,gen}-P^{n}_{j,recon}),
\label{equ:cov}
\end{equation}
where $P^{n}_{i,gen}$ and $P^{n}_{i,recon}$ stand for the generated and reconstructed values of the $i^{th}$ 
component of the vector $P$ from the $n^{th}$ event.\\
Due to the inherent symmetries of the covariance matrix $(cov(i,j) = cov(j,i))$ 
and since the measured protons are indistinguishable, there are only 12 independent 
values which determine the 6 $\times$ 6 error matrix unambiguously \cite{prc69}.\\
The covariance matrix (in units of MeV$^2$/c$^4$) determined for the $pp \to pp\eta^{\prime}$ reaction looks as follows:\\
\newpage
\begin{flushleft}
~~~~~~~~~~~~~~~~~~~~~~~~~~~~~$p_{1x}~~~~~p_{1y}~~~~~~p_{1z}~~~~~~p_{2x}~~~~~p_{2y}~~~~~~p_{2z}$\\
\end{flushleft} 
\begin{displaymath} 
\left[ 
\begin{array}{cccccc}
  10.3 &-0.1          &-19.6        &$~~$7.6     &-0.0        &-13.0\\
     - &$~~$6.2       &$~~~~$0.1    &$~~$-       &$~~~$0.1    &$~~~$0.0\\
     - &$~~$-         &$~~~$43.0    &$~~$-       &$~~~$-      &$~~$21.7\\
     - &$~~$-         &$~~~$-       &$~~$-       &$~~~$-      &$~~~$-\\
     - &$~~$-         &$~~~$-       &$~~$-       &$~~~$-      &$~~~$-\\
     - &$~~$-         &$~~~$-       &$~~$-       &$~~~$-      &$~~~$-\\
\end{array} 
\right]
\begin{tabular}{c}
$p_{1x}$\\
$p_{1y}$\\
$p_{1z}$\\
$p_{2x}$\\
$p_{2y}$\\
$p_{2z}$\\
\end{tabular}
\end{displaymath}
As it was already pointed out in the previous paragraph, 6 variables 
(2 times 3 components of the momentum vectors of the two protons) have been measured 
in the experiment. 
It was assumed in the analysis, that an event with the missing mass equal to the mass of the $\eta^{\prime}$ meson (within the experimental resolution)
corresponds to the $pp \to pp\eta^{\prime}$ reaction. 
Under this assumption only five of the kinematical variables are independent compared to six measured variables. 
Therefore, a kinematical fitting procedure can be applied to improve the effect of the limiting resolution. 
Hence, the protons momenta were varied demanding that the missing mass of the unregistered 
particle is exactly equal to the known mass of the $\eta^{\prime}$ meson and there has been chosen the momentum vector which 
was the closest to the vector determined from the experiment. The inverse of the covariance matrix
was used as a metric for the distance calculation. The kinematical fit clearly improves the effective resolution, 
what can be seen in figure \ref{fig:delta_p}. 
After the kinematical fit was performed for each event aside from the experimentally determined 
momentum vector: $P^{exp} = [p^{exp}_{1x},p^{exp}_{1y},p^{exp}_{1z},p^{exp}_{2x},p^{exp}_{2y},p^{exp}_{2z}]$, \\
the kinematically fitted momenta: $P^{fit} = [p^{fit}_{1x},p^{fit}_{1y},p^{fit}_{1z},p^{fit}_{2x},p^{fit}_{2y},p^{fit}_{2z}]$ were obtained.\\
The $P^{fit}$ vectors are more precise and assure that the determined distributions 
do not spread beyond the kinematical boundaries. \\
For the further analysis of events corresponding to the 
$\eta^{\prime}$ production, the distribution of the $\chi^{2}$ from the kinematical fit procedure was checked. 
In figure \ref{fig:chi2} the $\chi^{2}$ distribution as a function of the missing mass determined 
for the experimental momentum vectors is presented. 
Having each event described by the two vectors $P^{exp}$ and $P^{fit}$ and the $\chi^{2}$ of the kinematical fit, 
the variables $s_{pp}$, $s_{p\eta^{\prime}}$, $\psi$ and $|cos\theta^{*}_{\eta^{\prime}}|$ were evaluated by
using the kinematically fitted vectors $P^{fit}$ if the $\chi^{2}$--value of the fit 
for the event was lower than 1.5. However, if the $\chi^{2}$ was larger than this limitation, $P^{exp}$ were used, since 
those events correspond to the background with a missing mass lower or higher than 
the $\eta^{\prime}$ meson mass. For the multi-pion production background the kinematical fit procedure is not justified.
The evaluation of the differential distributions requires only events corresponding to the 
$pp \to pp\eta^{\prime}$ reaction. In order to select the number of those events from the multi-pion production reactions,  
a missing mass spectrum for each studied interval of the variables: 
$s_{pp}$, $s_{p\eta^{\prime}}$, $\psi$ and $|cos\theta^{*}_{\eta^{\prime}}|$ was determined. 
In the next section, the procedure of the background subtraction applied in the analysis will be discussed.

\section{Background subtraction}
\label{Background subtraction}
\begin{figure}[H]  
 \includegraphics[height=.34\textheight]{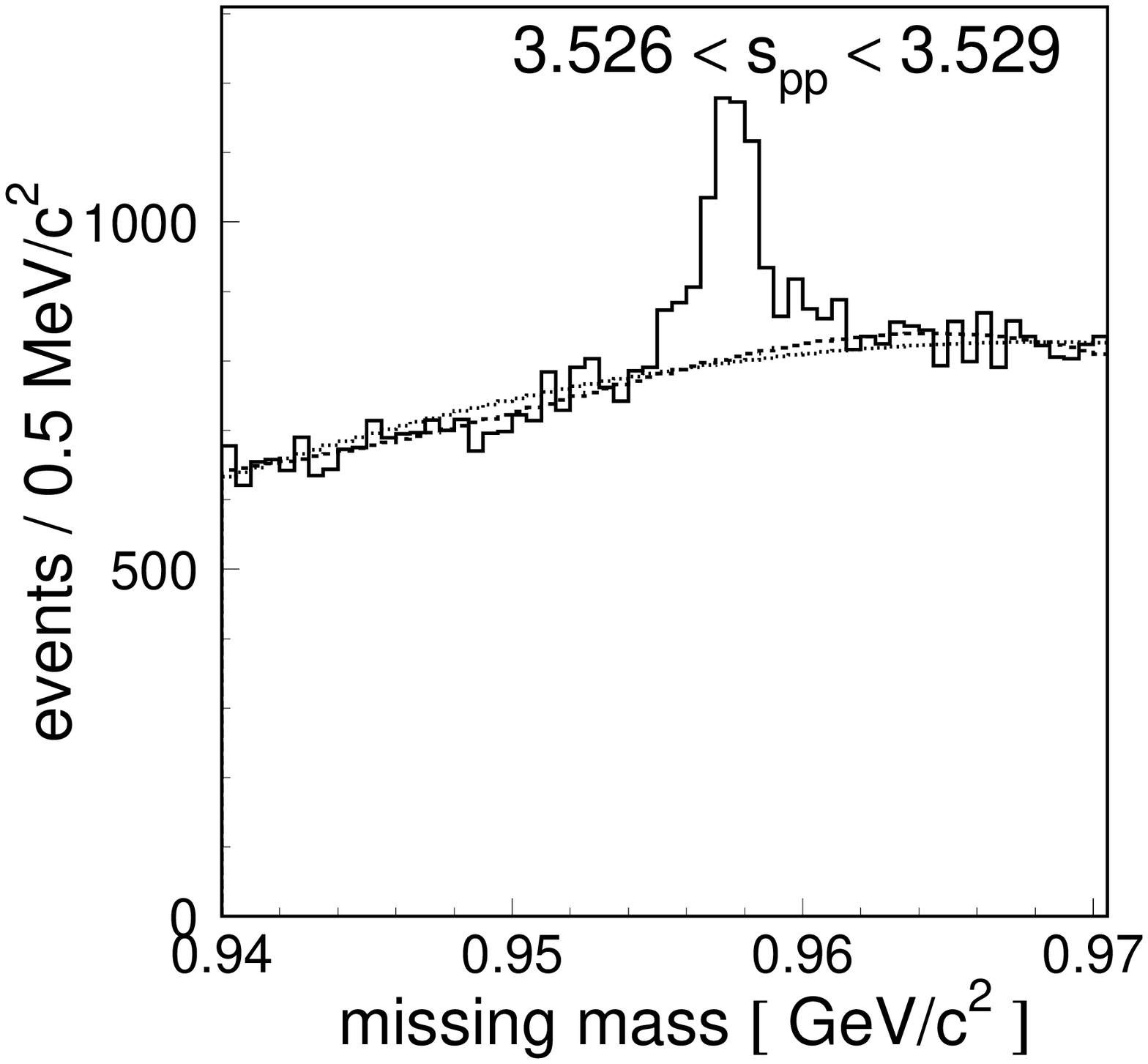}
  \includegraphics[height=.34\textheight]{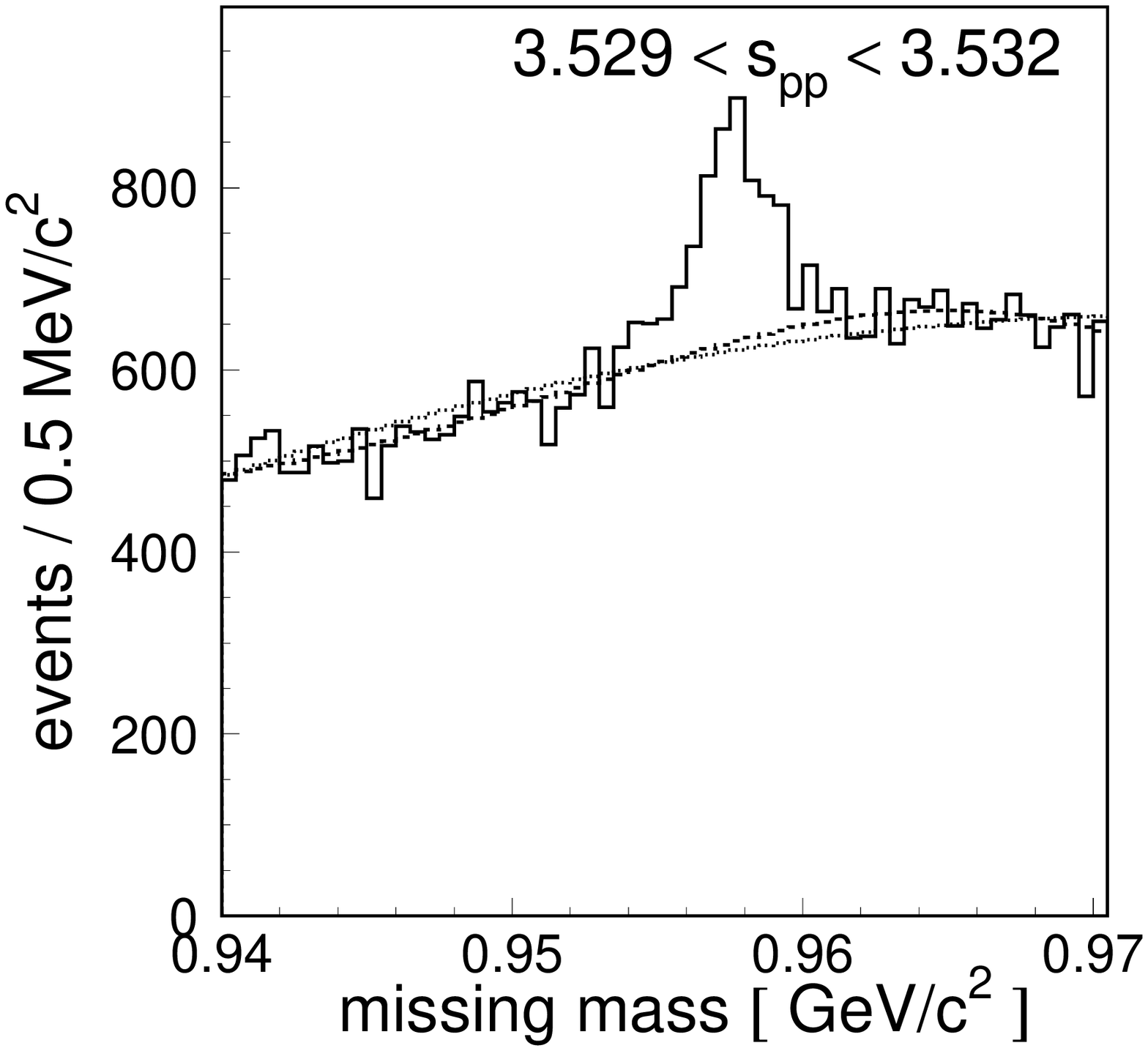}
          \caption{Examples of experimental missing mass spectra for two chosen intervals of 
	  $s_{pp}$. The dotted lines indicate the second order polynomial according to the formula \ref{equ:pol1} 
	  whereas the dashed lines show the result of the fit of two Gaussian distributions 
	  described by the formula \ref{equ:gauss1}. }
	  \label{fig:gauss_pol}	
\end{figure}

In the following section 
the method of the multi-pion background subtraction is presented as used in the analysis.\\
For the background free determination of the differential $s_{pp}$, $s_{p\eta^{\prime}}$, 
$\psi$ and $|cos\theta^{*}_{\eta^{\prime}}|$ distributions, first the distribution of the 
considered variable has been divided into a reasonable number of bins, and then for each bin a missing mass spectrum 
was produced 
and the number of the $pp \to pp\eta^{\prime}$ events was calculated for each interval of $s_{pp}$, $s_{p\eta^{\prime}}$, 
$\psi$ and $|cos\theta^{*}_{\eta^{\prime}}|$ separately.\\
The missing mass distributions include a smooth multi-pion production 
below and above the produced $\eta^{\prime}$ meson mass which allows to estimate 
the background corresponding to the multi-pion creation in the range of the $\eta^{\prime}$ meson signal 
using a polynomial fitting function, as it was already shown in section \ref{Identification of the meson}. 
In figure \ref{fig:gauss_pol}, examples of missing mass spectra for two intervals of 
the invariant proton-proton mass~$s_{pp}$ are presented.
In both panels of figure \ref{fig:gauss_pol} the dotted lines correspond to the polynomial fits described by 
the mathematical function:
\begin{equation}
  P(mm,a,b,c)~=~a+b \cdot mm~+~c \cdot mm^{2},
\label{equ:pol1}
\end{equation}
where $mm$ denotes the missing mass and $a$, $b$ and $c$ are the free parameters varied during the fit. 
Since not only a second order polynomial can be used as a good approximation of the 
background also a fit by the sum of two Gaussian distributions was performed.
The dashed lines in both panels of figure \ref{fig:gauss_pol} correspond to the fit with the function $G$ 
described by the following formula:
\begin{equation}
  G(mm,A_{1},B_{1},C_{1},A_{2},B_{2},C_{2})~=~G_{1}~(mm,A_{1},B_{1},C_{1})~+~G_{2}~(mm,A_{2},B_{2},C_{2}),
\label{equ:gauss1}
\end{equation}
where the terms $G_{1}$ and $G_{2}$ denote Gaussian distributions depending on the parameters 
$A_{1}$, $B_{1}$, $C_{1}$, $A_{2}$, $B_{2}$ and $C_{2}$ varied freely during the fit and $mm$ stands for the missing mass. 
The fits were performed using the functions of the missing mass (\ref{equ:pol1} and \ref{equ:gauss1}) in 
the whole range of missing mass outside of the $pp \to pp\eta^{\prime}$ signal. 
As it is seen, the fitted functions reproduce the background very well.\\

The smooth behaviour of the multi-pion background was verified by Monte Carlo simulations. 
But in order to use these Monte Carlo distributions 
as a reasonable description for the background an extensive time consuming simulation studies would be necessary. 
Therefore, a smooth function adjusted to the regions beside the $\eta^{\prime}$ peak was used instead. 
The sum of two Gaussian distributions was applied in the analysis. 
It should be stressed that the alternative description with the second 
order polynomial is in rather good agreement with the result obtained 
using formula \ref{equ:gauss1}.\\

\begin{figure}[H]  
  \includegraphics[height=.34\textheight]{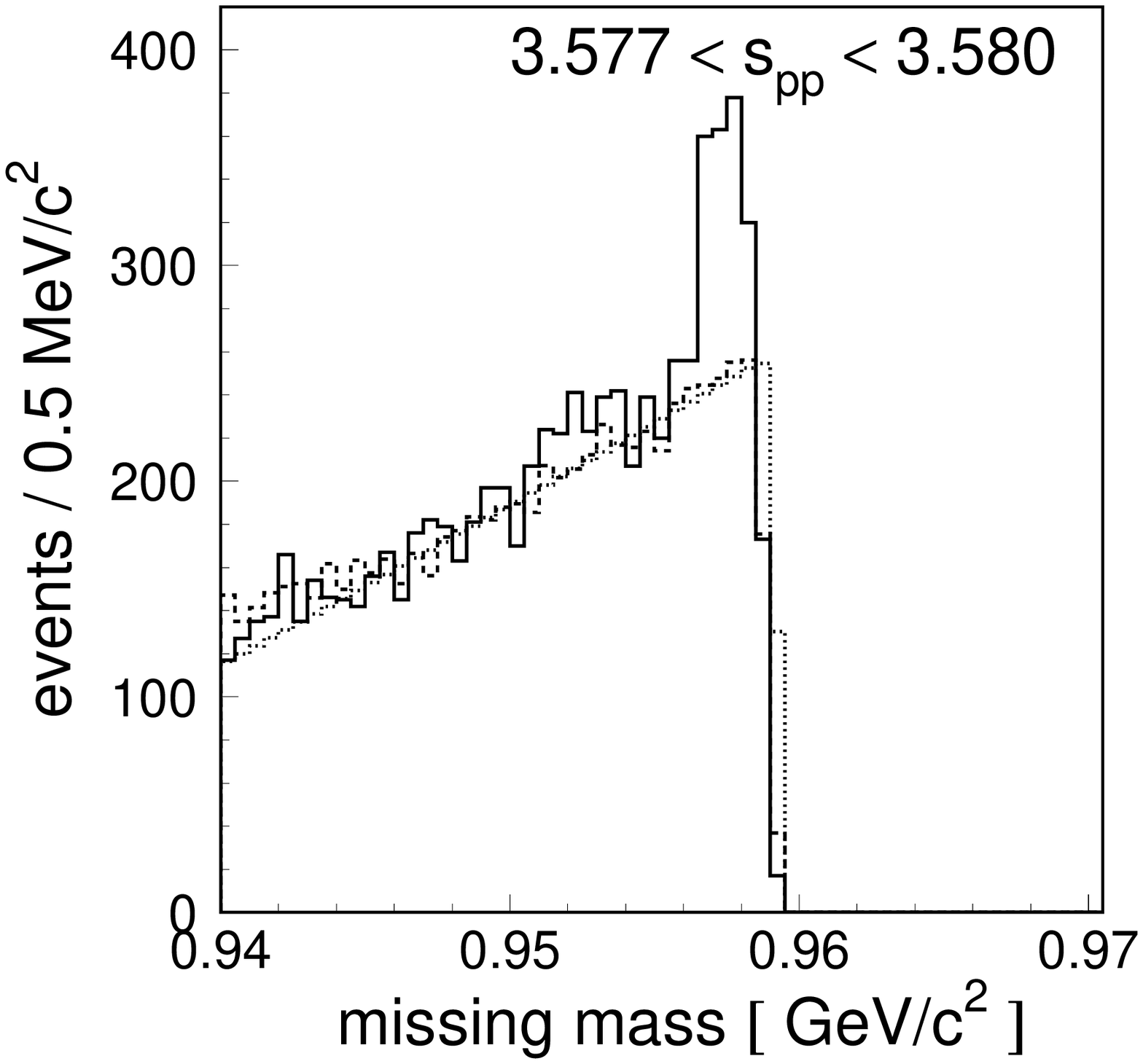}
  \includegraphics[height=.34\textheight]{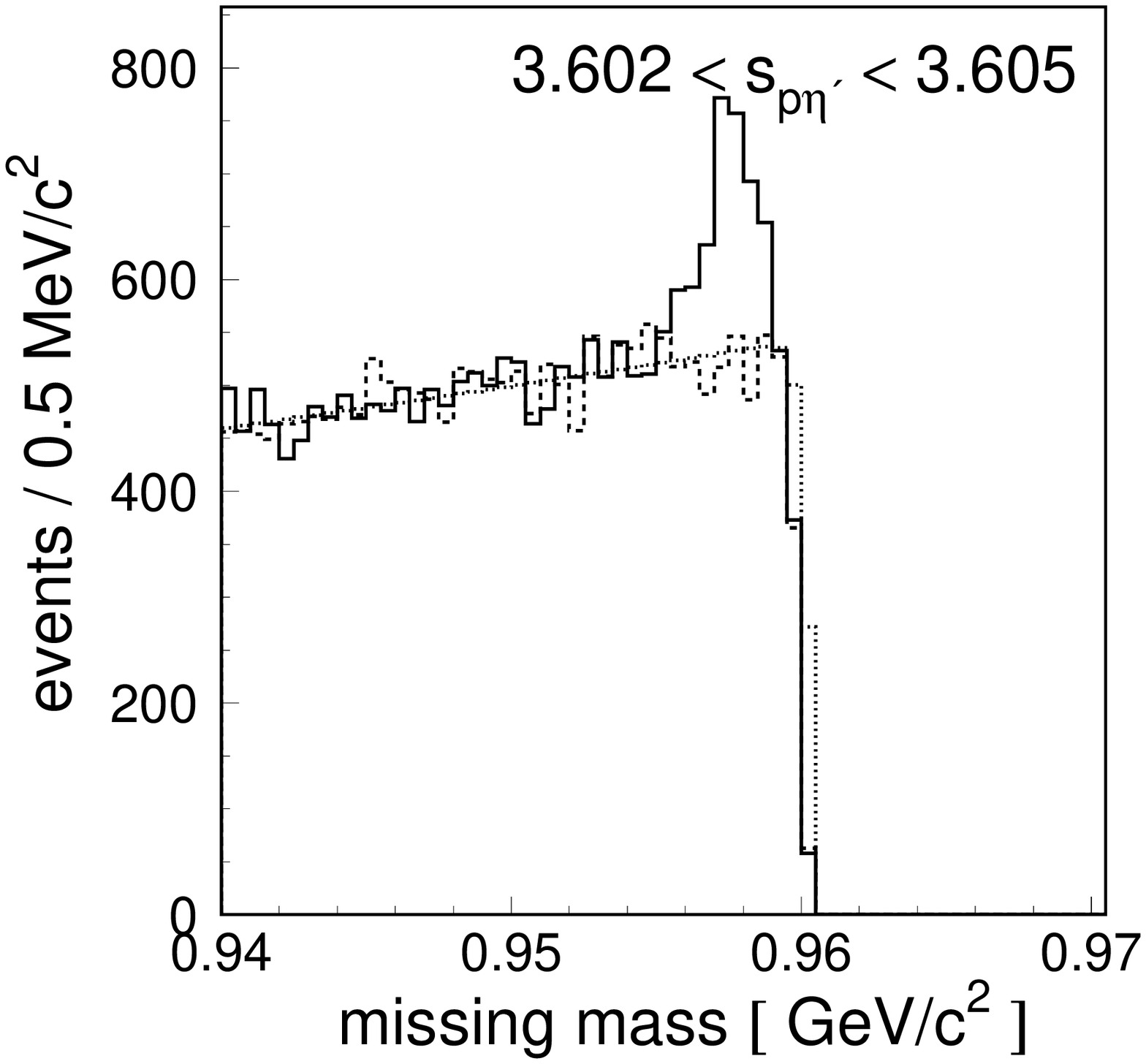}
          \caption{Examples of missing mass spectra determined experimentally for the $pp \to ppX$ 
	  reaction (solid lines) with superimposed Monte-Carlo simulations (dashed lines) including 
	  $pp \to pp2\pi\eta$, $pp \to pp3\pi$ and $pp \to pp4\pi$ reactions, 
	  fitted in amplitude to the experimental data.
	  Missing mass spectra for squared invariant proton-proton mass $s_{pp} \in [3.577;~3.580]~ $GeV$^{2}$/c$^{4}$
	   (left panel) and 
	  squared invariant proton-$\eta^{\prime}$ mass $s_{p\eta^{\prime}} \in [3.602;~3.605]~ $GeV$^{2}$/c$^{4}$ (right panel).
	  The dotted lines in both panels correspond to the fit of the formula \ref{equ:fermi}.}
	  \label{fig:exp_mc}	
\end{figure}

The situation is more complicated for missing mass spectra with the signal 
close to the kinematical limit (see e.g. Fig. \ref{fig:exp_mc}). In this case the shape of the background 
on the right side of the peak cannot be easily predicted.
Such spectra are obtained for kinematical regions of higher
squared invariant proton-proton masses 
$s_{pp}$ and relatively low squared invariant proton-$\eta^{\prime}$ masses 
$s_{p\eta^{\prime}}$. In order to describe the shape of the background 
in those regions, the $pp \to pp2\pi\eta$, $pp \to pp3\pi$ and $pp \to pp4\pi$ reactions\footnote{Those background
 reaction channels 
were chosen as a representation of possible multi-pion production background. Since, the $pp \to pp5\pi$ and 
$pp \to pp6\pi$ 
reactions simulations can be neglected and the $pp~\to~pp2\pi$ missing mass of two protons spectrum has similar 
shape as those for $3\pi$ and $4\pi$ \cite{marcin}, only simulations for listed reactions were performed.} 
have been simulated and the simulated events were analysed in the same way as it was done 
for the experimental data. The result of these simulations is compared to the experimental data in figure~\ref{fig:exp_mc}.
The simulations of the different reactions channels were performed with phase space 
distribution including the proton-proton final state interaction~\cite{swave, jpg}.\\
The simulated missing mass spectra were fitted to the data using the formula:
\begin{equation}
  S(mm,\alpha,\beta,\gamma)~=~\alpha \cdot f_{pp \to pp 2\pi \eta}(mm)~+~\beta \cdot f_{pp \to pp 3\pi}(mm)~+~\gamma \cdot f_{pp \to pp 4\pi}(mm),
\label{equ:mc}
\end{equation}
where $\alpha$, $\beta$ and $\gamma$ denote the free parameters varied during the fit procedure. 
The functions $ f_{pp \to pp2\pi\eta}$, $~~$ $f_{pp \to pp3\pi}$ $~~$ and$~~$  $f_{pp \to pp4\pi}$ 
correspond to the simulated missing mass distributions for the $pp \to pp2\pi\eta$,$~~$  $pp \to pp3\pi$ $~~$ 
and $~~$ $pp \to pp4\pi$$~~$  reactions.\\

In figure \ref{fig:exp_mc}, examples of missing mass spectra for 
squared invariant~proton-proton masses 
$s_{pp} \in [3.577;~3.580]~ $GeV$^{2}$/c$^{4}$ (left panel) and squared invariant~proton-$\eta^{\prime}$ masses 
$s_{p\eta^{\prime}} \in [3.602;~3.605]~ $GeV$^{2}$/c$^{4}$ (right panel) are presented.
In both examples, it is clearly seen that the simulations are in a good agreement with 
the experimental background distributions below the $\eta^{\prime}$ peak. Moreover the behaviour of the simulated background 
fits well to the kinematical limit of the missing mass distributions. 
Close to the kinematical boundary, the interaction in the final state is dominated 
by the proton-proton interaction \cite{jpg} and this interaction influences strongly 
the shape of the observed missing mass spectra. 
Therefore, it was taken into account in the calculations according to the formulas of reference \cite{swave}  
presented in details in~appendix~\ref{Parameterization of the proton-proton Final State Interaction}.\\
For the dynamics of the pion production it had been assumed, that pions are produced homogeneously over the phase space. 
As it was described in reference \cite{jpg}, the shape of the missing mass 
spectrum does not change significantly at the edge of the kinematical limit if one assumes resonant 
or direct pion production.\\
In order to raise the confidence to the estimation of the background behaviour near the kinematical boundary, 
those distributions were described in an independent way with a second order polynomial divided by the Fermi 
function for the description of the rapid slope at the end of the distributions.
To this end the following formula was applied:
\begin{equation}
  F(mm,a,b,c,d,g)~=~(a~+~b \cdot mm~+~c \cdot mm^{2})/(1~+~e^{(mm~-~d)/g}),
\label{equ:fermi}
\end{equation}
where $a$, $b$, $c$, $d$ and $g$ are the free parameters varied during the fit procedure.\\
The results are presented in figure \ref{fig:exp_mc} as dotted lines. 
It is seen that under the $\eta^{\prime}$ peak the result of formula \ref{equ:fermi} 
agrees well with the background determined 
from the simulations and that both reproduce the shape of the slope quite well.\\

\begin{figure}[H]  
  \includegraphics[height=.34\textheight]{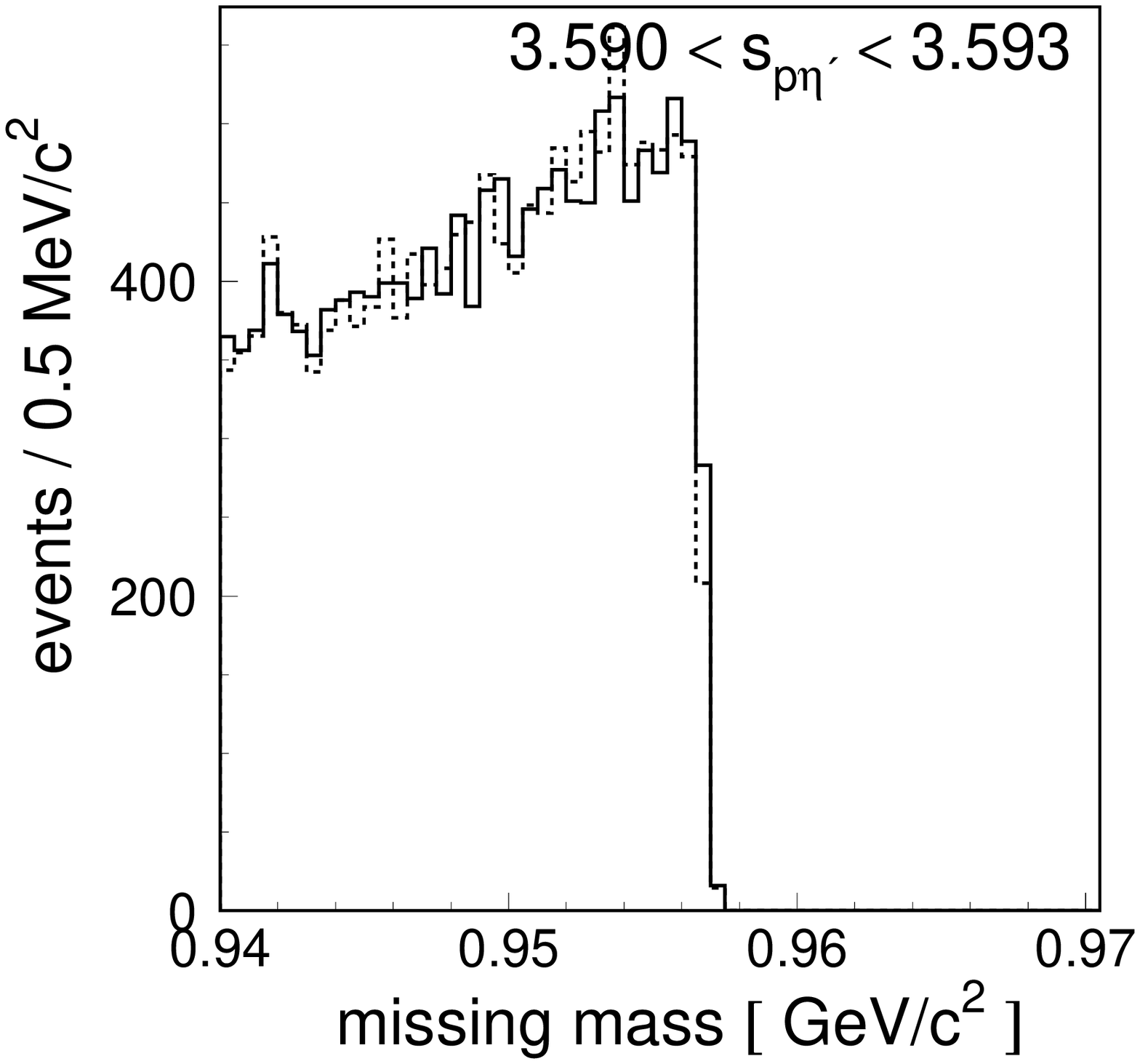}
  \includegraphics[height=.34\textheight]{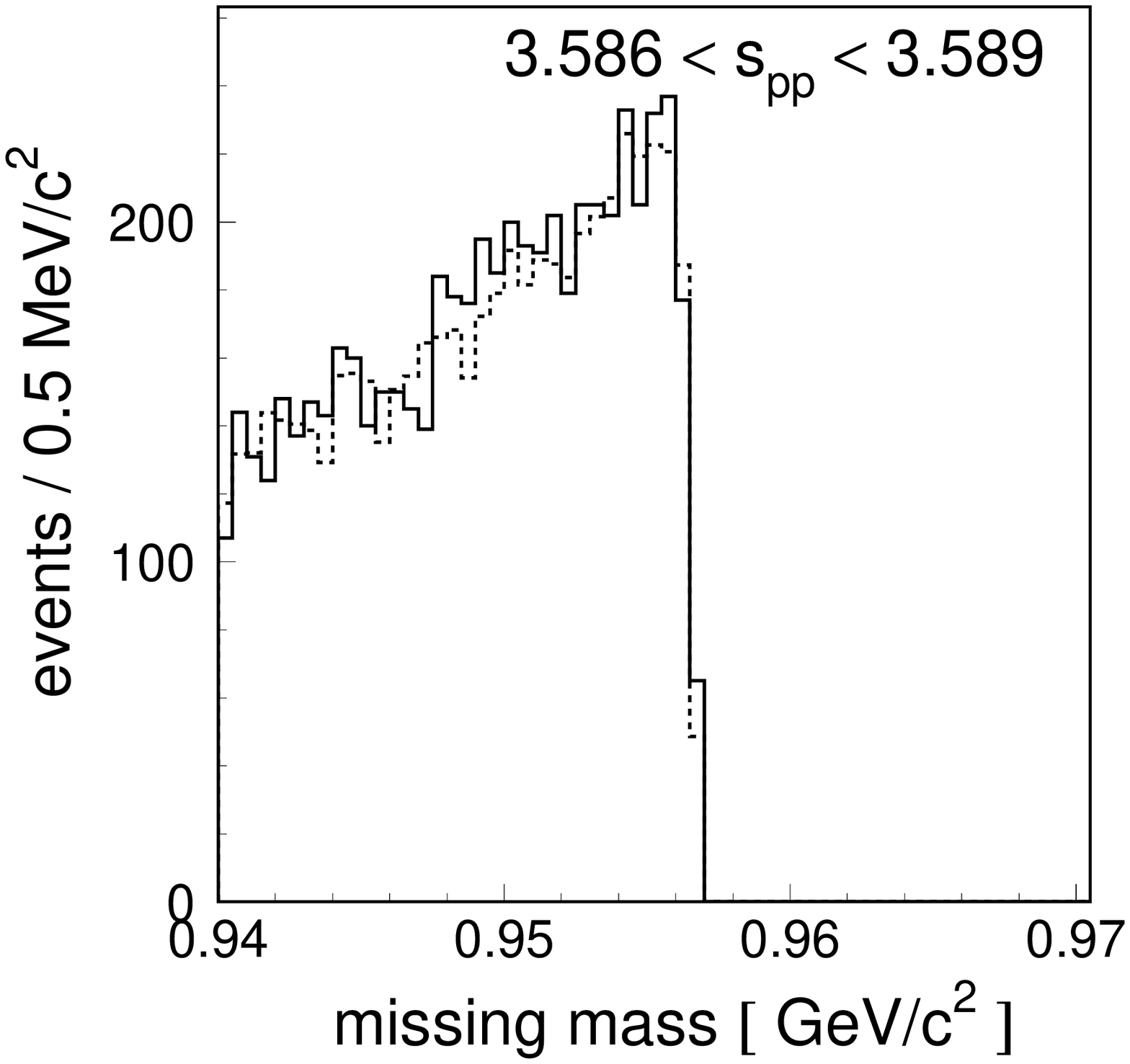}
          \caption{Missing mass spectra for low values of $s_{p\eta^{\prime}} \in [3.590;~3.593]~$GeV$^{2}$/c$^{4}$ (left) and high 
	  values of $s_{pp} \in [3.586;~3.589]~$GeV$^{2}$/c$^{4}$~(right). The dashed
	  lines correspond to the 
	  simulations of the $pp \to pp2\pi\eta$, $pp \to pp3\pi$ and $pp \to pp4\pi$
	  reactions fitted to the experimental distributions (solid lines) using only the amplitudes as free parameters.}
	  \label{fig:exp_pions}	
\end{figure}

In order to perform a further check of the background estimation, examples of the missing mass distributions 
for the regions of the squared invariant masses of proton-proton and proton-meson  
where the $\eta^{\prime}$ is not produced are presented. The missing mass distributions are shown in figure \ref{fig:exp_pions} and
represent the regions of low squared invariant masses of the proton-meson subsystem (left) 
and high squared invariant masses of the proton-proton subsystem (right). 
For such values of $s_{pp}$ or $s_{p\eta^{\prime}}$ the production of the $\eta^{\prime}$ meson is not possible 
because $s_{p\eta^{\prime}} < (m_{p} + m_{\eta^{\prime}})^{2}$ and $s_{pp}$ is too large leaving not enough energy for the 
$\eta^{\prime}$ meson creation.\\
The fitted simulations reproduce the background very well.
\newpage
\begin{figure}[H]  
  \includegraphics[height=.34\textheight]{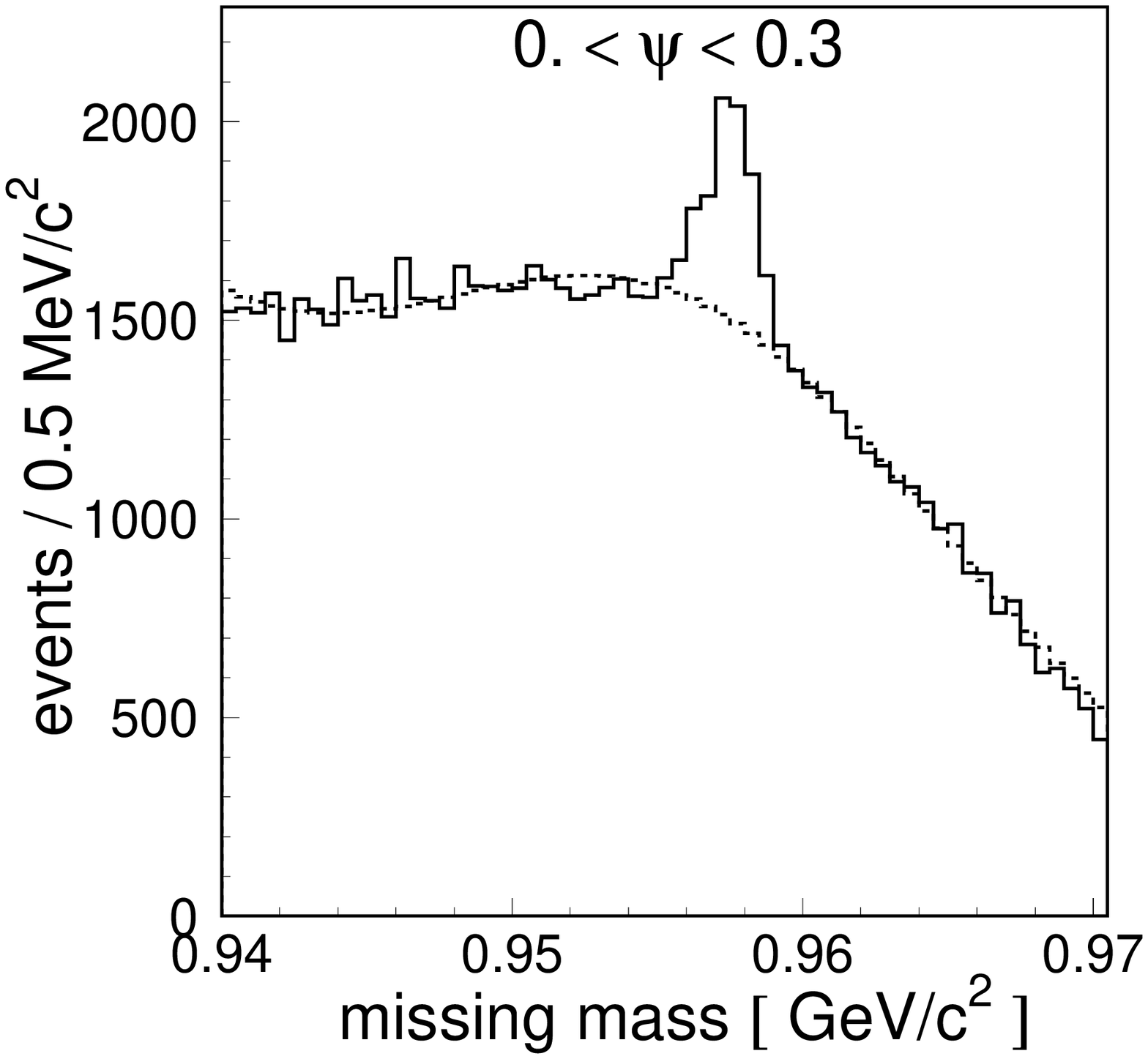}
  \includegraphics[height=.34\textheight]{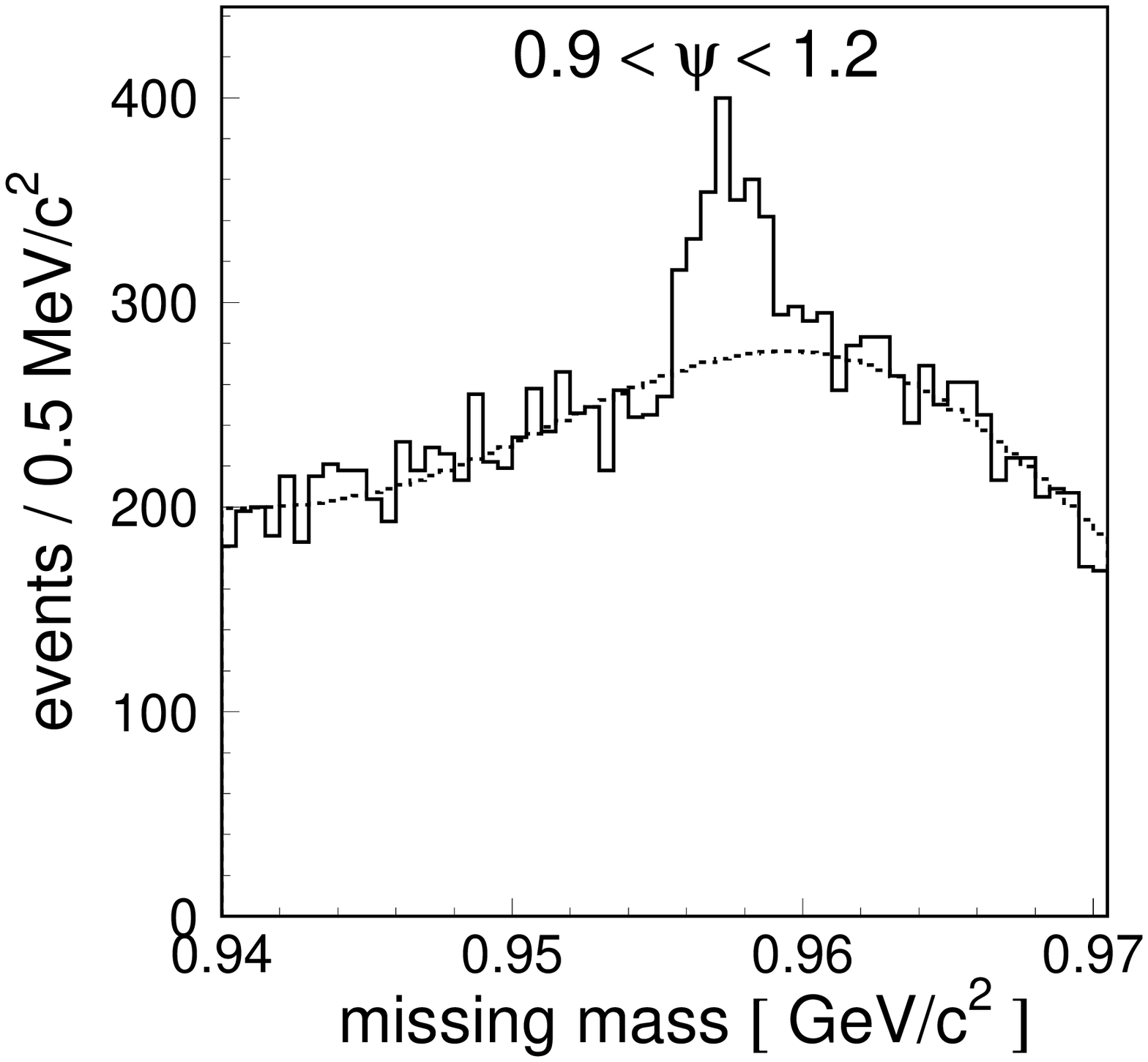}
        \caption{Examples of missing mass spectra determined for the first and fourth  bin 
	of the $\psi$ angle with a width of $0.3~radian$. 
	The dashed lines correspond to fits using the function described by formula \ref{equ:gauss1}.}	
	\label{fig:psia}
\end{figure} 
\vspace{-0.35cm}
\begin{figure}[H] 
  \includegraphics[height=.34\textheight]{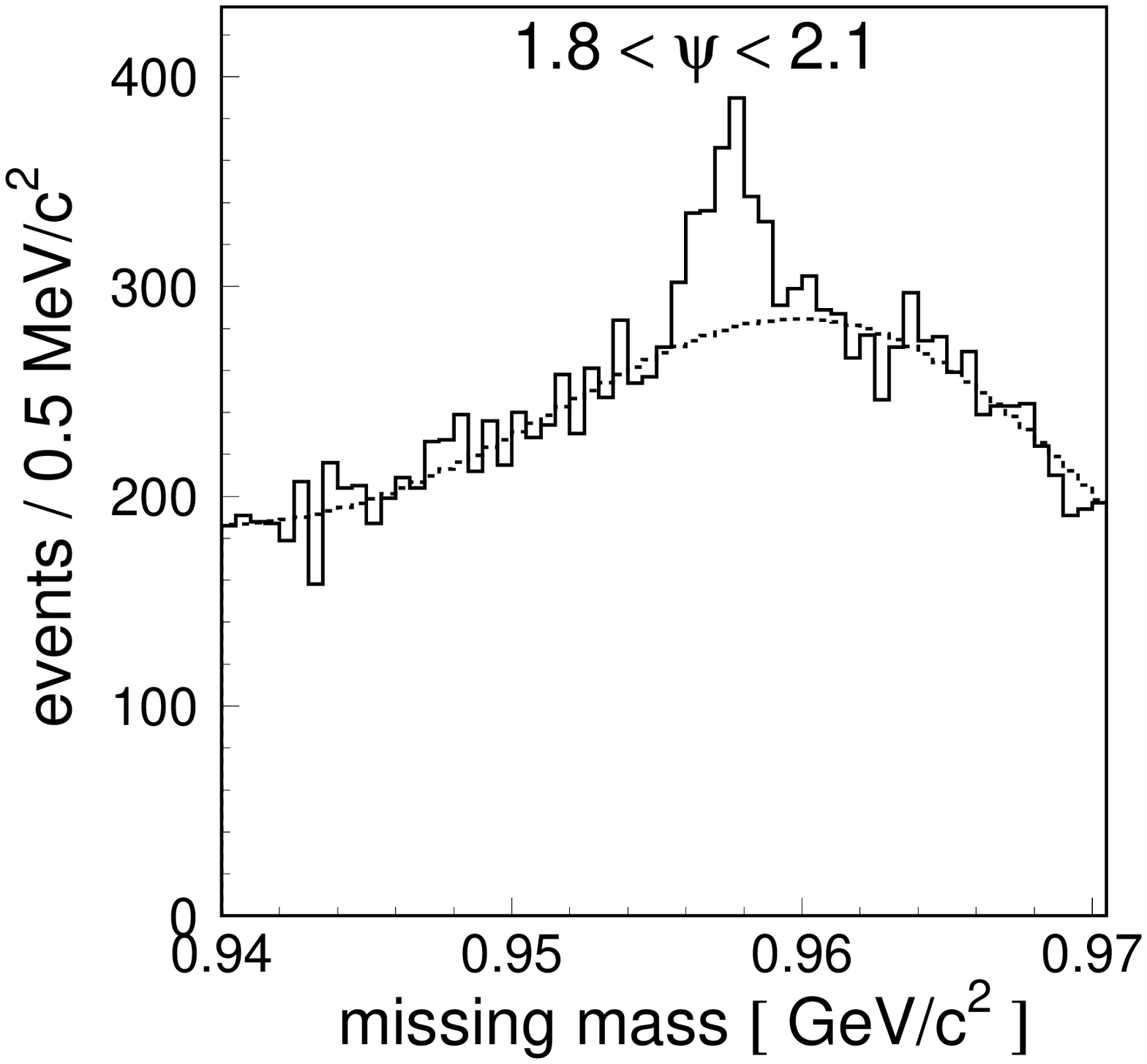}
  \includegraphics[height=.34\textheight]{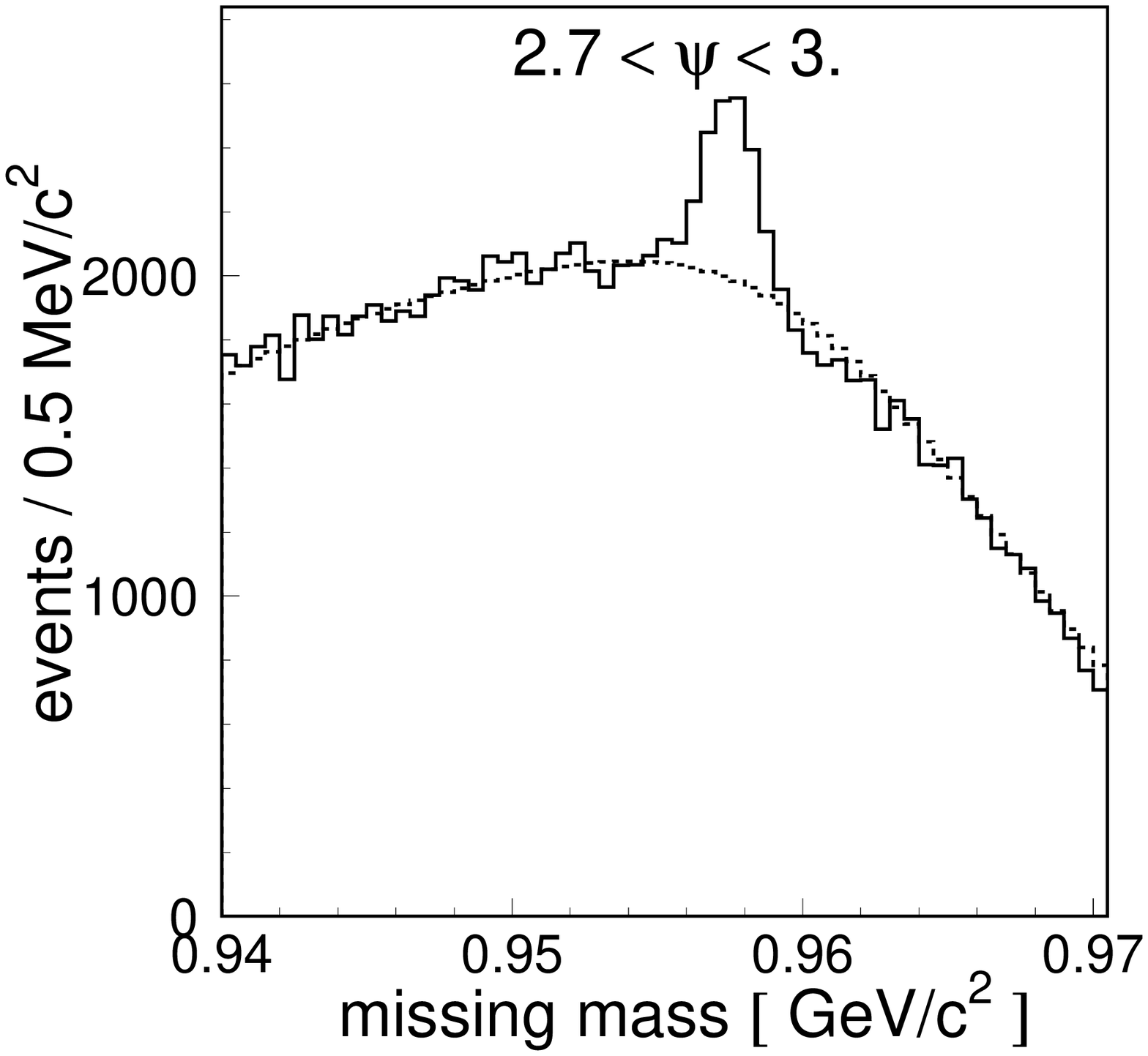}
        \caption{Examples of missing mass spectra determined for the seventh and tenth bin 
	of the $\psi$ angle with a width of $0.3~radian$. 
	The dashed lines correspond to fits employing the function described by formula \ref{equ:gauss1}.}	
	\label{fig:psib}
\end{figure} 
\newpage
\begin{figure}[H]  
 \includegraphics[height=.33\textheight]{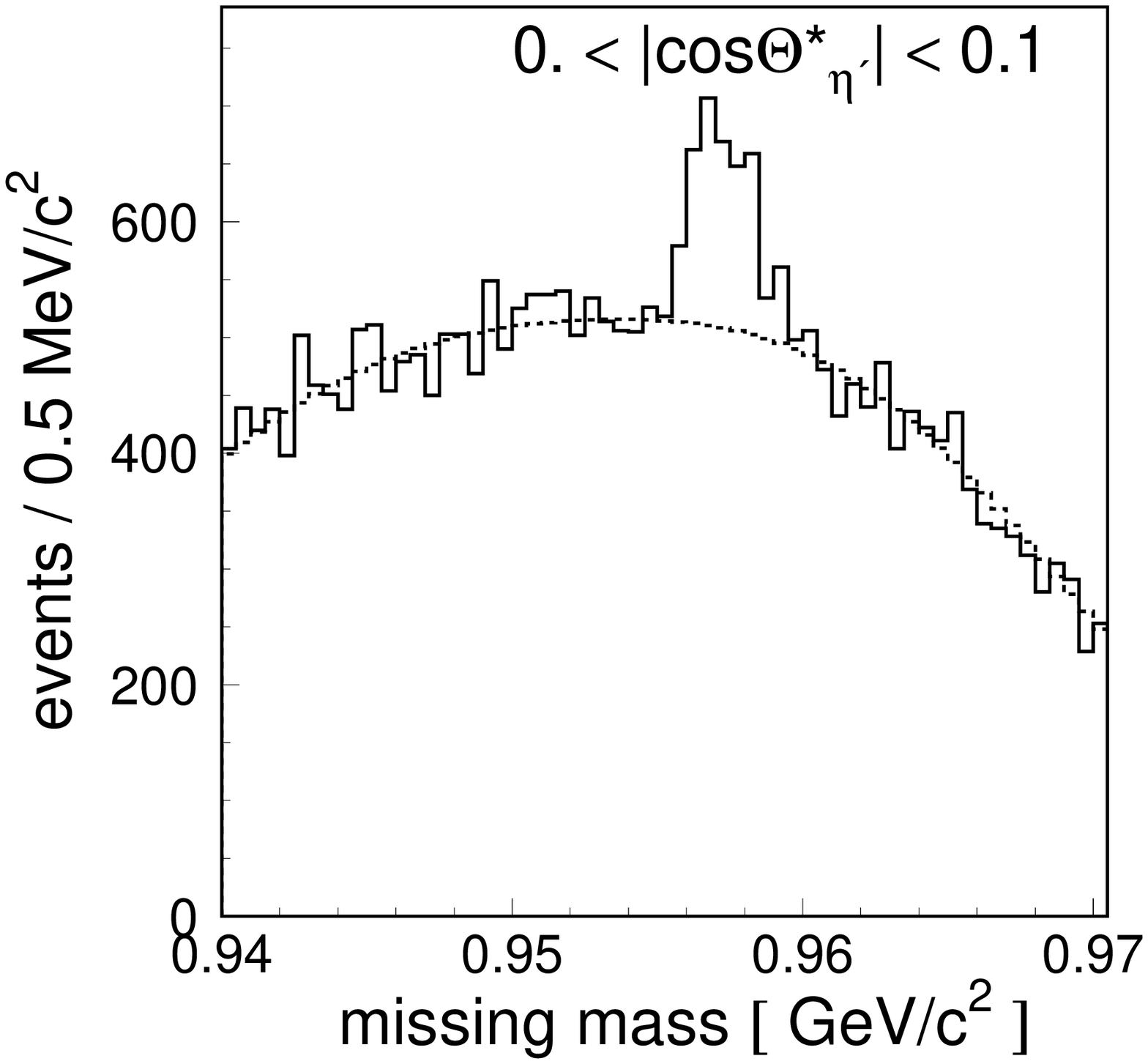}
 \includegraphics[height=.33\textheight]{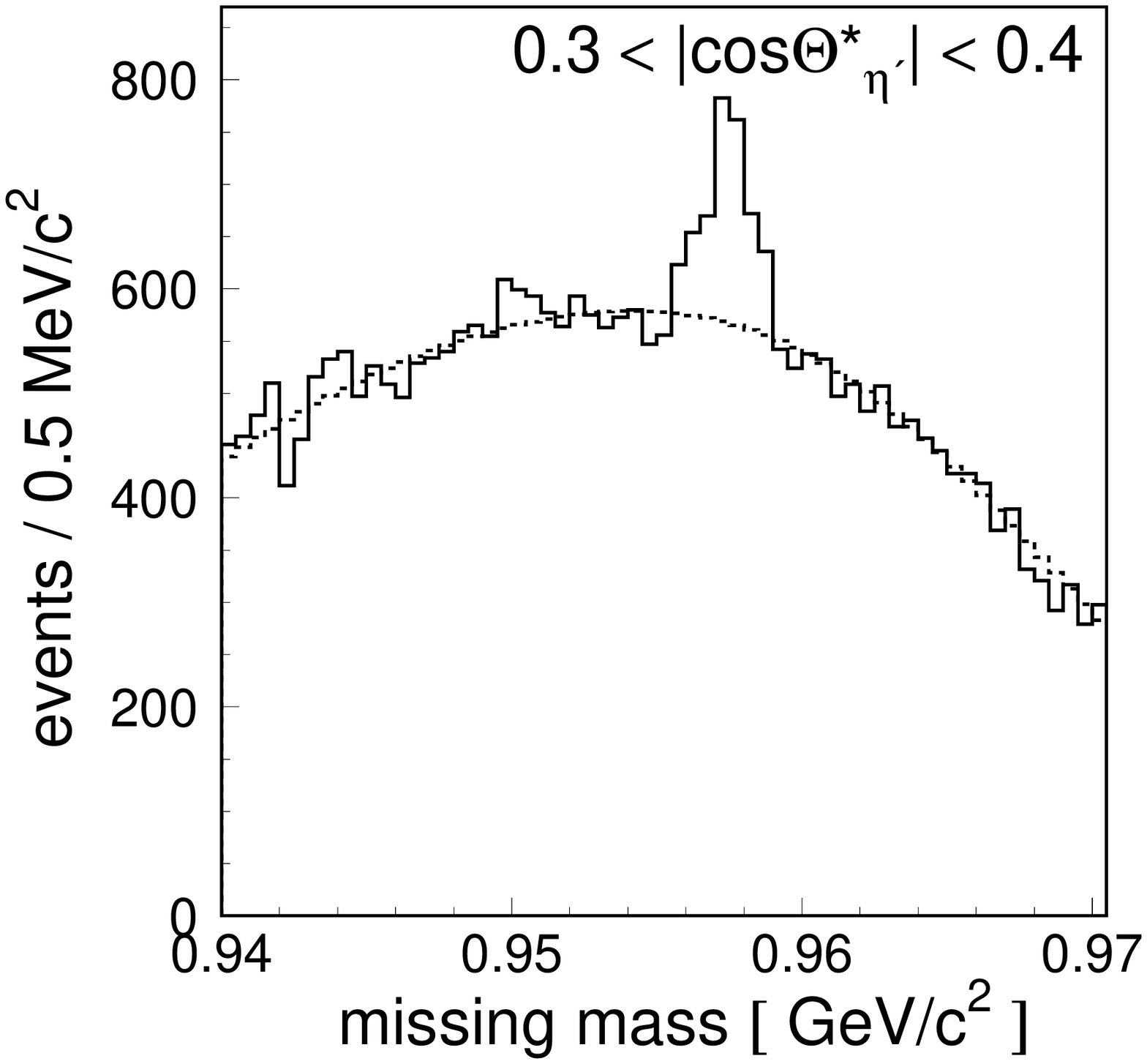}	
        \caption{Examples of missing mass spectra determined for the first and fourth bin
	 of $|cos\theta^{*}_{\eta^{\prime}}|$ with a width corresponding to $\Delta |cos\theta^{*}_{\eta^{\prime}}| = 0.1$.
	 The dashed lines correspond to fits according to the function described by formula \ref{equ:gauss1}.}	
	\label{fig:thetaa}
\end{figure}	
\begin{figure}[H]  	
  \includegraphics[height=.33\textheight]{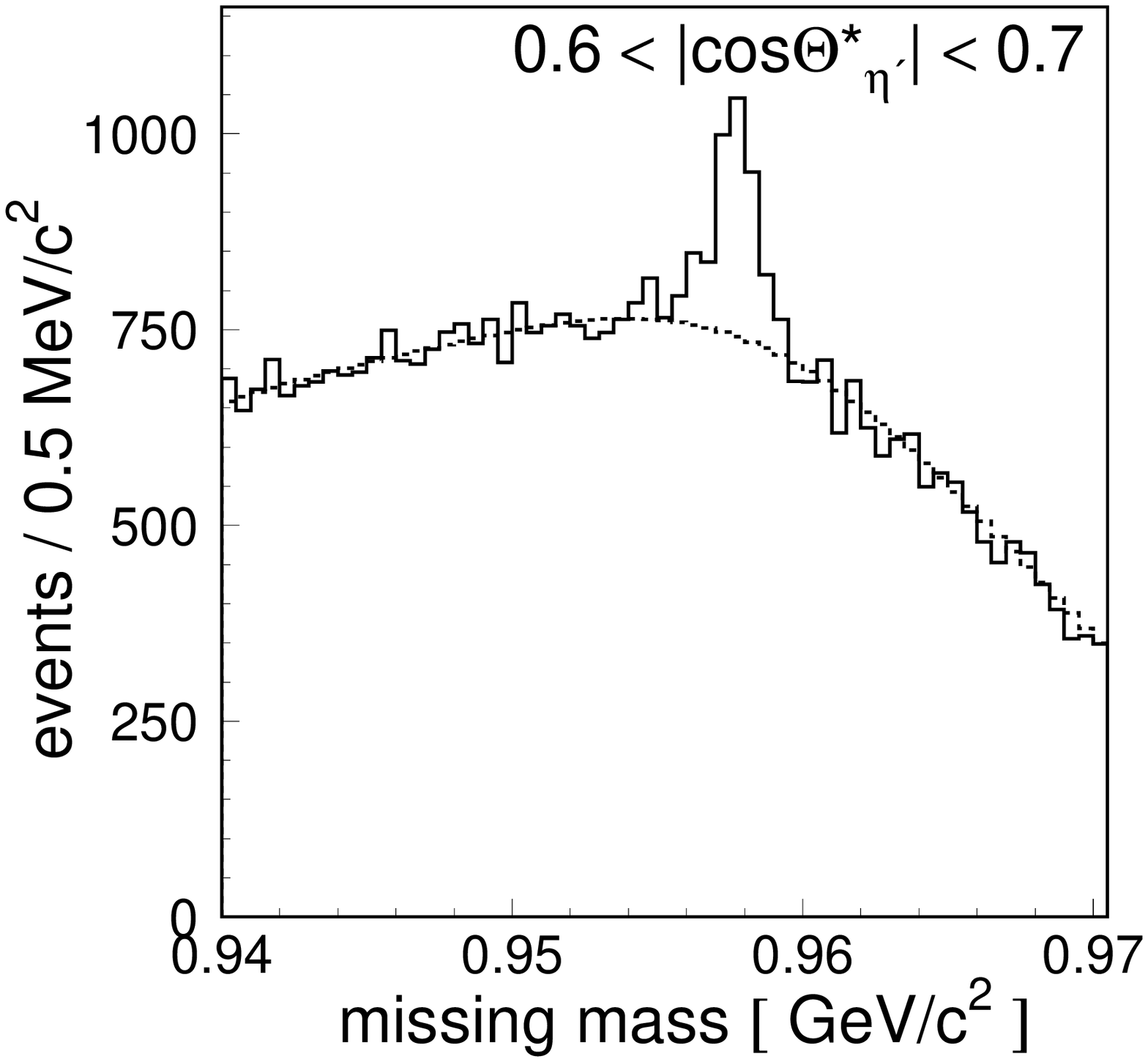}  
   \includegraphics[height=.33\textheight]{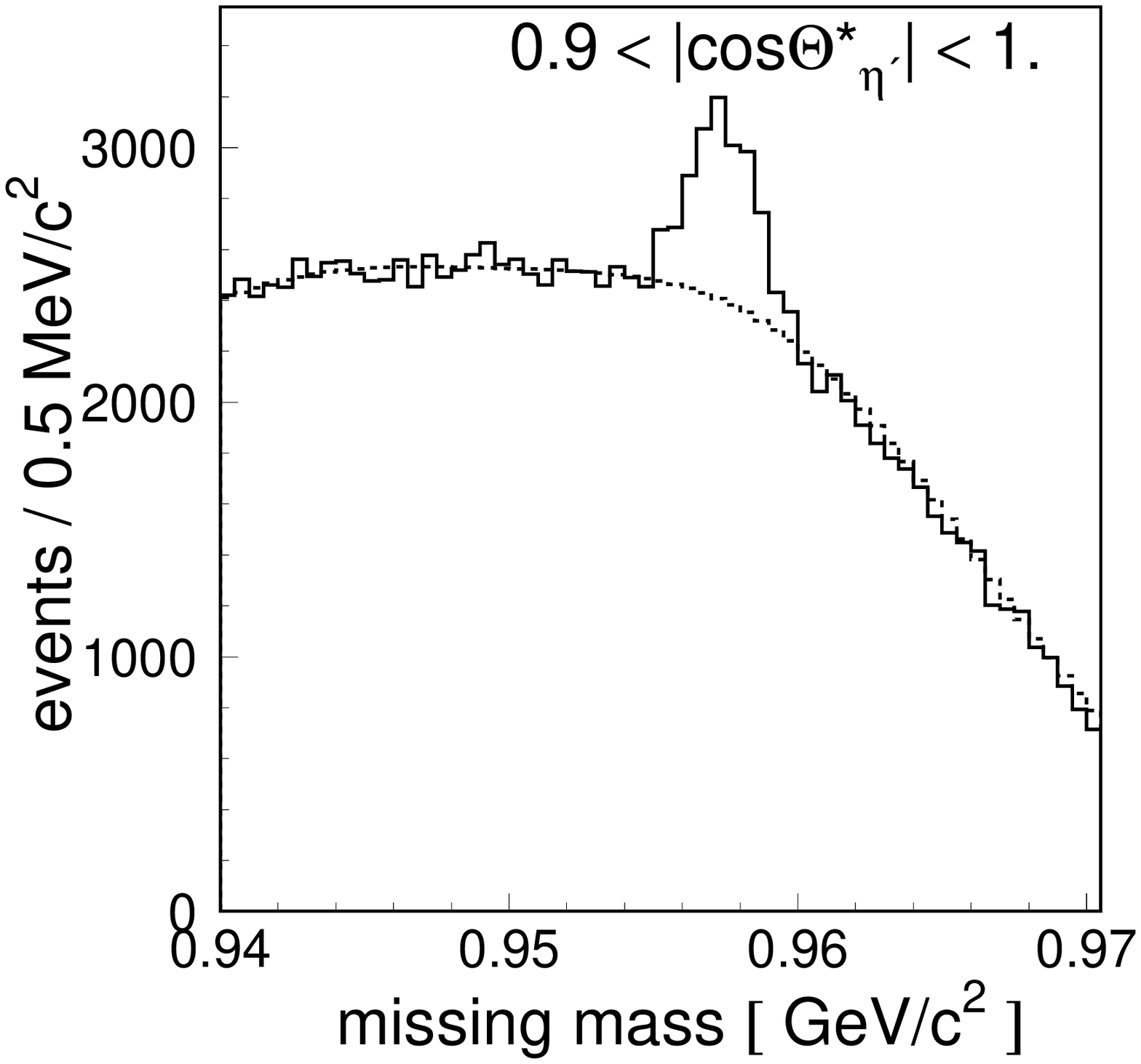} 
        \caption{Examples of missing mass spectra determined for the seventh and tenth bin
	 of $|cos\theta^{*}_{\eta^{\prime}}|$ with a width corresponding to $\Delta |cos\theta^{*}_{\eta^{\prime}}| = 0.1$.
	 The dashed lines correspond to fits using the function as described by formula \ref{equ:gauss1}.}	
	\label{fig:thetab}
\end{figure}	
	
In case of the angular distributions the background could be described as a polynomial or Gaussian 
in the whole angular range. This is because the signal from the $\eta^{\prime}$ meson is far from 
the kinematical limit for all spectra (see Fig. \ref{fig:psia}, \ref{fig:psib}, \ref{fig:thetaa} and \ref{fig:thetab}) 
and additionally the background varies smoothly on both sides of the peak. 
In this context it is also important to note that the width of the peak 
is due to the experimental resolution and since the 
background is smooth on both sides, no 
structures are being expected below the peak as it is verified by the simulation studies 
including relevant background channels.\\ 
Missing mass distributions for every third bin of $\psi$ and $|cos\theta^{*}_{\eta^{\prime}}|$ 
are presnted in figures \ref{fig:psia}, \ref{fig:psib}, \ref{fig:thetaa} and \ref{fig:thetab}, respectively, 
using formula \ref{equ:gauss1} for fitting to the  background.

\section{Systematic errors of background estimation}
\label{Systematic errors of background estimation}
The main contribution of the systematic uncertainty of the differential cross section determination comes from 
the uncertainty of the estimation of the yield of the $\eta^{\prime}$ events which in turn is due to the assumption of the shape of the background. 
In order to estimate those errors, the numbers of background events 
extracted under two different assumptions were compared. 
For the missing mass spectra with the signal far from the kinematical limit  
the background determined by Gaussian distributions with the background estimated by a second order polynomial 
were compared. 
Whereas, for the spectra close to the kinematical limit a comparison of the background determination by
Monte-Carlo simulations with a second order polynomial divided by the Fermi distribution was performed.  
The relative difference  
between determined numbers of events $N^{A}_{back}$ (two Gaussian distributions or Monte-Carlo simulations fit, respectively)
 and $N^{B}_{back}$ (polynomial fit) 
was used as an estimation for the systematic uncertainty of the differential cross sections evaluation.\\
The relative systematic errors, expressed in per cent were calculated in the following way:
\begin{equation}
\Delta_{back} =  \frac{N^{A}_{back} - N^{B}_{back}}{N^{A}_{back} + N^{B}_{back}} \cdot 100\%.
\label{equ:back}
\end{equation}
The averaged value of the systematic deviations in the various applied fit procedures calculated according to formula 
\ref{equ:back} are below $3\%$.\\
The systematic uncertainty of the differential cross sections evaluation was calculated using the formula:
\begin{equation}
\Delta_{sys} \Big( \frac{d\sigma}{df}(f) \Big) = \frac{d\sigma}{df}(f) \cdot \frac{N_{back}(f)}{N_{\eta^{\prime}}(f)} \cdot 0.03,
\label{equ:sys}
\end{equation}
where $f$ stands for both the squared invariant proton-proton mass 
$s_{pp}$ and the squared invariant proton-$\eta^{\prime}$ mass 
$s_{p\eta^{\prime}}$. $N_{back}$ and $N_{\eta^{\prime}}$ correspond to the average numbers of background and $\eta^{\prime}$ creation events, 
respectively. 
The averaged values of the systematic errors for the differential cross sections, calculated according to formula 
\ref{equ:sys}, are in the order of 0.32 $\mu b/ ($GeV$^{2}$/c$^{4})$ and  will be presented in details in the next chapter.

\chapter{Cross sections}
\label{Cross sections}
\markboth{\bf Chapter 10.}
         {\bf Cross sections}
	
This chapter is devoted to the derivation of the final results.\\
\section{Acceptance corrections}
\label{Acceptance corrections}
\begin{figure}[H]
 \includegraphics[height=.34\textheight]{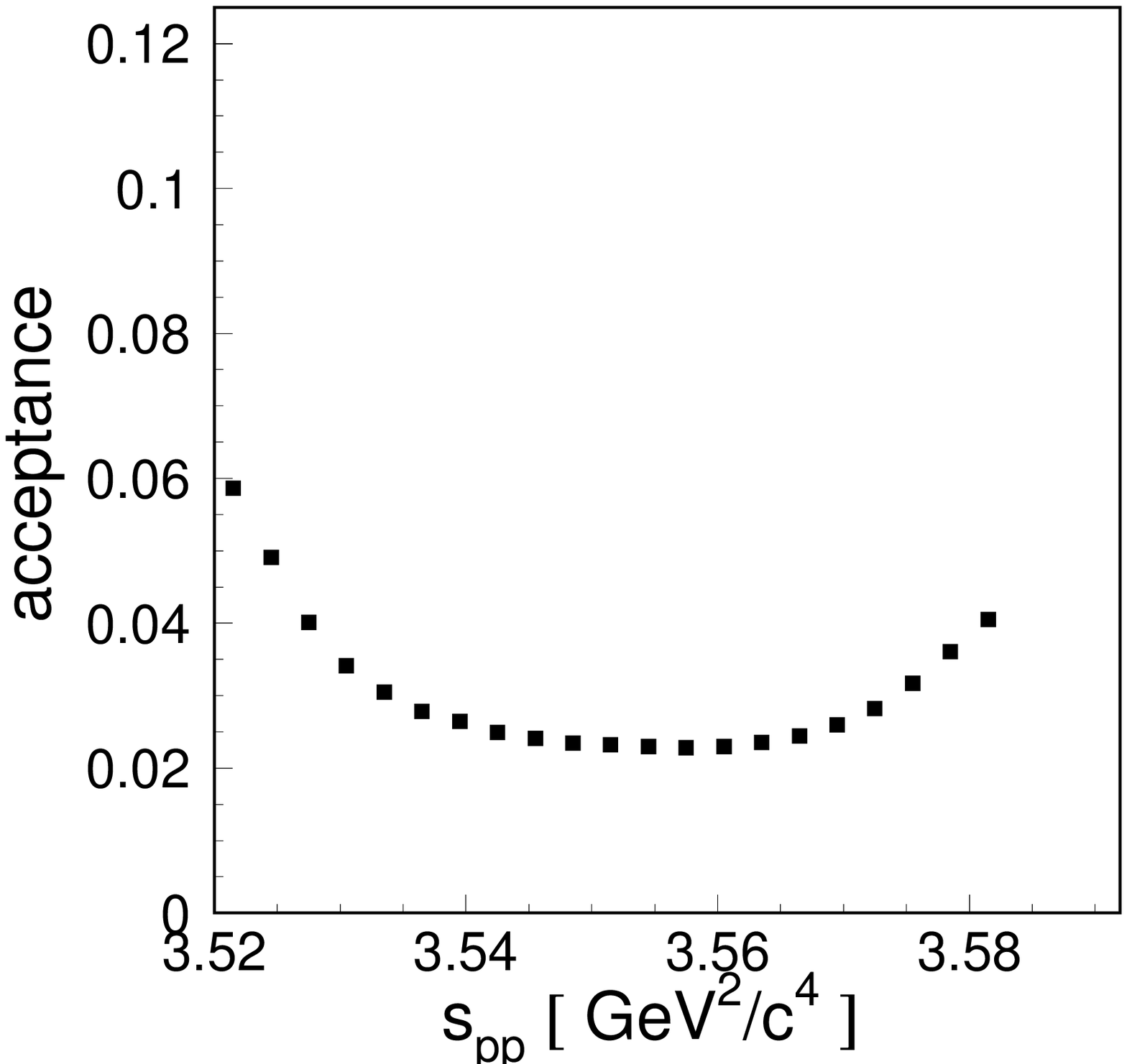}
  \includegraphics[height=.34\textheight]{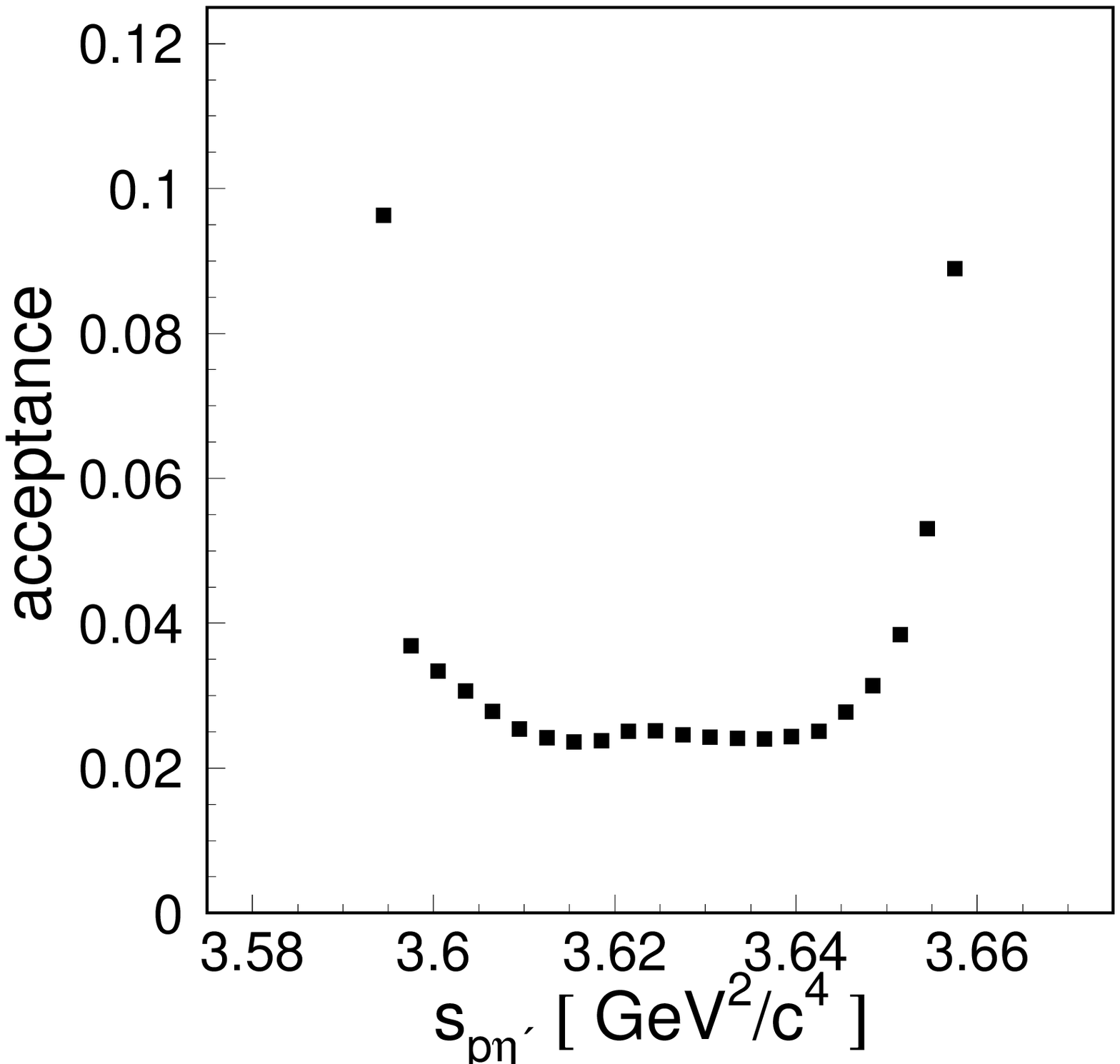}
        \caption{Geometrical acceptance of the COSY-11 detection setup determined for the $pp \to pp\eta^{\prime}$
	reaction, simulated for a beam momentum $P_{B}$~=~$3.260$ GeV/c, as a function of the squared
	invariant mass of the proton-proton subsystem (left), and the squared invariant mass of the proton-$\eta^{\prime}$ subsystem (right).}	 
\label{fig:acc_pp_peta}	
\end{figure}

\begin{figure}[H]
 \includegraphics[height=.34\textheight]{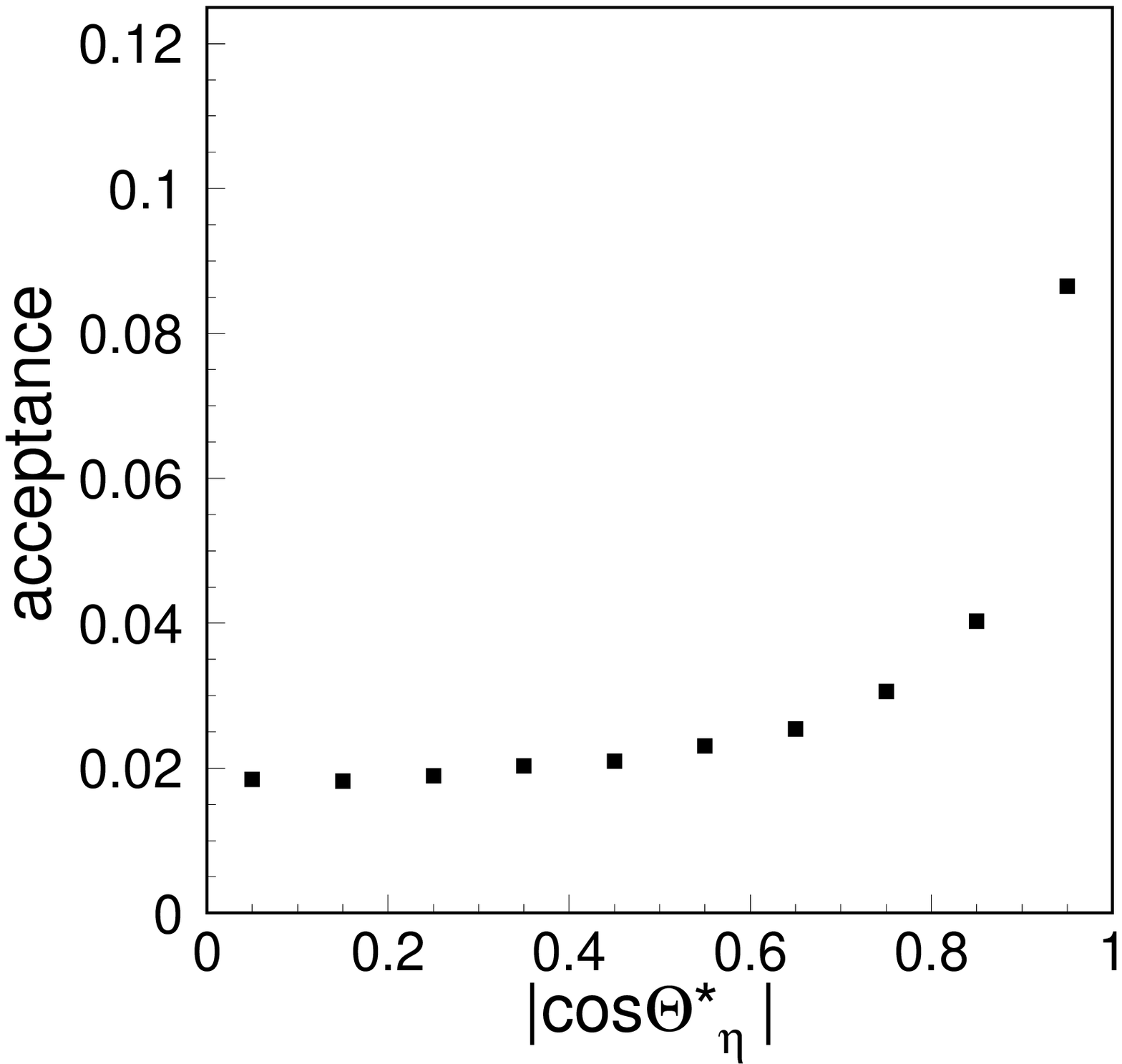}
  \includegraphics[height=.34\textheight]{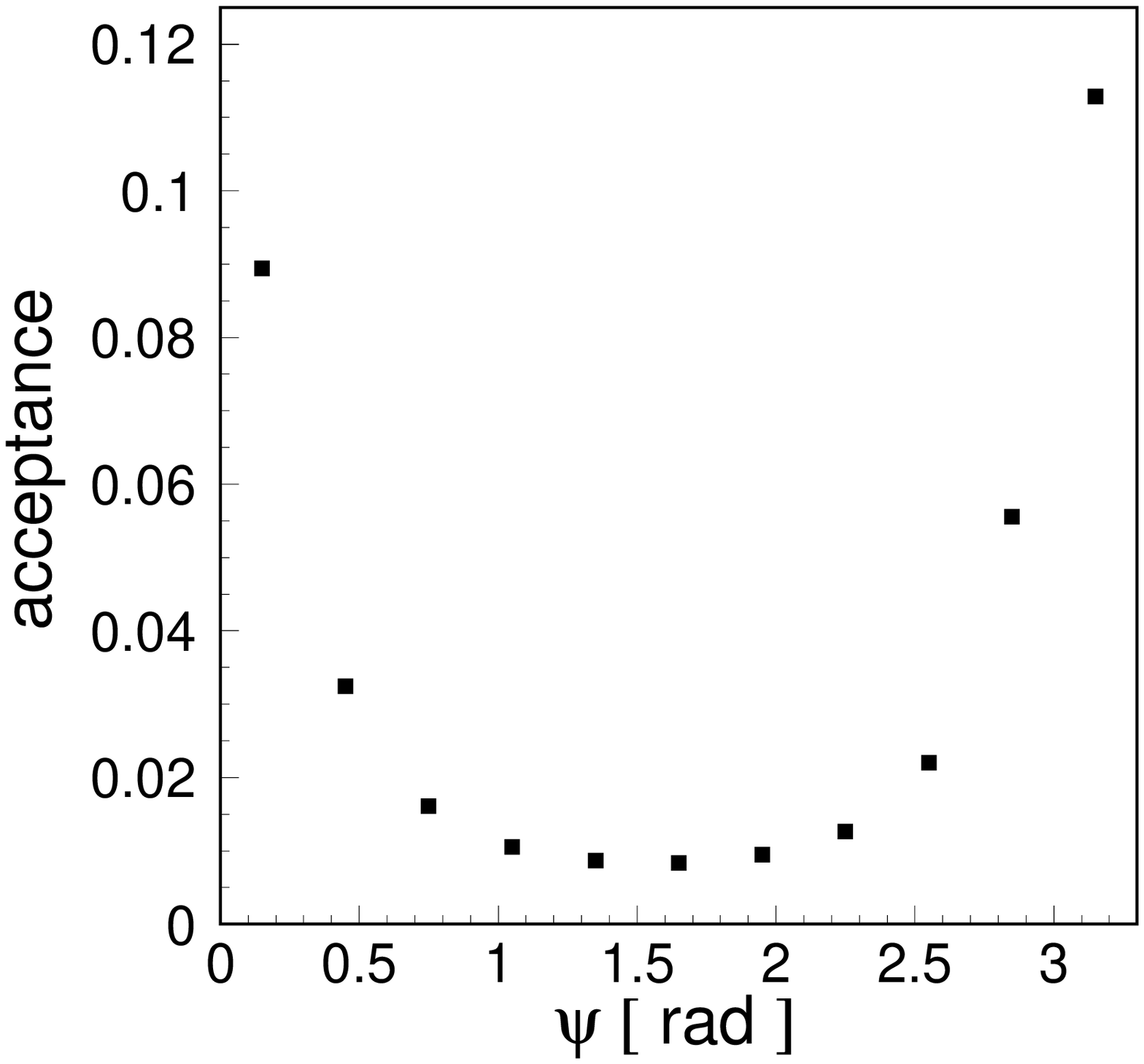}
        \caption{The COSY-11 detection acceptance as a function of the polar angle of the $\eta^{\prime}$ meson
	momentum vector $|cos\theta^{*}_{\eta^{\prime}}|$ (left), and for the $\psi$ angle describing the rotation of the
	reaction plane around the direction of the $\eta^{\prime}$ meson momentum (right).}
\label{fig:acc_cos_psi}		
\end{figure}

\begin{center}
 \begin{figure}[H]
 \includegraphics[height=.4\textheight]{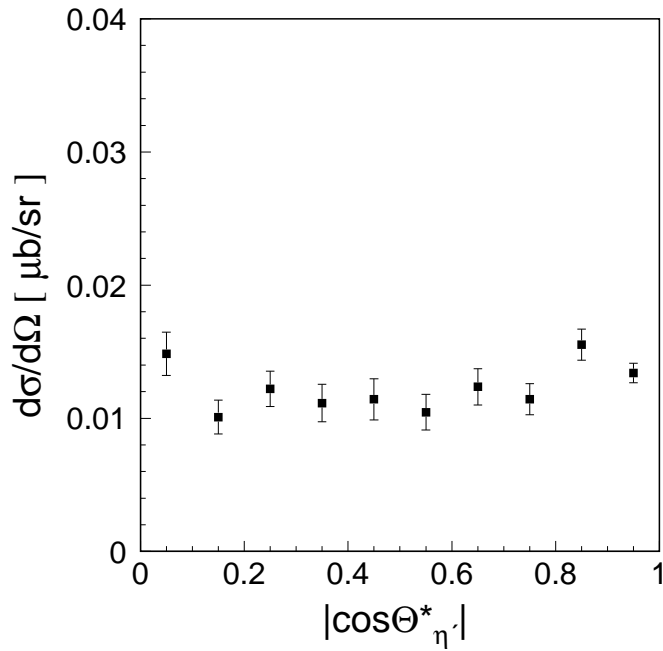}
        \caption{Distribution of $|cos\theta^{*}_{\eta^{\prime}}|$
	of the $\eta^{\prime}$ meson momentum vector in the centre-of-mass frame.
	The shown result was evaluated under the assumption that $\frac{d\sigma}{d\psi}$ is isotropic.}
	\label{fig:cos_theta}	
\end{figure}
\end{center}

\begin{center}
\begin{figure}[H]
  \includegraphics[height=.35\textheight]{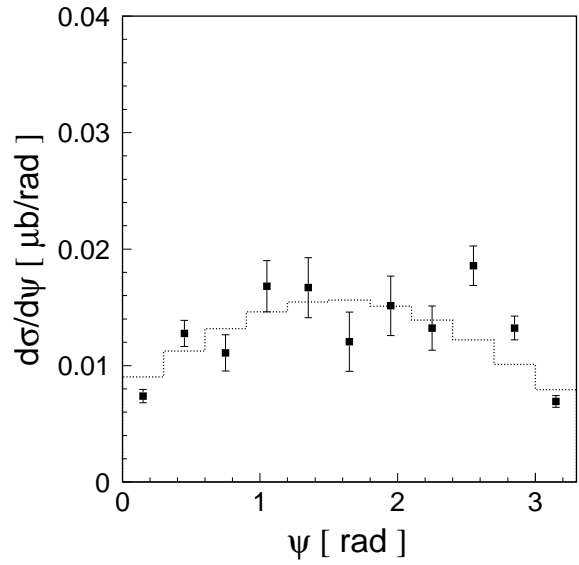}
        \caption{Distribution of the differential cross section as a function of the
	$\psi$ angle determined experimentally.
	The superimposed histogram corresponds to a fit by the function $y~$=$~a~$+$~b$~$\cdot |sin(\psi)|$, where $a~$=$~7.87~$ nb/rad and $b=7.77$ nb/rad.
	Since, in the analysis the advantage of the symmetry
	$\frac{d\sigma}{d\psi}(\psi)~=~\frac{d\sigma}{d\psi}(\psi~+~\pi)$ was taken,
	only the range of $\psi$ from 0 to $\pi$ is presented.}
	\label{fig:psi2}	
\end{figure}
\end{center}

The COSY-11 detector setup does not cover the
full $4\pi$ solid angle in the centre-of-mass system of the $pp \to pp\eta^{\prime}$ reaction at the proton beam momentum of $P_{B}$ = 3.260 GeV/c.
Therefore, in order to
study the differential cross sections, one has to perform acceptance corrections
for the measured distributions.\\
In general, the acceptance corrections should be done in the five dimensional phase space of e.g. $s_{pp}$, $s_{p\eta^{\prime}}$, $\phi$,
$\psi$ and $cos\theta^{*}_{\eta^{\prime}}$ (as depicted in chapter \ref{Definitions of observables}).\\
But one can safely assume that the dependence on the $\phi^{*}_{\eta^{\prime}}$ angle, the
centre-of-mass azimuthal angle
of the $\eta^{\prime}$ meson
momentum, is isotropic because of the axial symmetry of the initial unpolarized state. In addition, the identical particles
in the initial state imply that the distribution of  $cos\theta^{*}_{\eta^{\prime}}$ should be symmetrical around $90^{o}$.
This limits the variables for the differential acceptance to $s_{pp}$, $s_{p\eta^{\prime}}$, $\psi$ and $|cos\theta^{*}_{\eta^{\prime}}|$.\\
To determine the distributions of squared invariant masses $s_{pp}$ and $s_{p\eta^{\prime}}$,
the available range was divided into 22 bins. Number and width of the bins were chosen as a compromise between statistics and the
experimental resolution.
The width of the bins was chosen for the variables $s_{pp}$ and $s_{p\eta^{\prime}}$ to be 0.003~GeV$^{2}$/c$^{4}$.\\
In the case of angular distributions, the $|cos\theta^{*}_{\eta^{\prime}}|$ range was divided into 10 bins with a bin width
of 0.1, and  the $\psi$ range into 11 bins of 0.3 radian.\\
The acceptance correction of the data will be performed iteratively. First, it was assumed that the distributions
are determined by a homogeneous phase space occupation modulated by the $pp$-FSI.
Under this assumption the acceptance was calculated and the differential cross section for the variables $s_{pp}$, $s_{p\eta^{\prime}}$,
$\psi$ and $|cos\theta^{*}_{\eta^{\prime}}|$ was extracted. Next, using the derived distributions the acceptance
was calculated and the whole procedure was repeated until the differential distributions remain unaltered.\\
In order to calculate the differential acceptance it was assumed that the distribution over the $\psi$ angle
is isotropic like it was experimentally determined for the  $pp \to pp\rho$, $pp \to pp\omega$ and $pp \to pp\phi$ reactions \cite{balestra1, jim}.\\
The geometrical acceptance as a function of $s_{pp}$ and $s_{p\eta^{\prime}}$ is presented in the left and right panel of figure \ref{fig:acc_pp_peta},
respectively. In figure \ref{fig:acc_cos_psi}, the acceptance dependence for the $|cos\theta^{*}_{\eta^{\prime}}|$ and the
$\psi$ angular distributions is shown.\\
After performing the luminosity determination and having calculated the geometrical acceptance, one 
can evaluate angular distributions. In figure \ref{fig:cos_theta} the differential distribution of the cosine
of the $\theta^{*}_{\eta^{\prime}}$ polar angle is presented.
The distribution has an isotropic character within the statistical errors.\\
Then, the angular distribution of the $\psi$ angle was calculated. Unexpectedly, but similar to the
situation observed for such a distribution for the $pp \to pp\eta$ reaction \cite{prc69} measured at
the excess energy of 15.5 MeV, a significant anisotropic behaviour of the $\psi$ angle distribution is observed.
The $\frac{d\sigma}{d\psi}$ distribution determined experimentally is presented in figure \ref{fig:psi2}.
Such a shape cannot be explained by any background behaviour since $\frac{d\sigma}{d\psi}$ was extracted
for each bin separately, as it was mentioned earlier in section
\ref{Background subtraction}, and in fact
the distribution shown in figure \ref{fig:psi2} is background-free.
Systematic errors in the background subtraction are expected to be much smaller and can also
not cause such an anisotropy. The missing mass spectra used for the background subtraction could
satisfactorily well be reproduced for each bin in the phase space. For example see figures \ref{fig:psia} and \ref{fig:psib} in section
\ref{Background subtraction}.\\
Due to the anisotropic behaviour of the $\frac{d\sigma}{d\psi}$
the full procedure of acceptance correction for the $|cos\theta^{*}_{\eta^{\prime}}|$, $s_{pp}$ and
$s_{p\eta^{\prime}}$ distributions was performed once again
with the working assumption that the distribution $\frac{d\sigma}{d\psi}$
is as it was determined from the data. The procedure was repeated three times. After all, it was observed
that the output distributions are in good agreement with the input values.\\
It was also checked that the shape of the $\psi$ distribution assumed in the calculations
of the acceptance does not influence the shape of $s_{pp}$, $s_{p\eta^{\prime}}$ and $|cos\theta^{*}_{\eta^{\prime}}|$ distributions.
\newpage
\vspace{15.0cm}
\begin{center}
 \begin{figure}[H]
 \includegraphics[height=.35\textheight]{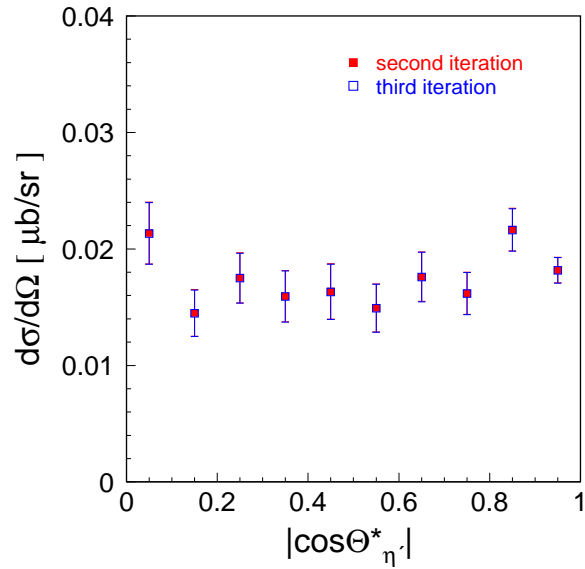}
        \caption{Distribution of the $|cos\theta^{*}_{\eta^{\prime}}|$ of the $\eta^{\prime}$ meson momentum vector
	in the centre-of-mass frame. The distributions
	after the second (closed squares) and third (open squares) iterations are compared (for details see text). The distributions are nearly identical
	and therefore the points can be hardly distinguished.}	
	\label{fig:cos_iter}
\end{figure}
\end{center}
\begin{center}
 \begin{figure}[H]
 \includegraphics[height=.39\textheight]{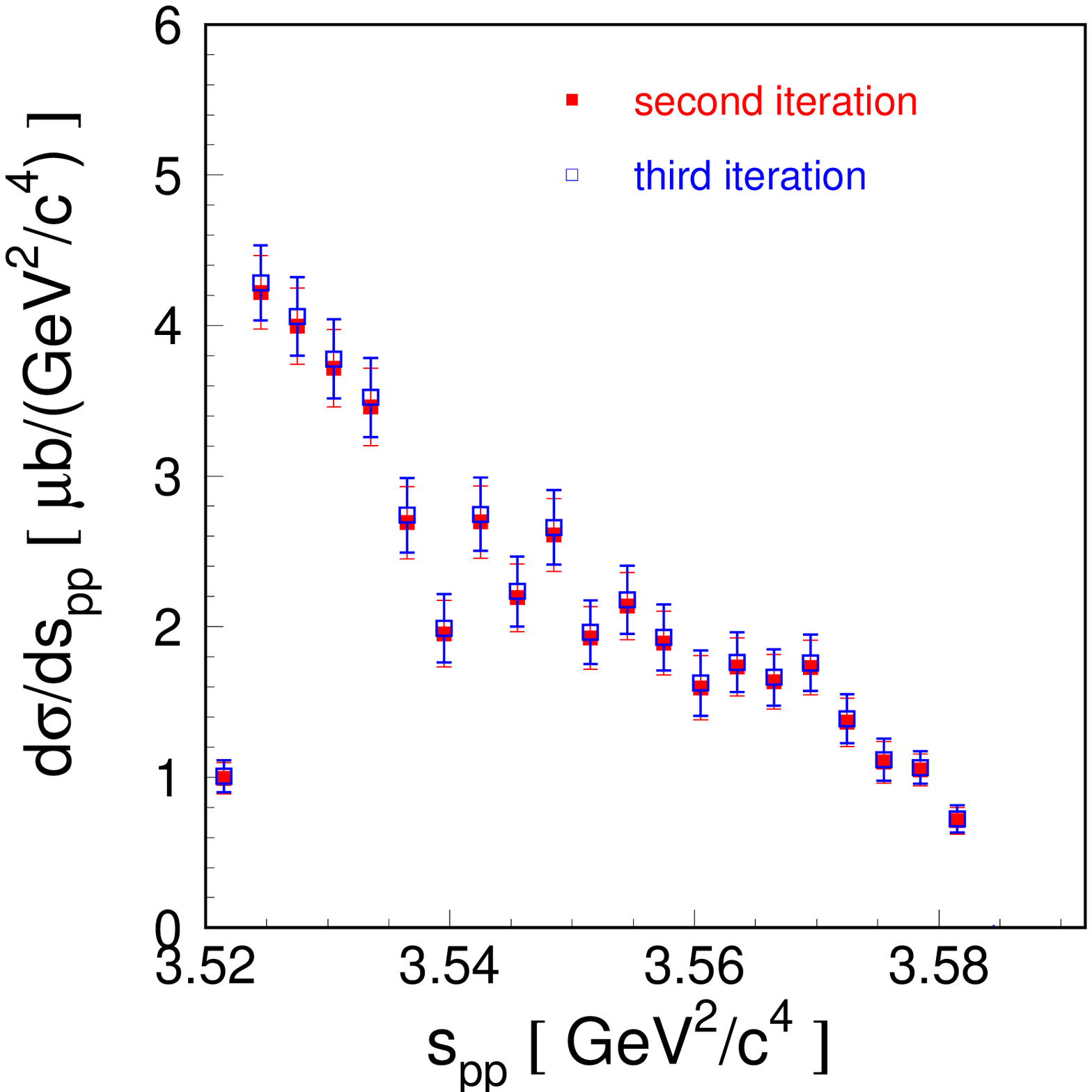}
        \caption{Comparison of the invariant mass $s_{pp}$ distributions
	after the second (showed as closed squares) and third (presented as open squares) iterations
	assuming different assumption about the $\psi$ distribution.}	
	\label{fig:pp_iter}
\end{figure}
\end{center}
\vspace{-1.5cm}
\begin{center}
\begin{figure}[H]
 \includegraphics[height=.39\textheight]{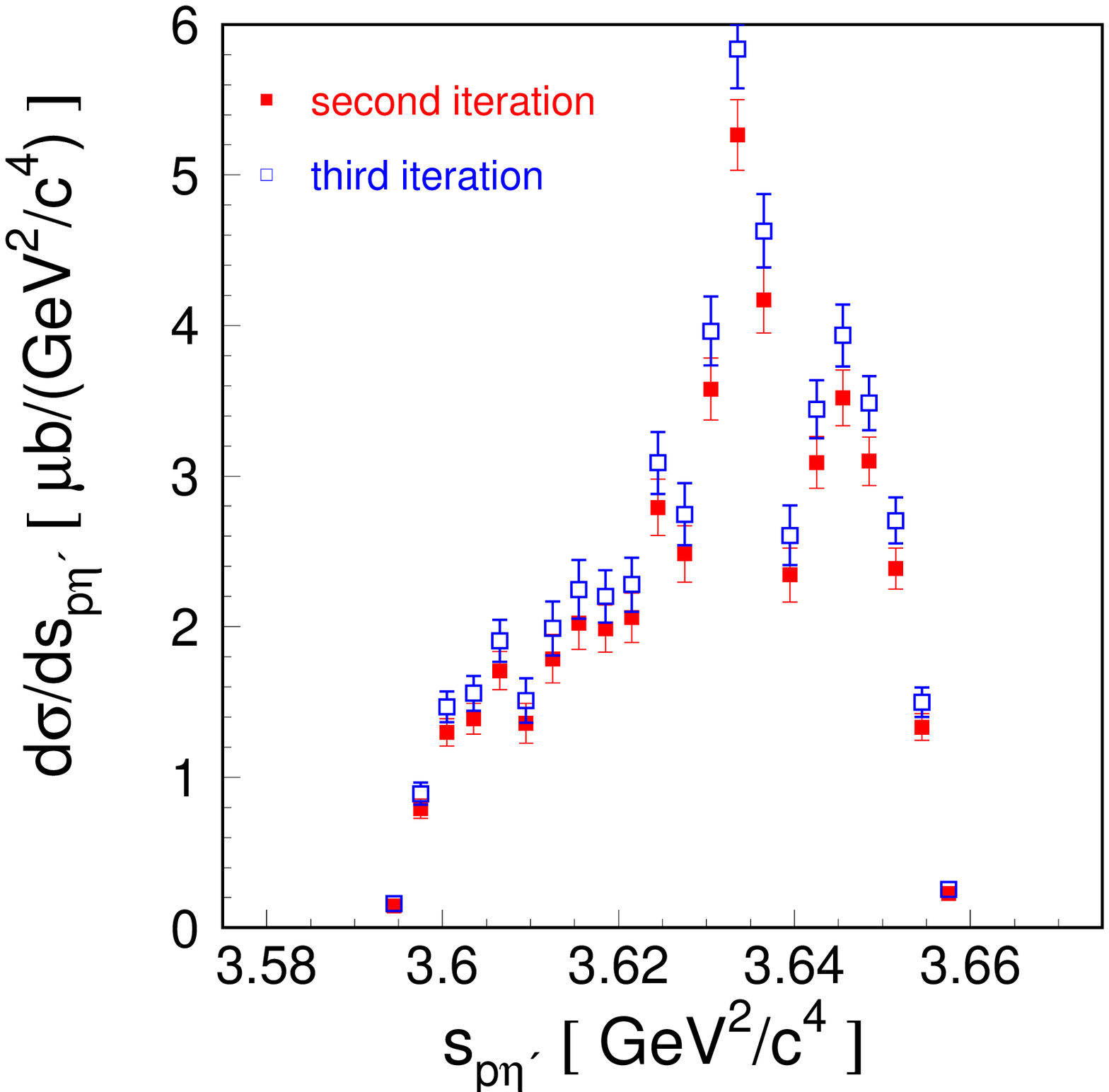}
        \caption{Comparison of the invariant mass $s_{p\eta^{\prime}}$ distributions
	after the second (showed as closed squares) and third (presented as open squares) iterations
	assuming different assumption about the $\psi$ distribution.}	
	\label{fig:peta_iter}
\end{figure}
\end{center}
\newpage
As an example figure \ref{fig:cos_iter} presents the $|cos\theta^{*}_{\eta^{\prime}}|$ spectrum 
determined assuming a $\psi$ distribution obtained from the second and the third iteration.\\
Similarly like in case of the distribution of the cross section as a function of $|cos\theta^{*}_{\eta^{\prime}}|$,
the differential cross sections as a function of $s_{pp}$ and $s_{p\eta^{\prime}}$  were extracted.
First, as it was mentioned, the distributions were determined with the assumption of
an isotropic behaviour of the angular distributions of polar and azimuthal angle
of the $\eta^{\prime}$ meson momentum vector in the centre-of-mass frame, and also an isotropic behaviour
of the $~\psi$ angle distribution. \\

Then, again using the experimental parameterization of the
distribution of the cross section as a function of the $\psi$ angle
in the centre-of-mass frame, the
invariant mass $s_{pp}$ and $s_{p\eta^{\prime}}$ distributions were determined. Results of the second and third iterations
are shown in figures \ref{fig:pp_iter} and \ref{fig:peta_iter}.\\
Again, it can be observed that the shapes of the distributions after the second and third iterations
are in agreement.

\section{Total and differential cross sections}
\label{Total and differential cross sections}
In this section, the results of the total cross section and differential cross sections
determined for the $pp \to pp\eta^{\prime}$ reaction are presented.\\

\begin{center}
\begin{figure}[H]
 \includegraphics[height=.5\textheight]{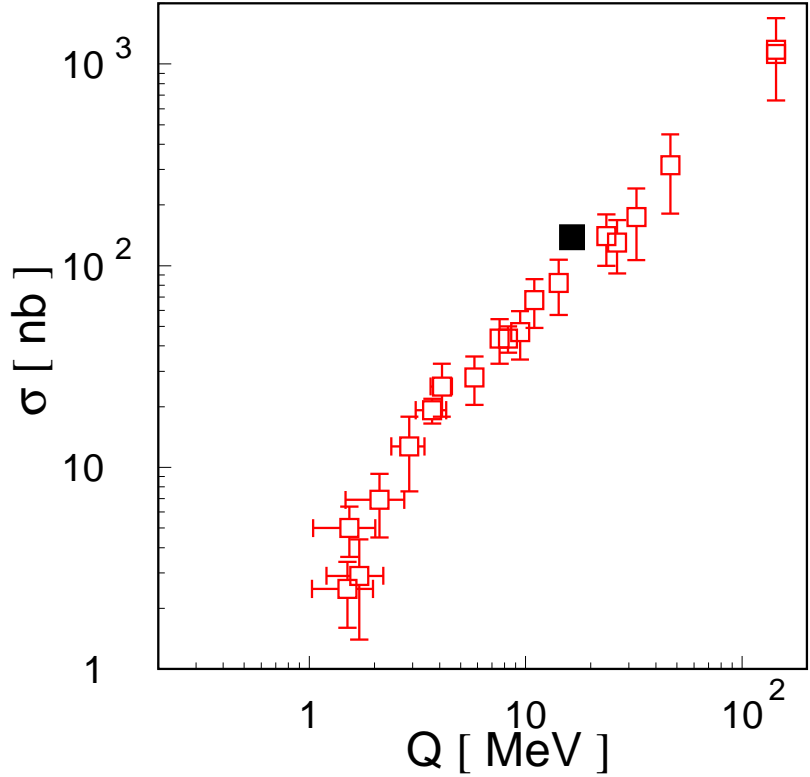}
        \caption{Excitation function for the $pp \to pp\eta^{\prime}$ reaction.
       The earlier measurements \cite{balestra, wurz, pm80, b474, khoukaz, hibou}
       are presented as open squares, the closed black square depicts the total cross section determined
       for the excess energy Q~=~16.4 MeV (the result of this thesis). The error bars
       denote the total uncertainty (statistical and systematic).}	
	\label{fig:result_cross}
\end{figure}
\end{center}

The total cross section for the $pp \to pp\eta^{\prime}$ reaction, determined at an excess energy of Q = 16.4 MeV,
is presented in figure \ref{fig:result_cross} and amounts to: $139~\pm3$ nb. In the calculations
the final distribution of the $s_{pp}$, $s_{p\eta^{\prime}}$, $\psi$ and $|cos\theta^{*}_{\eta^{\prime}}|$ were taken into account.\\
Though, the shapes of distributions $s_{pp}$, $s_{p\eta^{\prime}}$ and $|cos\theta^{*}_{\eta^{\prime}}|$ are independent of the $\psi$ distribution,
the total cross section depends quite significantly on the shape of $\frac{d\sigma}{d\psi}$ \cite{prc69} which is assumed
for the acceptance calculations.
The obtained value agrees within the errors to the previously determined
cross section values, however, it  is slightly higher than the former data.
This is due to the fact that in the previous analyses \cite{pm80, b474, khoukaz} the acceptance was calculated
only approximately disregarding the differential distributions of the $\psi$ angle
which was not established due to lack of statistics.
A similar effect of an increase in the cross section values after taking into account
the differential distribution of $\psi$ was obtained in the case of the $\eta$ meson \cite{prc69}.\\
The systematic error of the total cross section contains a systematic error of the luminosity
determination which amounts to 3$\%$ as it was evaluated in \cite{habil} and the uncertainty from the
 $\frac{d\sigma}{d\psi}$ distribution, since varying the parameters of the function $\frac{d\sigma}{d\psi} = a + b \cdot |sin(\psi)|$ 
 (within their uncertainties), 
 they change the total cross section by 2.5 nb. The averaged systematic error of the background
 subtraction (like pointed out in section \ref{Systematic errors of background estimation}) amounts to 5 nb.\\
Thus, the systematic error of the total cross section adds up to 12 nb. Summarizing,
 the total cross section for the $pp \to pp\eta^{\prime}$ reaction determined at the excess energy of 16.4 MeV
 is equal to ${\bf \sigma = (139~\pm3~\pm12)~nb}$ where the indicated errors are statistical and systematic, respectively.\\

\begin{center}
 \begin{figure}[H]
 \includegraphics[height=.4\textheight]{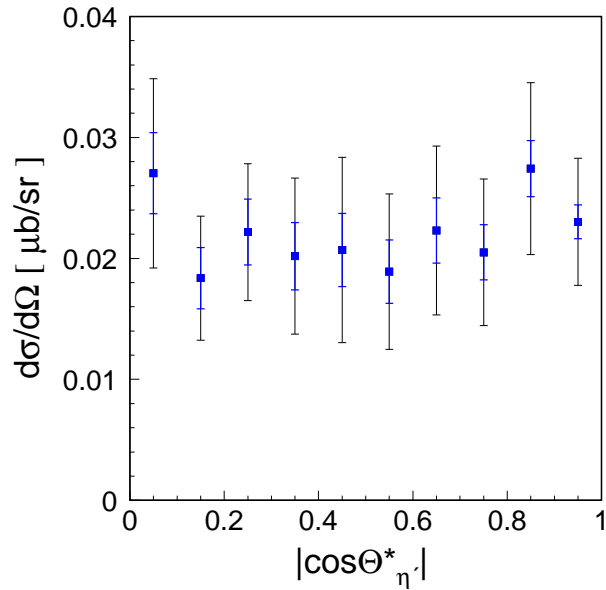}
        \caption{Distribution of the polar angle of the $\eta^{\prime}$ meson emission in the centre-of-mass system.	
	The error bars represent statistical and total (statistical $+$ systematic) uncertainties.}	
	\label{fig:cos_sys}
\end{figure}
\end{center}

\begin{center}
 \begin{figure}[H]
 \includegraphics[height=.4\textheight]{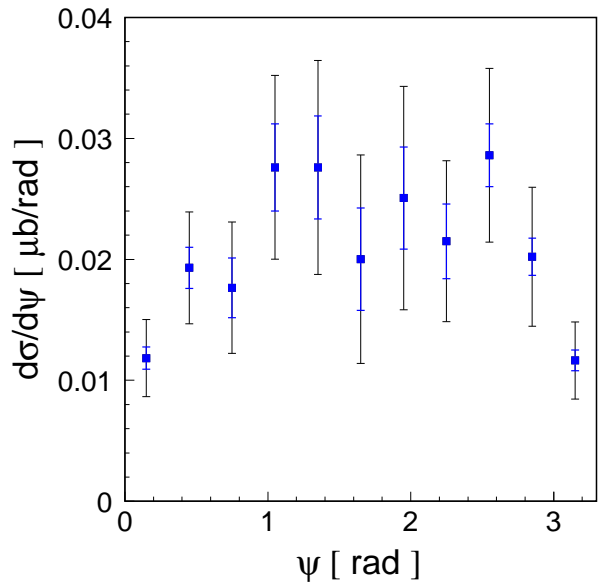}
        \caption{Differential cross section as a function of the $\psi$ angle obtained after the last iteration.
	The error bars represent statistical and total (statistical $+$ systematic) uncertainties.}	
	\label{fig:psi_sys}
\end{figure}
\end{center}

\begin{figure}[H]
 \includegraphics[height=.34\textheight]{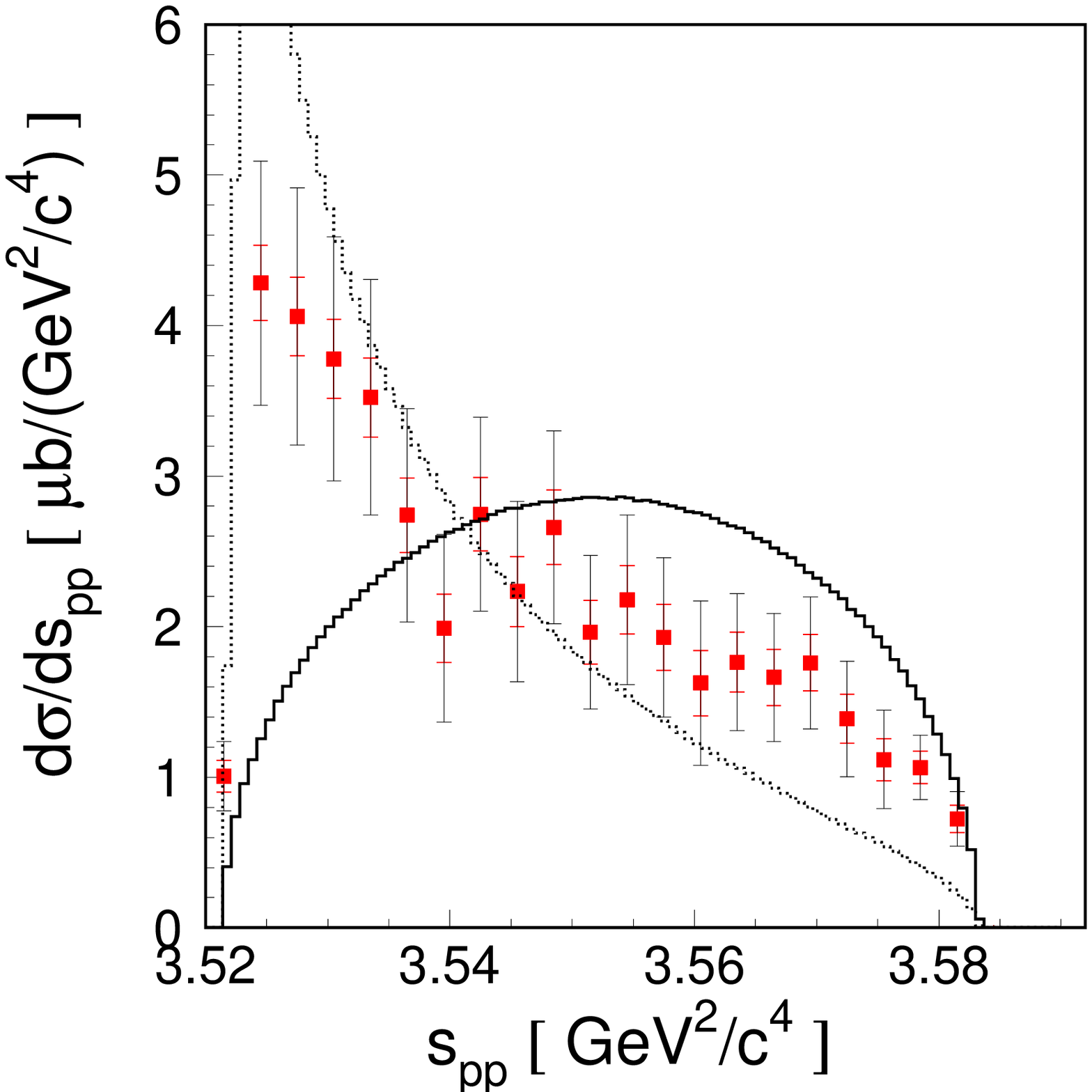}
  \includegraphics[height=.34\textheight]{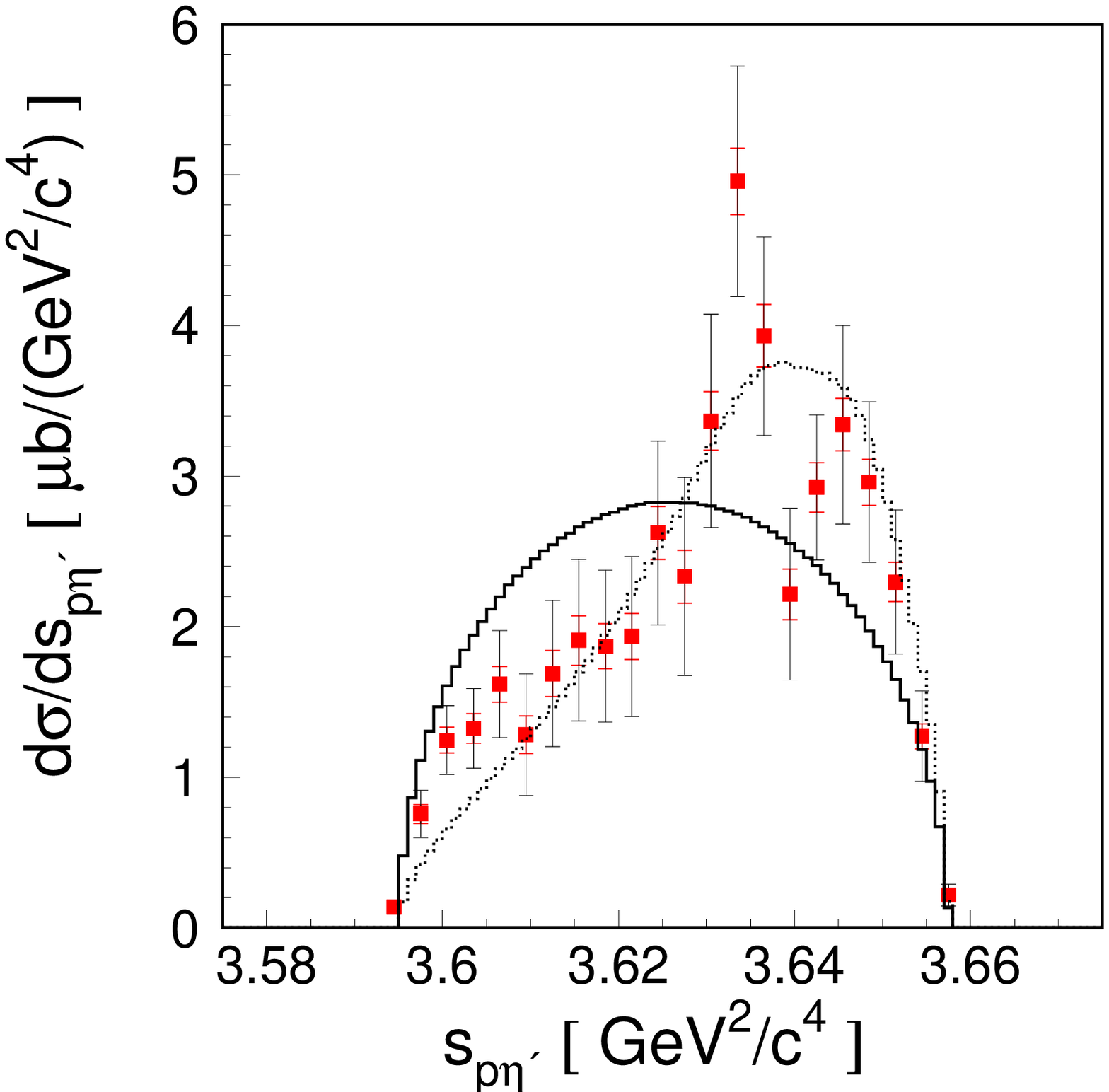}
        \caption{Distributions of the squared proton-proton ($s_{pp}$)
	and proton-$\eta^{\prime}$ ($s_{p\eta^{\prime}}$) invariant masses, determined experimentally
	for the $pp \to pp\eta^{\prime}$ reaction at the excess energy of Q = 16.4 MeV (closed squares).
	The experimental data are
	compared to the expectation under the assumption of a homogeneously populated phase space
	which is presented as solid lines. The integrals of the phase space weighted
	by the proton-proton scattering amplitude - FSI$_{pp}$ are marked as dotted histograms.
	The results of the simulations were normalized to the data in amplitude.}
	\label{fig:norm_pp_peta}	
\end{figure}

In the following tables and figures, the values of the differential cross sections as functions of the
observables $|cos\theta^{*}_{\eta^{\prime}}|$, $\psi$, $s_{pp}$, and $s_{p\eta^{\prime}}$ defined earlier 
in section \ref{Definitions of observables} are presented.\\
In all tables and figures statistical and systematic errors are included. \\
The distribution of the polar angle is listed in table \ref{tab_cos} and presented in figure  \ref{fig:cos_sys}.\\
The distribution of the $\psi$ angle is listed in table \ref{tab_psi} and presented in figure  \ref{fig:psi_sys}.\\
In  tables \ref{tab_pp} and \ref{tab_peta}, the results for the differential cross sections of the invariant
masses $pp$ and $p\eta^{\prime}$ are presented, respectively. The listed values of both distributions
are shown in figure \ref{fig:norm_pp_peta}.\\
\newpage
\begin{center}

\end{center}
\begin{table}[H]
\begin{center}
\begin{tabular}{c c}
\hline
  $|cos\theta^{*}_{\eta^{\prime}}|$ & $\frac{d\sigma}{d\Omega}(|cos\theta^{*}_{\eta^{\prime}}|)~[\mu b/sr]$\\
\hline
0.05 & 0.0270 $\pm~$ 0.0034$_{stat}$ $\pm~$ 0.0045$_{sys}$\\
0.15 & 0.0184 $\pm~$ 0.0025$_{stat}$ $\pm~$ 0.0026$_{sys}$\\
0.25 & 0.0222 $\pm~$ 0.0027$_{stat}$ $\pm~$ 0.0029$_{sys}$\\
0.35 & 0.0202 $\pm~$ 0.0028$_{stat}$ $\pm~$ 0.0037$_{sys}$\\
0.45 & 0.0207 $\pm~$ 0.0030$_{stat}$ $\pm~$ 0.0046$_{sys}$\\
0.55 & 0.0189 $\pm~$ 0.0026$_{stat}$ $\pm~$ 0.0038$_{sys}$\\
0.65 & 0.0223 $\pm~$ 0.0027$_{stat}$ $\pm~$ 0.0043$_{sys}$\\
0.75 & 0.0205 $\pm~$ 0.0023$_{stat}$ $\pm~$ 0.0038$_{sys}$\\
0.85 & 0.0274 $\pm~$ 0.0023$_{stat}$ $\pm~$ 0.0048$_{sys}$\\
0.95 & 0.0230 $\pm~$ 0.0014$_{stat}$ $\pm~$ 0.0039$_{sys}$\\
\hline
\end{tabular}
\caption{The differential cross sections as a function of $|cos\theta^{*}_{\eta^{\prime}}|$ for the $pp \to pp\eta^{\prime}$ reaction measured at Q = 16.4 MeV.}
\label{tab_cos}
\end{center}
\end{table}
\begin{center}

\end{center}
\vspace{-1.0cm}
\begin{table}[H]
\begin{center}
\begin{tabular}{c c}
\hline
  $\psi~[rad]$  & $\frac{d\sigma}{d\psi}~[\mu b/rad]$\\
\hline
0.15 & 0.0118 $\pm~$ 0.0009$_{stat}$ $\pm~$ 0.0023$_{sys}$\\
0.45 & 0.0193 $\pm~$ 0.0017$_{stat}$ $\pm~$ 0.0029$_{sys}$\\
0.75 & 0.0177 $\pm~$ 0.0025$_{stat}$ $\pm~$ 0.0030$_{sys}$\\
1.05 & 0.0276 $\pm~$ 0.0036$_{stat}$ $\pm~$ 0.0040$_{sys}$\\
1.35 & 0.0276 $\pm~$ 0.0043$_{stat}$ $\pm~$ 0.0046$_{sys}$\\
1.65 & 0.0200 $\pm~$ 0.0042$_{stat}$ $\pm~$ 0.0044$_{sys}$\\
1.95 & 0.0251 $\pm~$ 0.0042$_{stat}$ $\pm~$ 0.0050$_{sys}$\\
2.25 & 0.0215 $\pm~$ 0.0031$_{stat}$ $\pm~$ 0.0036$_{sys}$\\
2.55 & 0.0286 $\pm~$ 0.0026$_{stat}$ $\pm~$ 0.0046$_{sys}$\\
2.85 & 0.0202 $\pm~$ 0.0016$_{stat}$ $\pm~$ 0.0042$_{sys}$\\
3.15 & 0.0117 $\pm~$ 0.0009$_{stat}$ $\pm~$ 0.0024$_{sys}$\\
\hline
\end{tabular}
\caption{Differential cross sections as a function of the $\psi$ angle determined for the $pp \to pp\eta^{\prime}$ reaction measured at Q = 16.4 MeV. 
Since, $\frac{d\sigma}{d\psi}(\psi)~=~\frac{d\sigma}{d\psi}(\psi~+~\pi)$, the full range for $\psi$ angle is assumed from 0 to $\pi$.}
\label{tab_psi}
\end{center}
\end{table}
\newpage

\begin{table}
\begin{center}
\begin{tabular}{c l}
\hline
  $s_{pp}~[GeV^{2}/c^{4}]$  & $\frac{d\sigma}{ds_{pp}}~[\mu $b/GeV$^{2}$/c$^{4}]$\\
\hline
3.5215 & 1.01 $\pm~$ 0.11$_{stat}$ $\pm~$ 0.12$_{sys}$\\
3.5245 & 4.28 $\pm~$ 0.25$_{stat}$ $\pm~$ 0.56$_{sys}$\\
3.5275 & 4.06 $\pm~$ 0.26$_{stat}$ $\pm~$ 0.59$_{sys}$\\
3.5305 & 3.78 $\pm~$ 0.26$_{stat}$ $\pm~$ 0.55$_{sys}$\\
3.5335 & 3.52 $\pm~$ 0.26$_{stat}$ $\pm~$ 0.52$_{sys}$\\
3.5365 & 2.74 $\pm~$ 0.25$_{stat}$ $\pm~$ 0.46$_{sys}$\\
3.5395 & 1.99 $\pm~$ 0.23$_{stat}$ $\pm~$ 0.40$_{sys}$\\
3.5425 & 2.75 $\pm~$ 0.25$_{stat}$ $\pm~$ 0.40$_{sys}$\\
3.5455 & 2.23 $\pm~$ 0.23$_{stat}$ $\pm~$ 0.37$_{sys}$\\
3.5485 & 2.66 $\pm~$ 0.25$_{stat}$ $\pm~$ 0.39$_{sys}$\\
3.5515 & 1.96 $\pm~$ 0.21$_{stat}$ $\pm~$ 0.30$_{sys}$\\
3.5545 & 2.18 $\pm~$ 0.23$_{stat}$ $\pm~$ 0.34$_{sys}$\\
3.5575 & 1.93 $\pm~$ 0.22$_{stat}$ $\pm~$ 0.31$_{sys}$\\
3.5605 & 1.62 $\pm~$ 0.22$_{stat}$ $\pm~$ 0.33$_{sys}$\\
3.5635 & 1.76 $\pm~$ 0.20$_{stat}$ $\pm~$ 0.26$_{sys}$\\
3.5665 & 1.66 $\pm~$ 0.19$_{stat}$ $\pm~$ 0.24$_{sys}$\\
3.5695 & 1.76 $\pm~$ 0.19$_{stat}$ $\pm~$ 0.26$_{sys}$\\
3.5725 & 1.39 $\pm~$ 0.16$_{stat}$ $\pm~$ 0.22$_{sys}$\\
3.5755 & 1.12 $\pm~$ 0.14$_{stat}$ $\pm~$ 0.19$_{sys}$\\
3.5785 & 1.07 $\pm~$ 0.11$_{stat}$ $\pm~$ 0.11$_{sys}$\\
3.5815 & 0.72 $\pm~$ 0.09$_{stat}$ $\pm~$ 0.09$_{sys}$\\
3.5845 & 0.013 $\pm~$ 0.004$_{stat}$ $\pm~$ 0.002$_{sys}$\\
\hline
\end{tabular}
\caption{Distribution of the squared invariant mass of the proton-proton system ($s_{pp}$),
determined for the $pp \to pp\eta^{\prime}$ reaction measured at Q = 16.4 MeV. }
\label{tab_pp}
\end{center}
\end{table}

\begin{table}
\begin{center}
\begin{tabular}{c l}
\hline
  $s_{p\eta^{\prime}}~[GeV^{2}/c^{4}]$  & $\frac{d\sigma}{ds_{p\eta^{\prime}}}~[\mu $b/GeV$^{2}$/c$^{4}]$\\
\hline
3.5945 & 0.14 $\pm~$ 0.02$_{stat}$ $\pm~$ 0.02$_{sys}$\\
3.5975 & 0.76 $\pm~$ 0.06$_{stat}$ $\pm~$ 0.09$_{sys}$\\
3.6005 & 1.25 $\pm~$ 0.09$_{stat}$ $\pm~$ 0.14$_{sys}$\\
3.6035 & 1.32 $\pm~$ 0.10$_{stat}$ $\pm~$ 0.17$_{sys}$\\
3.6065 & 1.62 $\pm~$ 0.12$_{stat}$ $\pm~$ 0.24$_{sys}$\\
3.6095 & 1.28 $\pm~$ 0.13$_{stat}$ $\pm~$ 0.28$_{sys}$\\
3.6125 & 1.69 $\pm~$ 0.15$_{stat}$ $\pm~$ 0.33$_{sys}$\\
3.6155 & 1.91 $\pm~$ 0.16$_{stat}$ $\pm~$ 0.37$_{sys}$\\
3.6185 & 1.87 $\pm~$ 0.15$_{stat}$ $\pm~$ 0.36$_{sys}$\\
3.6215 & 1.94 $\pm~$ 0.15$_{stat}$ $\pm~$ 0.38$_{sys}$\\
3.6245 & 2.62 $\pm~$ 0.18$_{stat}$ $\pm~$ 0.44$_{sys}$\\
3.6275 & 2.33 $\pm~$ 0.18$_{stat}$ $\pm~$ 0.48$_{sys}$\\
3.6305 & 3.37 $\pm~$ 0.19$_{stat}$ $\pm~$ 0.51$_{sys}$\\
3.6335 & 4.96 $\pm~$ 0.22$_{stat}$ $\pm~$ 0.54$_{sys}$\\
3.6365 & 3.93 $\pm~$ 0.21$_{stat}$ $\pm~$ 0.45$_{sys}$\\
3.6395 & 2.21 $\pm~$ 0.17$_{stat}$ $\pm~$ 0.40$_{sys}$\\
3.6425 & 2.93 $\pm~$ 0.16$_{stat}$ $\pm~$ 0.32$_{sys}$\\
3.6455 & 3.34 $\pm~$ 0.18$_{stat}$ $\pm~$ 0.48$_{sys}$\\
3.6485 & 2.96 $\pm~$ 0.15$_{stat}$ $\pm~$ 0.38$_{sys}$\\
3.6515 & 2.30 $\pm~$ 0.13$_{stat}$ $\pm~$ 0.35$_{sys}$\\
3.6545 & 1.27 $\pm~$ 0.08$_{stat}$ $\pm~$ 0.22$_{sys}$\\
3.6575 & 0.22 $\pm~$ 0.03$_{stat}$ $\pm~$ 0.04$_{sys}$\\
\hline
\end{tabular}
\caption{Distribution of the squared invariant mass of the proton-$\eta^{\prime}$ system ($s_{p\eta^{\prime}}$),
determined for the $pp \to pp\eta^{\prime}$ reaction measured at Q = 16.4 MeV. }
\label{tab_peta}
\end{center}
\end{table}

In figure \ref{fig:norm_pp_peta} experimentally determined distributions are compared to the
expectation under the assumption of a homogeneously populated phase space, presented in the
figure  as solid lines. One can see that, for both distributions
the data do not agree with the calculations. In the same figure
the experimental spectra are compared also to the integrals of the phase space weighted
by the proton-proton scattering amplitude - FSI$_{pp}$ which are depicted as dotted histograms.
Also the calculations including only the FSI$_{pp}$ do not fit to the data satisfactorily.
One can see that the comparison depends on the way of normalization which will
be discussed later.\\

\section{Comparison with results for the $\eta$ meson production}
\begin{figure}[H]
 \includegraphics[height=.3\textheight]{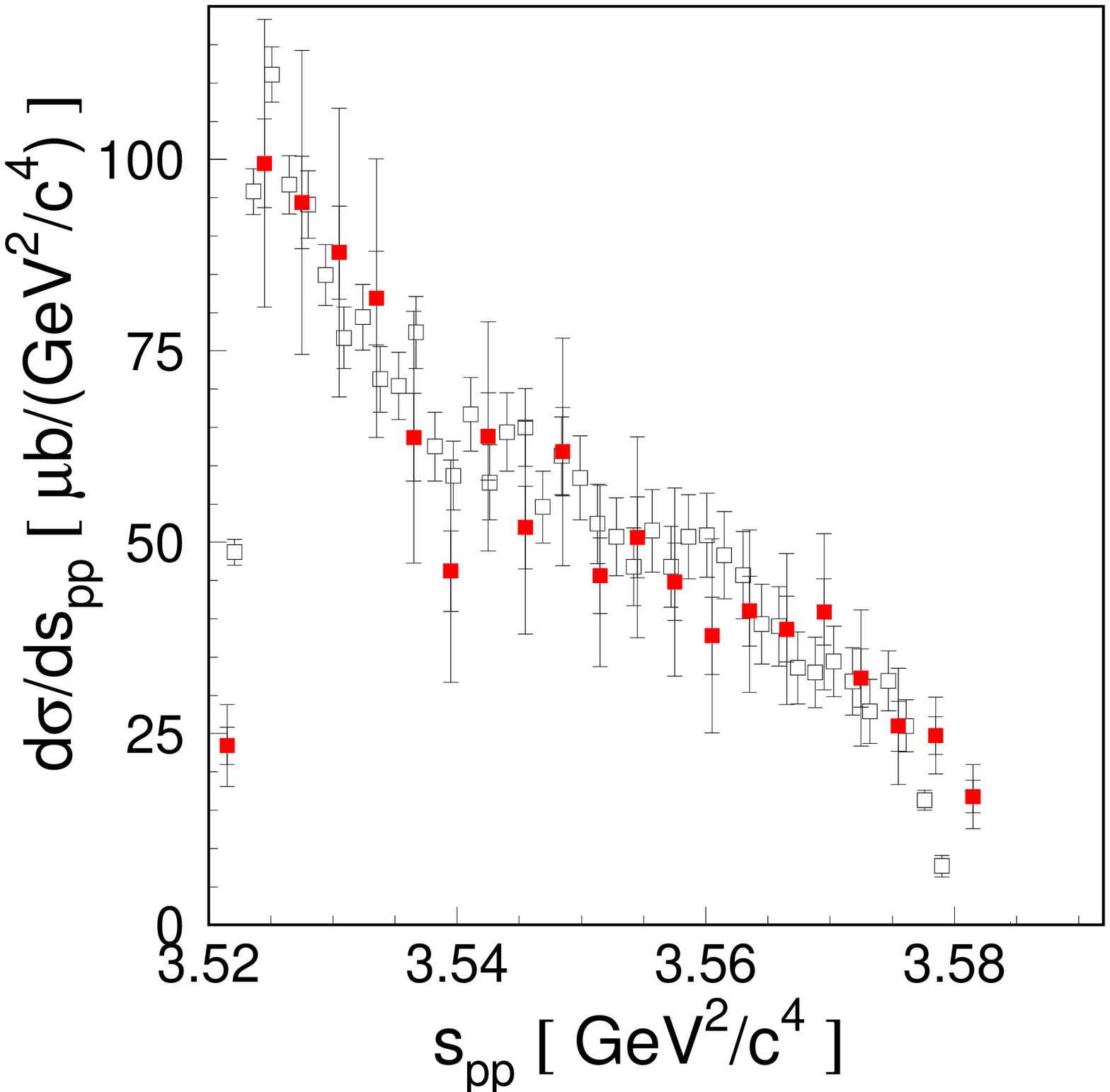}
  \includegraphics[height=.3\textheight]{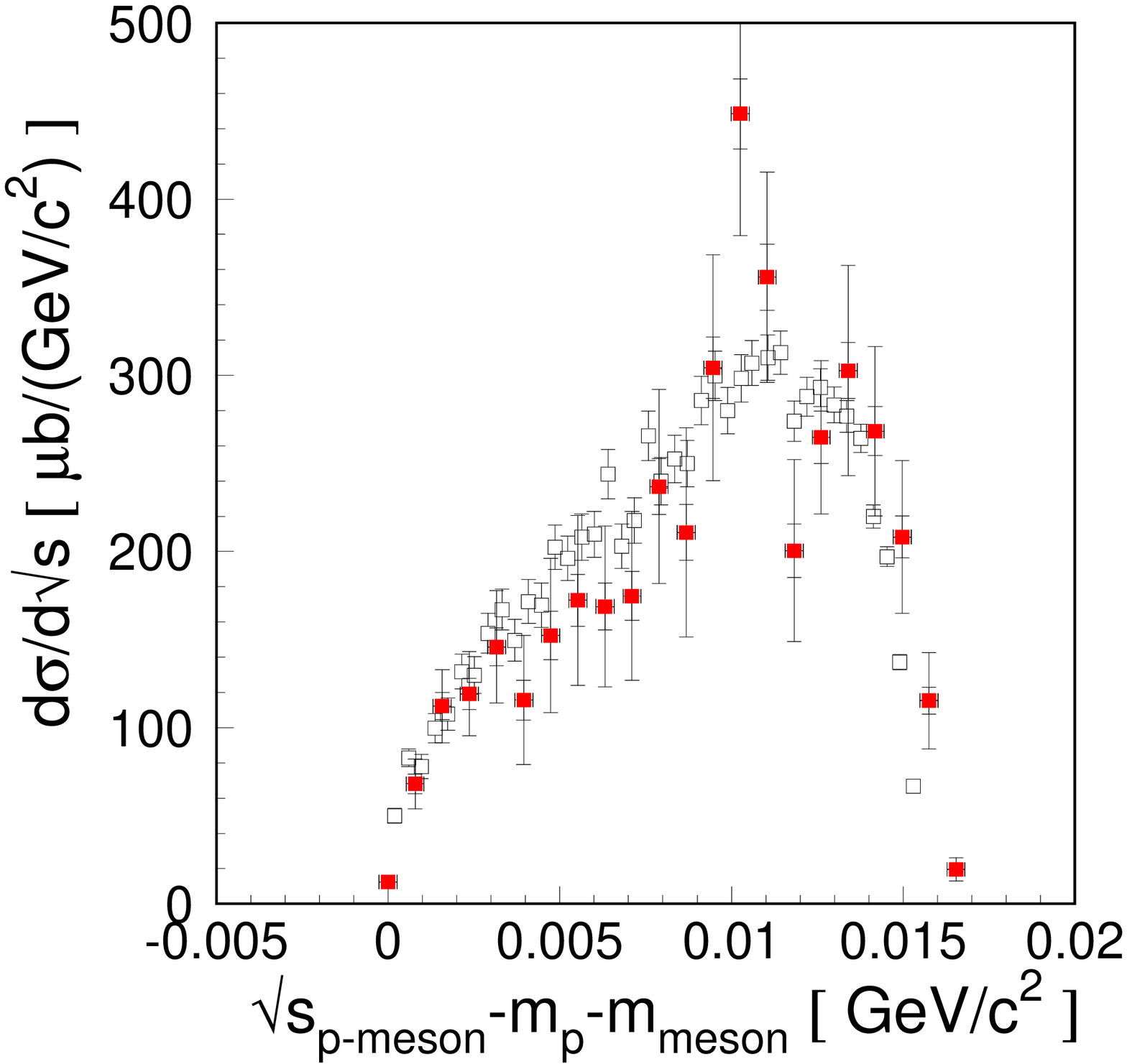}
        \caption{The comparison of the distributions of the square of the proton-proton invariant mass ($s_{pp}$) (left panel)
	and of proton-meson kinetic energy ($\sqrt{s_{p-meson}}~-~m_{p}~-~m_{meson}$) (right panel). The distributions for the $pp \to pp\eta^{\prime}$
	reaction (closed squares)
	were normalized to the distributions for the $pp \to pp\eta$ reaction (open squares) with respect to the total cross section.}
	\label{fig:comparison}	
\end{figure}

The comparison of the distributions of the square of the proton-proton ($s_{pp}$)
and proton-meson ($s_{p-meson}$) invariant masses between the $pp \to pp\eta^{\prime}$ and the $pp \to pp\eta$ reactions 
is presented in figure \ref{fig:comparison}.
For the proton-meson system the comparison was performed for the
kinetic energy of the proton-meson system ($\sqrt{s_{p-meson}}~-~m_{p}~-~m_{meson}$)
and not as a function of $s_{p-meson}$ because the range of the $s_{p\eta}$ and $s_{p\eta^{\prime}}$ are different
due to the different masses of the $\eta$ and $\eta^{\prime}$ mesons.
But the range of ($\sqrt{s_{p-meson}}~-~m_{p}~-~m_{meson}$) is the same since
the measurements for the $\eta$ and $\eta^{\prime}$ production were by purpose performed at about
the same excess energy.
The differential distributions as a function of $s_{pp}$ are compared directly, because we
compare the same system at the same energy range.\\
In both panels it is seen that the experimental points indicating the $pp \to pp\eta$
measurement (open squares) are in agreement with those from the $pp \to pp\eta^{\prime}$ reaction (closed squares) within the error bars.
It is unexpected that the shapes do not differ, showing the same enhancement
at the same values of the square of the proton-proton ($s_{pp}$) invariant mass.\\
The total cross sections as a function of excess energy for the $pp \to pp\eta$ and $pp \to pp\eta^{\prime}$
reactions and the comparison of the ratios of the production amplitudes for both reactions
and the production amplitude for the $pp \to pp\pi^{0}$ reaction showed that the interaction
within the proton-$\eta$ should have been stronger than the proton-$\eta^{\prime}$ system interaction.\\
Therefore, if indeed the $\eta^{\prime}$-proton interaction is much smaller than the $\eta$-proton
as inferred from the excitation function, then the spectra presented here
rather exclude the hypothesis that the enhancement (see figures and discussion in section \ref{invariant mass distributions}) is due to the interaction
of the meson and the proton. Moreover, based on those distributions it is not possible
to disentangle between the hypothesis of the admixture of higher partial waves during the $\eta$ production or
the energy dependence of the production amplitude.\\

\section{$s_{pp}$ and $s_{p\eta^{\prime}}$ distributions in view of theoretical predictions}
\label{distributions in view of theoretical predictions}
\begin{figure}[H]
  \includegraphics[height=.34\textheight]{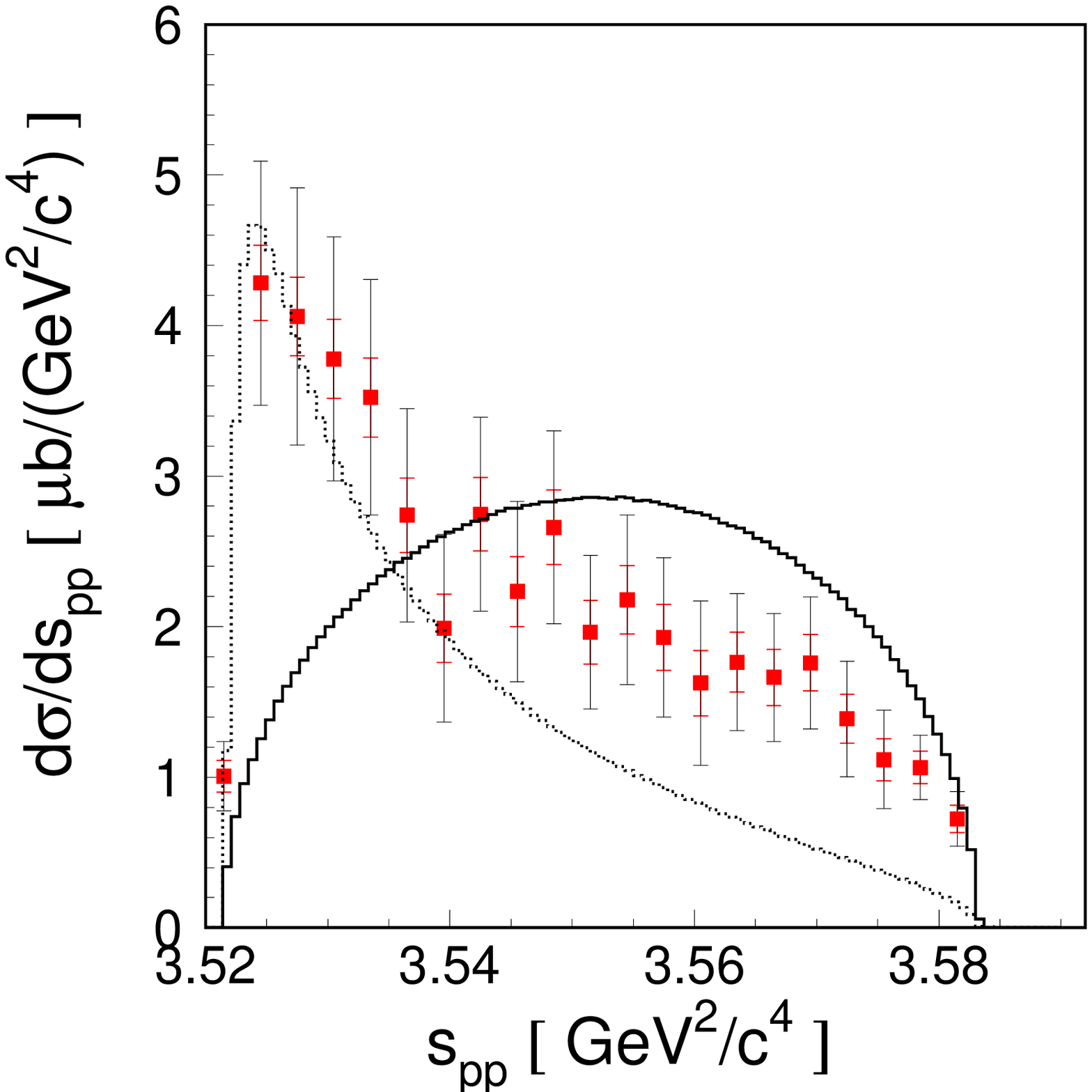}
  \includegraphics[height=.34\textheight]{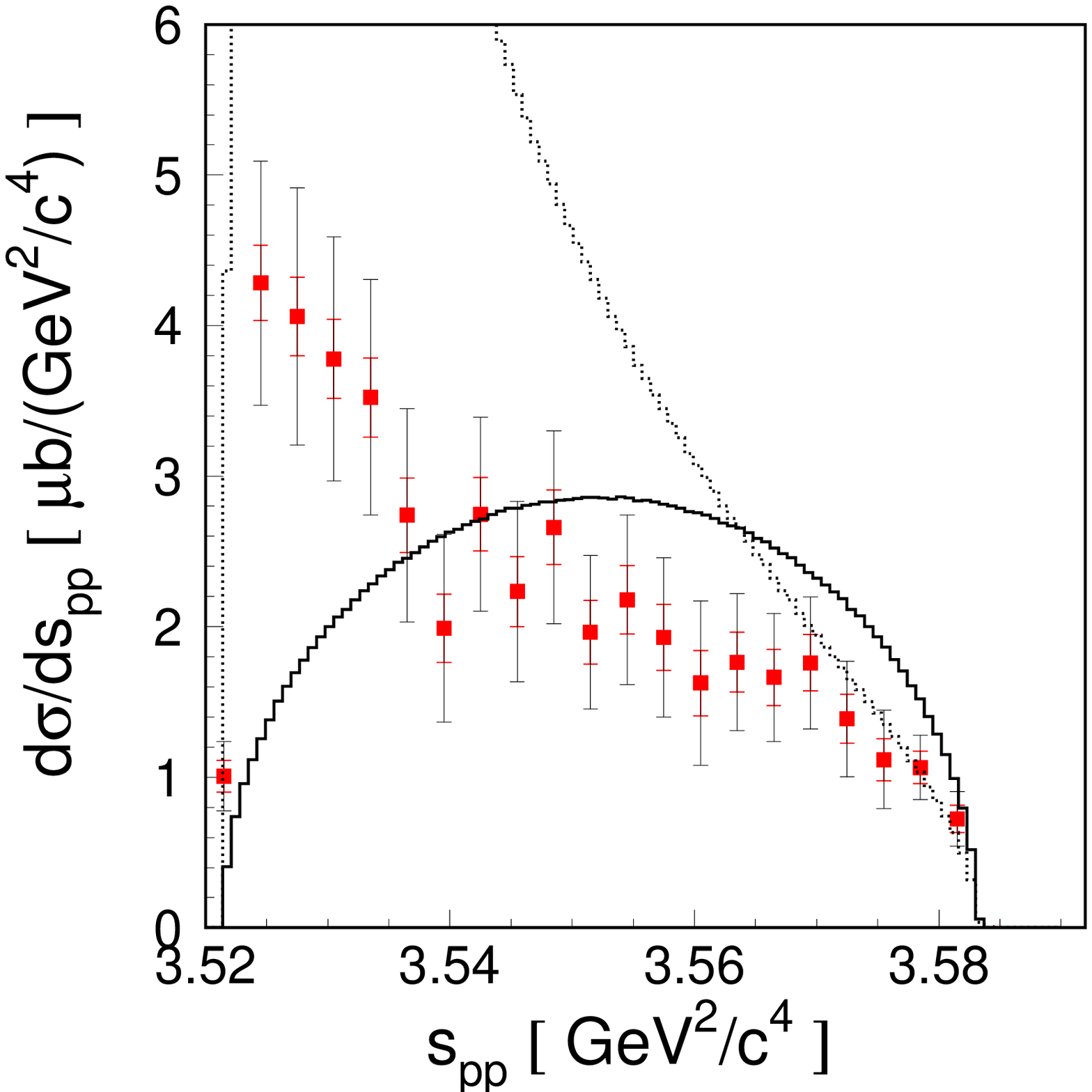}
        \caption{
	Distribution of the square of the proton-proton ($s_{pp}$)
	 invariant mass determined experimentally for
  the $pp \to pp\eta^{\prime}$ reaction (closed squares). The integrals of the phase space weighted by a
  square of the proton-proton on-shell scattering amplitude
(dotted lines)-FSI$_{pp}$ have been normalized arbitrarily to the lower values of $s_{pp}$ in the
left panel and to the higher values of $s_{pp}$ in the right panel.
The expectations
under the assumption of a homogeneously populated phase space are shown as solid lines.}	
\label{fig:pp_normal}
\end{figure}

\begin{figure}[H]
  \includegraphics[height=.34\textheight]{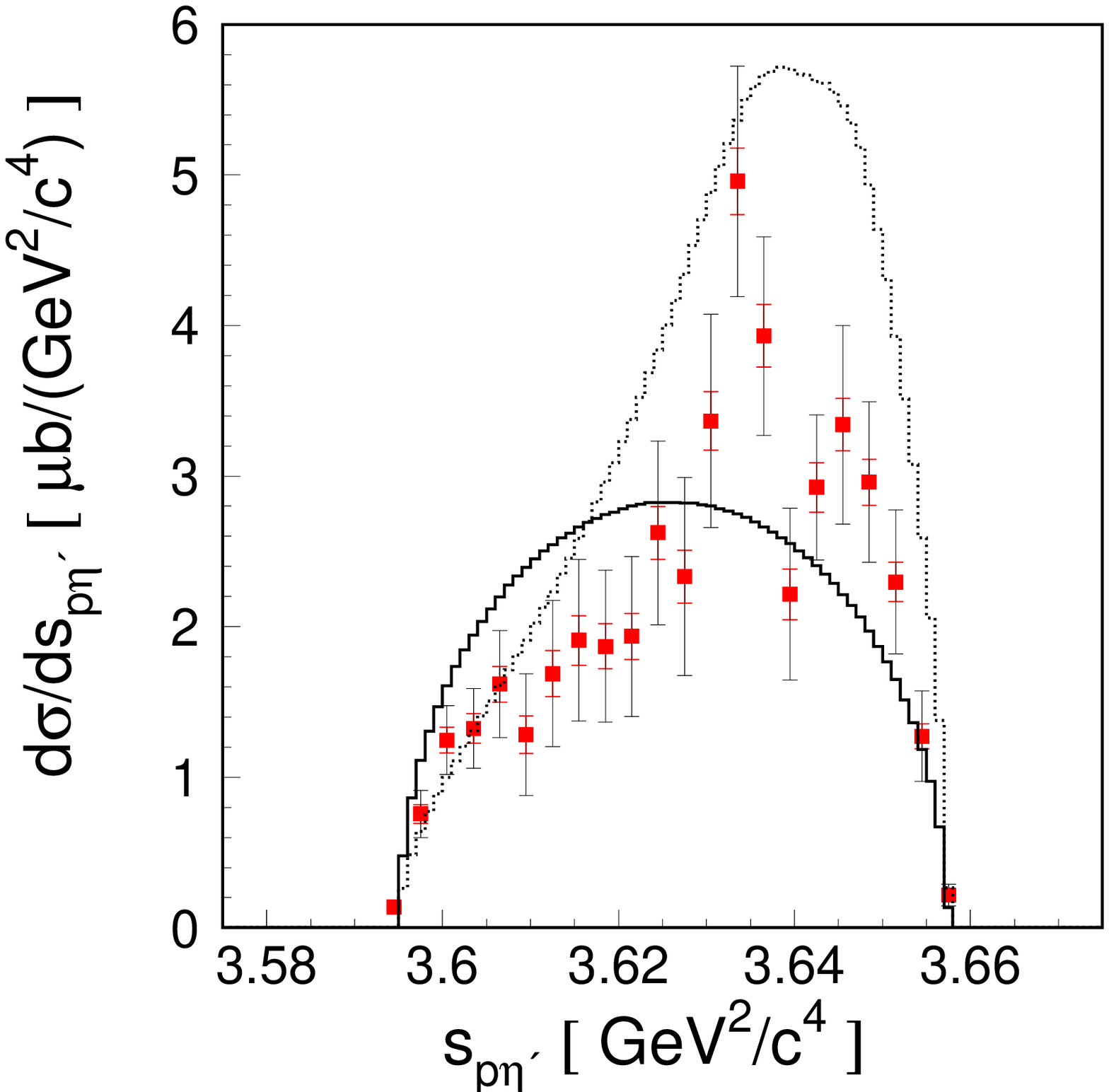}
  \includegraphics[height=.34\textheight]{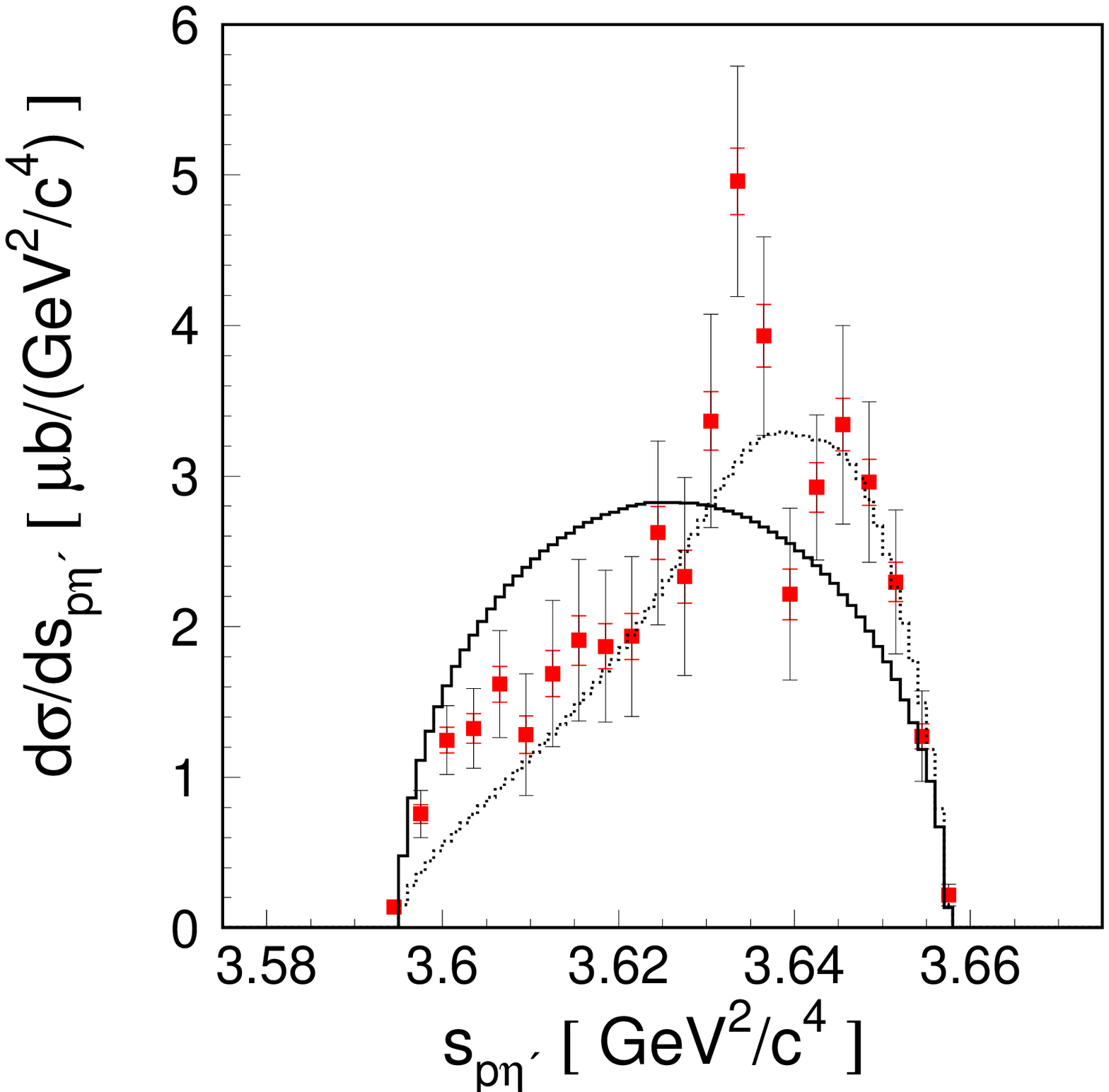}
        \caption{
	Distribution of the square of the proton-$\eta^{\prime}$ ($s_{p\eta^{\prime}}$)
	 invariant mass determined experimentally for
  the $pp \to pp\eta^{\prime}$ reaction (closed squares). The integrals of the phase space weighted by a
  square of the proton-proton on-shell scattering amplitude
(dotted lines)-FSI$_{pp}$ normalized arbitrarily to the lower values of $s_{p\eta^{\prime}}$ in the
left panel and to the higher values of $s_{p\eta^{\prime}}$ in the right panel.
The expectations
under the assumption of the homogeneously populated phase space are shown as solid lines.}	
\label{fig:peta_normal}
\end{figure}

The interaction should manifest itself in the regions where particles possess small relative velocities
as it was mentioned in previous sections.\\
In figure \ref{fig:pp_normal} the comparison of the experimentally determined cross sections as a function of
the squared invariant mass of the proton-proton system ($s_{pp}$) to theoretical calculations is presented. One can see the strong deviation
of the experimental spectrum from the phase space predictions (solid lines), normalized in amplitude.
It can be also seen that the inclusion of the FSI$_{pp}$ \cite{swave} improves the agreement
but still there is a significant discrepancy between the data and the calculations\footnote{A detailed description of the proton-proton FSI
parameterization based on the proton-proton on-shell amplitude is presented in appendix
\ref{Parameterization of the proton-proton Final State Interaction}}.
Here, it is also presented that two different possible normalizations of the
model calculations do not agree with the shape of the experimental distribution. The left panel shows the fit to the
lower values of the proton-proton invariant mass. The right one depicts the fitting to the
high energy region of  $s_{pp}$.\\
These different normalizations are presented for the invariant mass of the proton-$\eta^{\prime}$ subsystem
 in figure \ref{fig:peta_normal}.\\
Inspecting figures \ref{fig:pp_normal} and \ref{fig:peta_normal}, one can conclude that similarly to the case of
the $pp \to pp\eta$ reaction, the proton-proton on-shell interaction is not sufficient to explain the enhancement
seen in both distributions, independent of the applied normalization.\\

\begin{figure}[H]
  \includegraphics[height=.34\textheight]{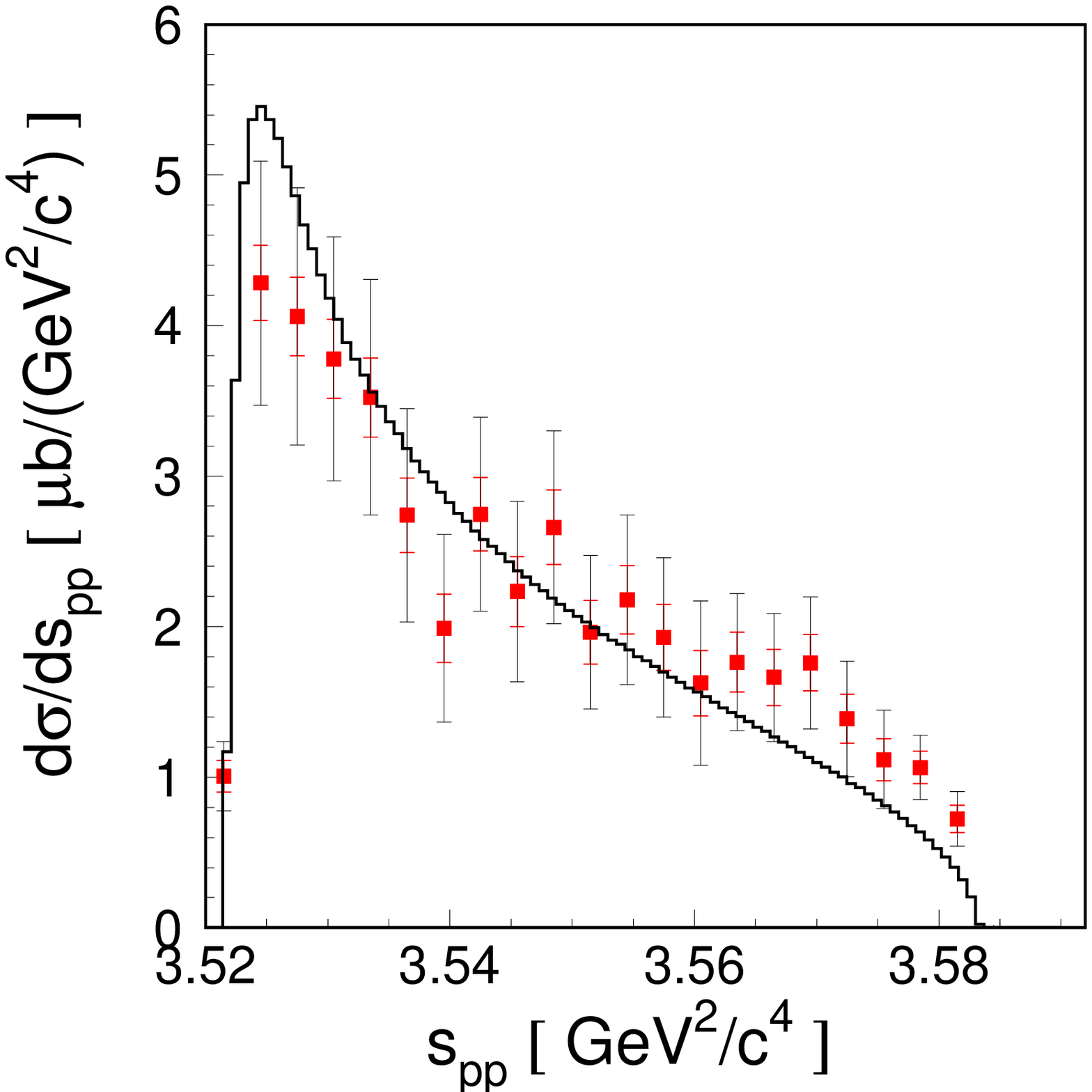}
  \includegraphics[height=.34\textheight]{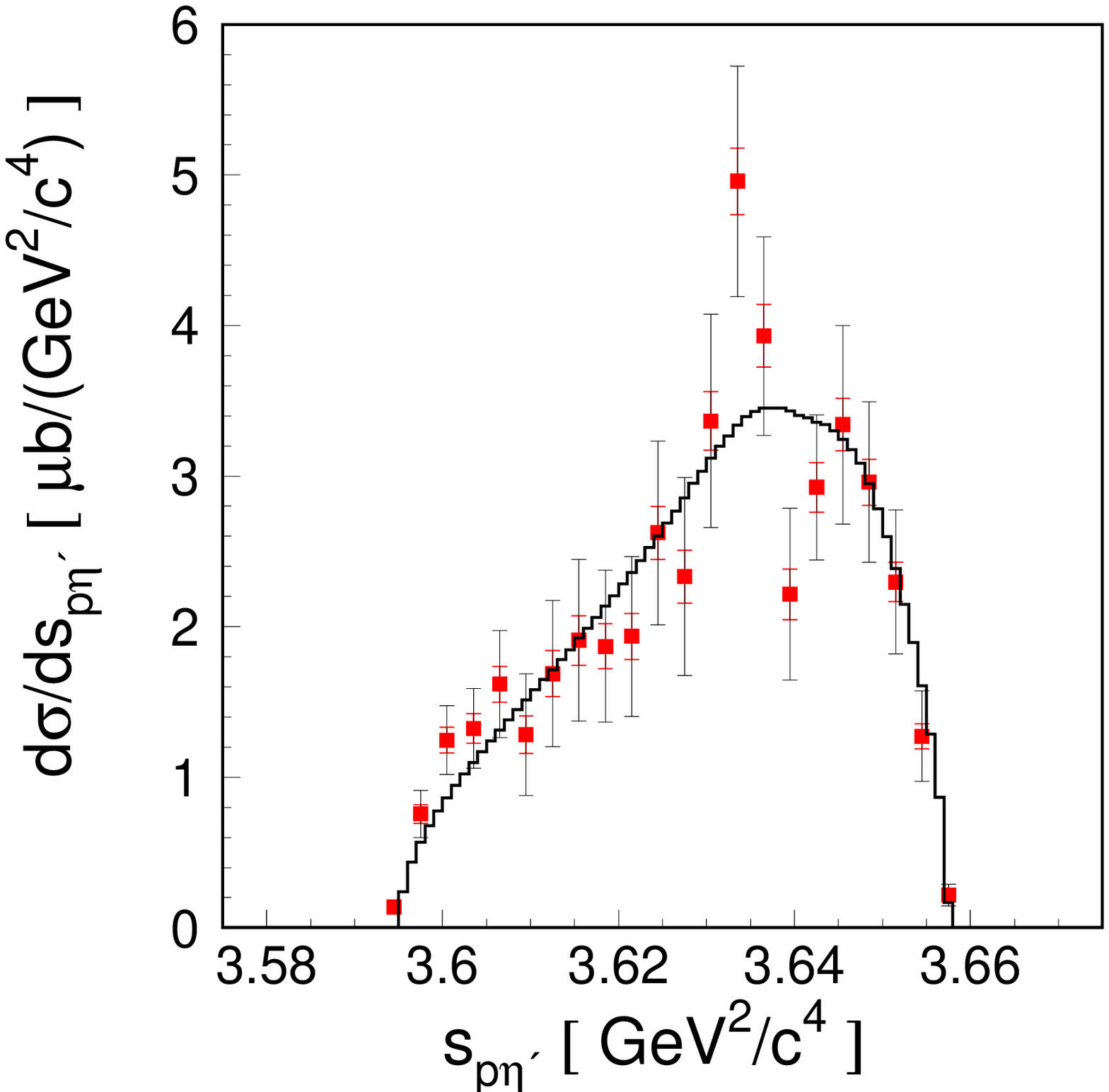}
        \caption{(Left) Distribution of the squared proton-proton ($s_{pp}$)
	 invariant mass determined experimentally for
  the $pp \to pp\eta^{\prime}$ reaction at the excess energy Q = 16.4 MeV (closed squares).
  (Right)  Distribution of the squared proton-$\eta^{\prime}$ ($s_{p\eta^{\prime}}$)
	 invariant mass determined experimentally for
  the $pp \to pp\eta^{\prime}$ reaction at the excess energy of 16.4 MeV (closed squares).
  The experimental data are compared to the calculations under the assumption of the $^{3}P_{0} \to ^{1}\!\!S_{0}s$
  transition according to the model of V. Baru et al., as described in text
  and in \cite{baru, vbaru}.}	
\label{fig:baru}
\end{figure}

\begin{figure}[H]
  \includegraphics[height=.34\textheight]{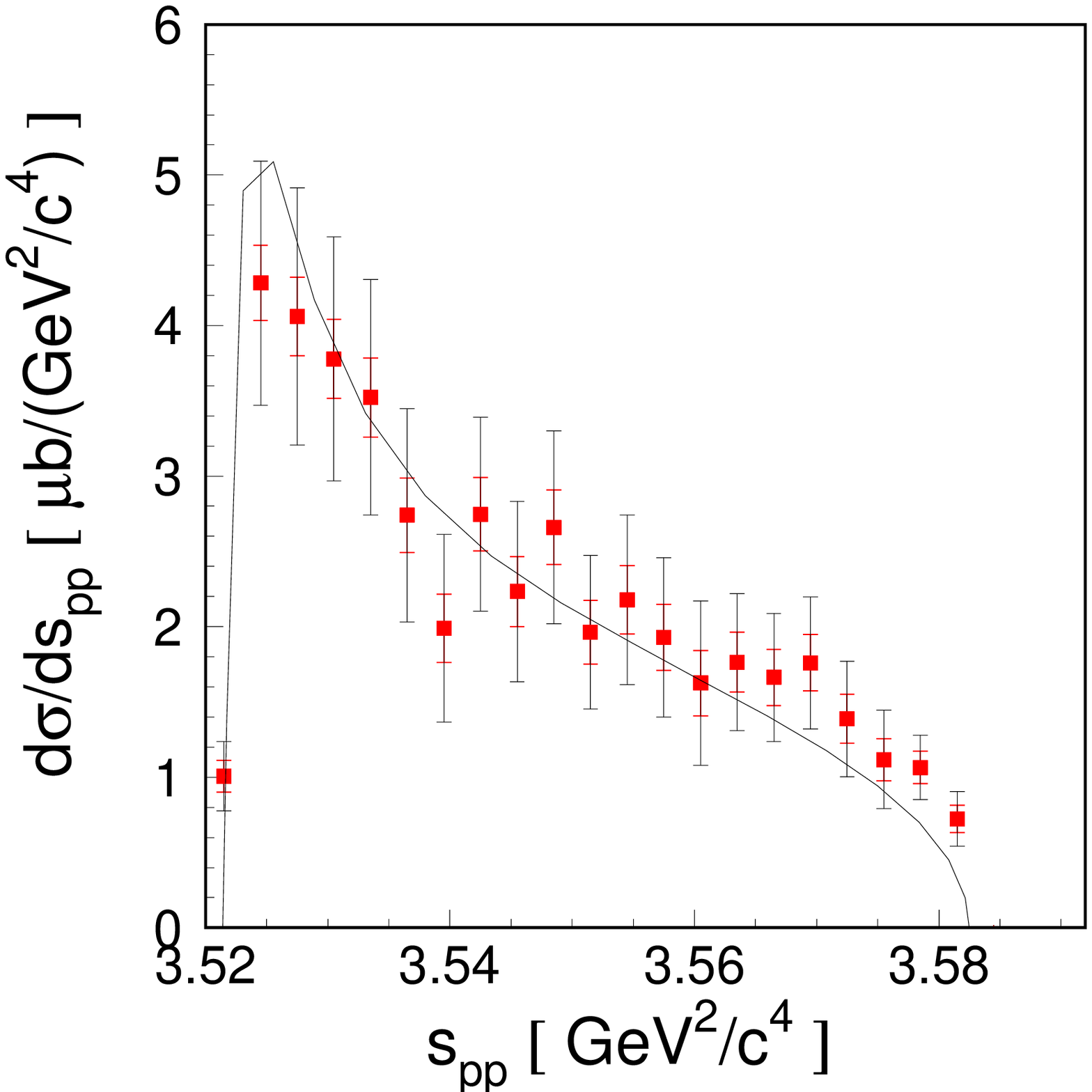}
  \includegraphics[height=.34\textheight]{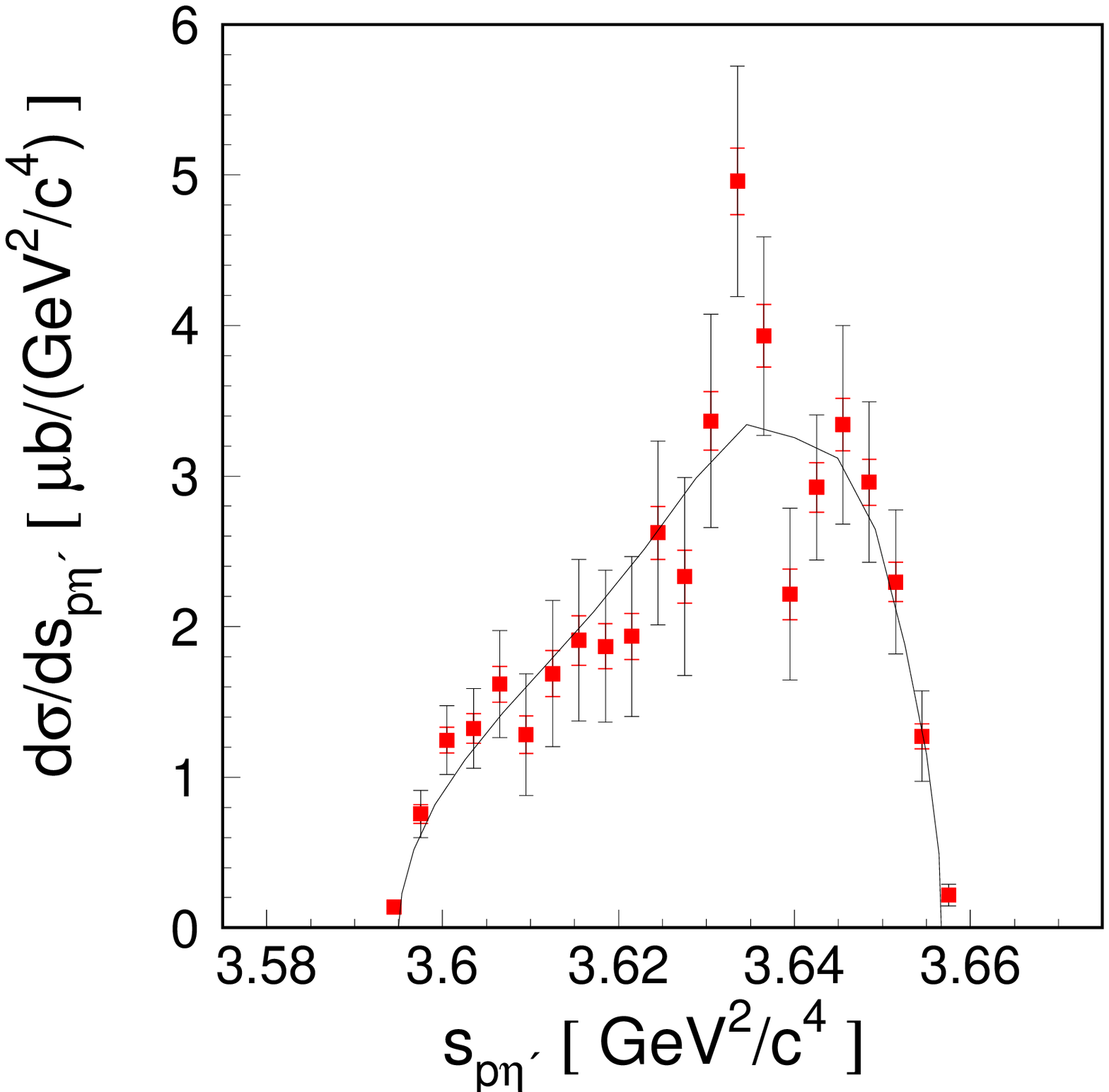}
        \caption{(Left) Distribution of the squared proton-proton ($s_{pp}$)
	 invariant mass determined experimentally for
  the $pp \to pp\eta^{\prime}$ reaction at the excess energy Q = 16.4 MeV (closed squares).
  (Right)  Distribution of the squared proton-$\eta^{\prime}$ ($s_{p\eta^{\prime}}$)
	 invariant mass determined experimentally for
  the $pp \to pp\eta^{\prime}$ reaction at the excess energy of 16.4 MeV (closed squares).
  The experimental distributions are compared to the calculations based on the assumption that S and P partial waves can
  contribute as it is described by K. Nakayama et al. in \cite{naka, kanzo_c11, naka1, naka2}.}	
\label{fig:kanzo}
\end{figure}

\begin{figure}[H]
  \includegraphics[height=.34\textheight]{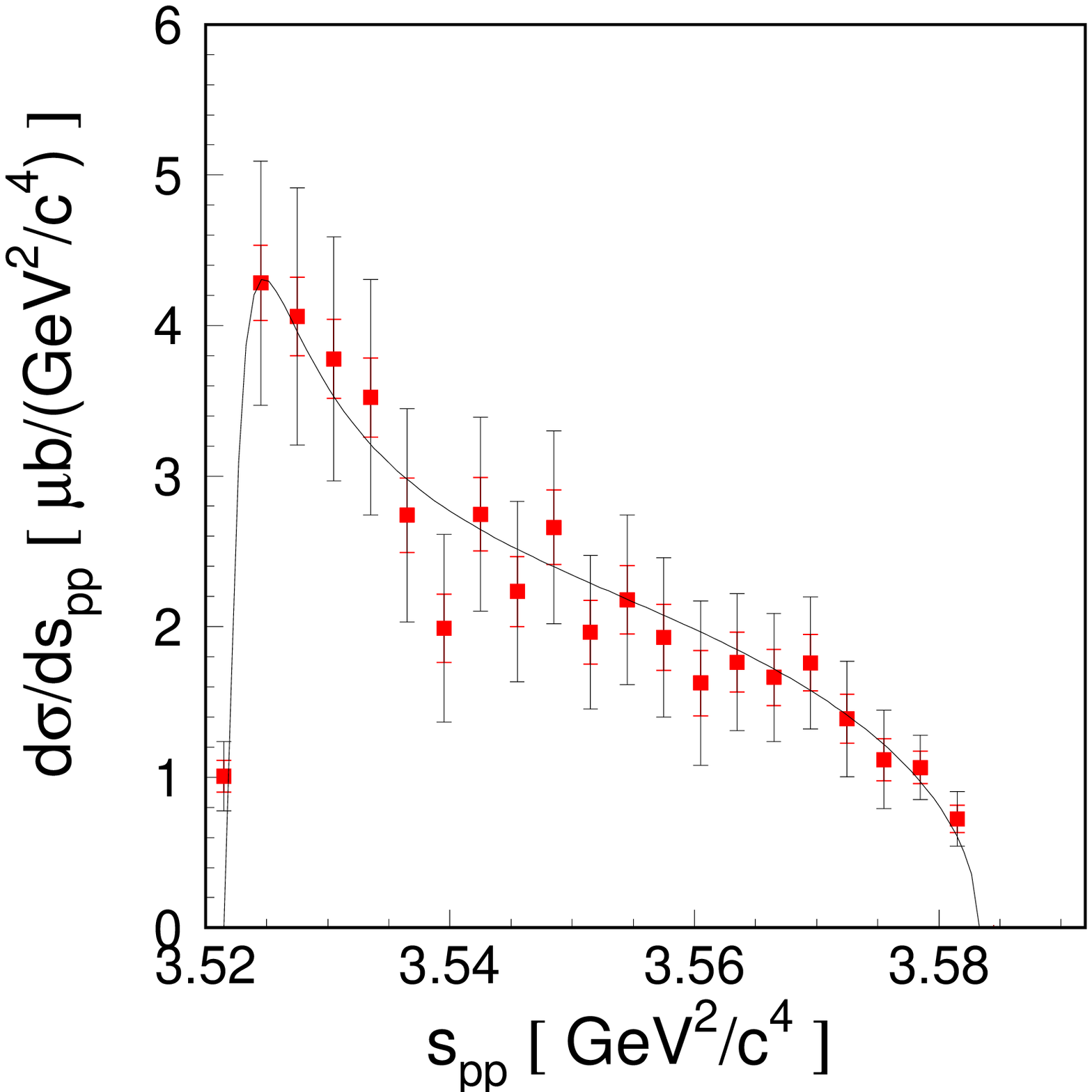}
  \includegraphics[height=.34\textheight]{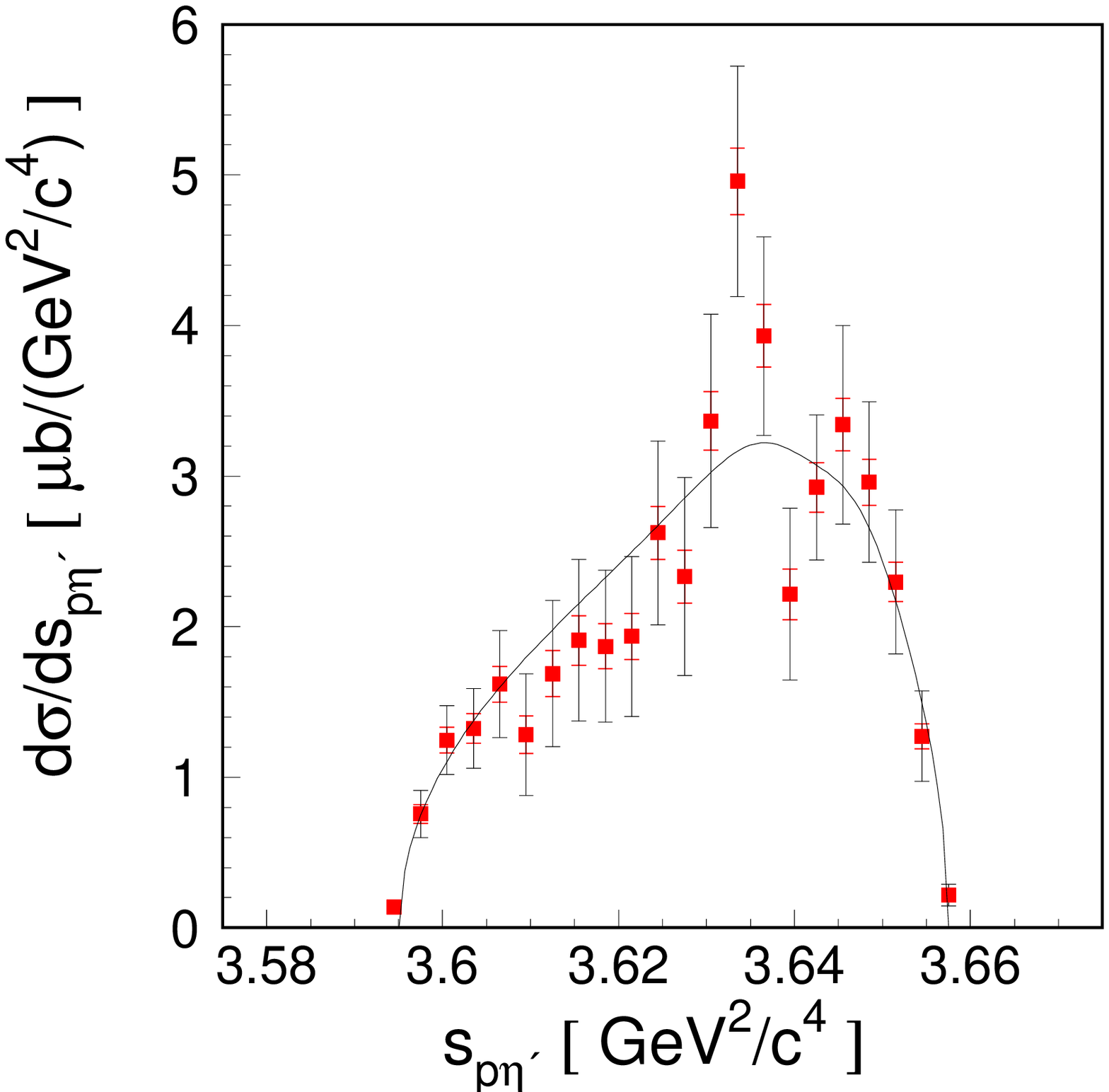}
        \caption{(Left) Distribution of the squared proton-proton ($s_{pp}$)
	 invariant mass determined experimentally for
  the $pp \to pp\eta^{\prime}$ reaction at the excess energy Q = 16.4 MeV (closed squares).
  (Right)  Distribution of the squared proton-$\eta^{\prime}$ ($s_{p\eta^{\prime}}$)
	 invariant mass determined experimentally for
  the $pp \to pp\eta^{\prime}$ reaction at the excess energy of 16.4 MeV (closed squares).
  The experimental data are compared to the calculations performed assuming a linear energy dependence of the production
  amplitude as proposed by A. Deloff \cite{deloff, deloff1}.}	
\label{fig:deloff}
\end{figure}

A better description is obtained when instead of the on-shell proton-proton amplitude
the proton-proton FSI is parameterized by the Jost function \cite{habil, jost}.\\
The experimental
distributions of $s_{pp}$ and $s_{p\eta^{\prime}}$ are compared to the theoretical model proposed by
V. Baru and collaborators~\cite{baru} in figure \ref{fig:baru}.\\
The model is based on calculations in which the overall transition matrix element
can be factorized into the primary production and the proton-proton final state interaction
expressed as the inverse of the Jost function derived from the Bonn potential \cite{habil, baru, vbaru}.
The production amplitude is calculated using the one boson exchange model (OBE) and is nearly constant in the calculations.
One can see that the model normalized to the total cross section, hardly fit to the experimental data at the range of lower and
higher $s_{pp}$ values. On the other hand, the $s_{p\eta^{\prime}}$ distribution is rather well reproduced by the model.\\
A better agreement with the experimental data can be achieved by the implementation of
contributions from the higher partial waves in the theoretical calculations.
In figure \ref{fig:kanzo} the distributions of the squared masses of proton-proton ($s_{pp}$) and proton-$\eta^{\prime}$ 
($s_{p\eta^{\prime}}$)
subsystems are compared to the calculations of K. Nakayama et al. \cite{naka}.\\
The calculations result from a combined analysis (based on the effective Lagrangian approach) of the production of $\eta$ and $\eta^{\prime}$ mesons
in photo- and hadro-induced reactions\footnote{The general overview of the calculations proposed by K. Nakayama and group
is discussed in appendix~\ref{Formalism kanzo}.} \cite{naka, naka1, naka2, kanzo_c11}.
The authors assumed in the calculations that the $^{1}S_{0} \to ^{3}\!\!P_{0}s$ transition
contributes beside the  $^{3}P_{0} \to ^{1}\!\!S_{0}s$
transition at the excess energy of 16.4 MeV.\\
Interestingly, the authors claim that for the $pp \to pp\eta^{\prime}$ reaction the dominant production mechanism for the $NN \to NN\eta^{\prime}$
reaction is due to the excitation of the intermediate $S_{11}$ resonance \cite{kanzo_c11}.
Figure  \ref{fig:kanzo} shows that both distributions are well reproduced by the theoretical calculations.
On the other hand, as proposed by A. Deloff, one can explain the enhancement seen in the distributions
by an energy dependent production amplitude, which in the calculations of
V. Baru et al. \cite{baru, vbaru} and of K. Nakayama et al. \cite{naka, kanzo_c11, naka1, naka2} was constant.\\
In figure \ref{fig:deloff} the experimentally determined spectra
of the squared proton-proton invariant mass ($s_{pp}$) and proton-$\eta^{\prime}$ ($s_{p\eta^{\prime}}$) system 
is presented and compared
to calculations by A.~Deloff \cite{deloff, deloff1}. The calculations are based on the assumption that
the production amplitude changes linearly with energy\footnote{The phenomenological model of
the differential cross section parameterization assuming the linear energy dependence
of the production amplitude is presented in general in appendix \ref{Linear dependence}.}.
The calculations based on the standard on-shell approximation
were modified
allowing the linear energy dependence of the $^{3}P_{0} \to ^{1}\!\!S_{0}s$ partial wave amplitude.
Other partial wave transitions were neglected in the model.\\

In conclusion to this chapter one has to stress that from the above presented considerations, the determined invariant
mass distributions ($s_{pp}$ and $s_{p\eta^{\prime}}$) strongly deviate from the predictions based on the
homogeneous population of events over the phase space. Also, the  parameterization of the proton-proton on-shell interaction
is not sufficient to explain the enhancement seen in the determined distributions.\\

Furthermore, one can see that both calculations assuming a significant contribution of P-wave in the
final state
(K. Nakayama), or the model assuming
a linear energy dependence of the production amplitude by A. Deloff, reproduce the data
within the error bars quite well.
Therefore, on the basis of the presented invariant mass distributions, it is impossible
to disentangle which of the discussed models is more appropriate.\\
On the other hand, it is clearly seen that the calculations assuming only $^{3}P_{0} \to ^{1}\!\!S_{0}s$ transition,
independent of the energy, underestimate
the experimental data at the higher values and overestimate them at the lower values of $s_{pp}$ as it is
presented in figure \ref{fig:baru}.

Moreover, independently from the theoretical calculations, one can conclude from the comparison
of $s_{pp}$ and $\sqrt{s_{p-meson}}$ presented in figure \ref{fig:comparison} that the observed enhancement
is not caused by a proton-meson interaction, since the strength of proton-$\eta$ and proton-$\eta^{\prime}$ interaction
is different but the enhancement in both cases is the same.\\

\chapter{Summary}
\label{Summary}
\markboth{\bf Chapter 11.}
         {\bf Summary}

Differential distributions of the squared proton-proton ($s_{pp}$) and proton-$\eta^{\prime}$
($s_{p\eta^{\prime}}$) invariant masses, as well as angular distributions for the $pp \to pp\eta^{\prime}$ reaction 
have been evaluated at an excess energy of Q~=~16.4 MeV.
The data were measured by the COSY-11 collaboration at the accelerator COSY.\\

The theoretical framework aiming to explain the unexpected 
enhancement observed in the invariant mass distributions determined for the $pp \to pp\eta$ 
reaction was presented and it was tried to 
verify one of the three challenging theories describing the observed behaviour by:\\[-0.8cm]
\begin{itemize}
\item 
a significant role of {\bf proton-$\eta$ interaction} in the final state,\\[-0.8cm]
\item 
an {\bf admixture of higher partial waves} in the produced \\
proton-proton-meson system,\\[-0.8cm]
\item
an {\bf energy dependence of the production amplitude}.
\end{itemize}
$~$\\[-0.8cm]
On a basis of only $pp \to pp\eta$ data, it was impossible to judge which 
model could describe the effects. Therefore, in order to verify those hypotheses 
by a comparison of the invariant mass spectra for the production of two different mesons,  
the analysis of the $pp \to pp\eta^{\prime}$ reaction, presented in detail in this thesis, was performed.\\
The reaction was measured using the COSY-11 detector setup designed especially for close to threshold 
measurements. The experiment was performed at a beam momentum of $P_{B} = 3.260$ GeV/c, 
which corresponds to an excess energy of Q~=~16.4~MeV.\\
A short description of the synchrotron COSY together 
with a closer look at the detectors of the COSY-11 apparatus used for the measurement was presented 
and its calibration and adjustment was described at length.\\
In the analysed experiment two charged particles were registered. Their identification as protons was based on 
the independent evaluation of their velocities and momenta. The decay products of the $\eta^{\prime}$ meson were not measured, therefore 
the number of $pp \to pp\eta^{\prime}$ events was determined using the
missing mass technique.\\

For the first time ever high statistics invariant mass $s_{pp}$ and $s_{p\eta^{\prime}}$ spectra
and angular distributions, determined in the close to kinematical threshold region
for the $pp \to pp\eta^{\prime}$ reaction were presented. The distributions were corrected for the geometrical acceptance of the COSY-11 apparatus 
and were determined free from the background of the multi-pion production. 
Besides the differential distributions also 
the total cross section for the $pp \to pp\eta^{\prime}$ reaction has been established, 
determined at an excess energy of Q~=~16.4~MeV to be: $~~$ ${\bf{\sigma}}{\bf{~=~(139~\pm3~\pm12)~nb}}$.  \\

The evaluated invariant mass distributions were {\bf similar to those determined for the $\eta$ meson production. 
This result together with the shape of the excitation functions for the 
$pp \to pp\eta$ and $pp \to pp\eta^{\prime}$ reactions, 
excludes the hypothesis postulating that 
the observed enhancement is caused by the interaction between proton and meson.}\\
Furthermore, the determined $s_{pp}$ and $s_{p\eta^{\prime}}$ differential distributions 
do not allow to judge between the proposed theoretical models. 
{\bf Both, the hypothesis that an admixture of higher partial waves in the produced proton-proton-meson system and 
the proposal that an energy dependence of the production amplitude leads to the observed
shape of the $s_{pp}$ and $s_{p\eta^{\prime}}$ distributions are consistent with the data 
at the achieved level of accuracy.}\\

The results presented in this thesis may be useful for the verification of other recently published theoretical 
models applied for the proton-proton-meson interaction, for example the calculations 
proposed by S. Ceci et al. in~\cite{ceci, ceci1}.\\
\\
In order to disentangle the possible explanations for the observed differential distributions more experimental data are
required. The data base could be extended by studying the excess energy dependence of the shape 
of the $s_{pp}$ and $s_{p\eta^{\prime}}$ distributions, as it was done for the $pp \to pp\eta$ reaction.  
Such studies would allow for a direct comparison of the differential 
distributions behaviour for the $\eta^{\prime}$ production with already determined quantities 
for the $pp \to pp\eta$ reaction, in a similar manner as described in this thesis.\\

A more sensitive test, whether higher partial waves 
are indeed important for the studied reactions, could be 
gained from polarization observables. Therefore, measurement of analysing powers 
and spin correlation coefficients would be desirable  for a better 
understanding of the dynamics and interaction in the $pp\eta$ and $pp\eta^{\prime}$ systems.\\
 
  \renewcommand{\theequation}{A.\arabic{equation}}
  \renewcommand{\thefigure}{A.\arabic{figure}} 
  \renewcommand{\thetable}{A.\arabic{table}} 
  \setcounter{equation}{0}  
  \setcounter{figure}{0}  
  \setcounter{table}{0}  
 
\appendix 
\chapter{Pseudoscalar mesons}
\label{Pseudoscalar mesons}
\markboth{\bf Appendix A}
         {\bf Appendix A}

Mesons are bound states built from a quark $q$ and an antiquark $\bar{q}^{\prime}$ according to the Quantum Chromodynamics, abbreviated as QCD.  
The quarks $q$ and $\bar{q}^{\prime}$ can be of the same or a different flavour.
Since quarks possess the spin 1/2, they can build a singlet state with the total spin $J = 0$ or triplet states with the 
total spin $J = 1$. 

A positive parity is assigned to the quarks whereas a negative parity is assigned to the antiquarks. Using that 
convention for the bound state of $q\bar{q}^{\prime}$ and denoting by $L$ the orbital angular momentum of the 
system, the parity of the meson built out of the quark pairs is $P = (-1)^{L + 1}$.

\begin{center}  
\begin{figure}[H]
\includegraphics[width=.4\textheight]{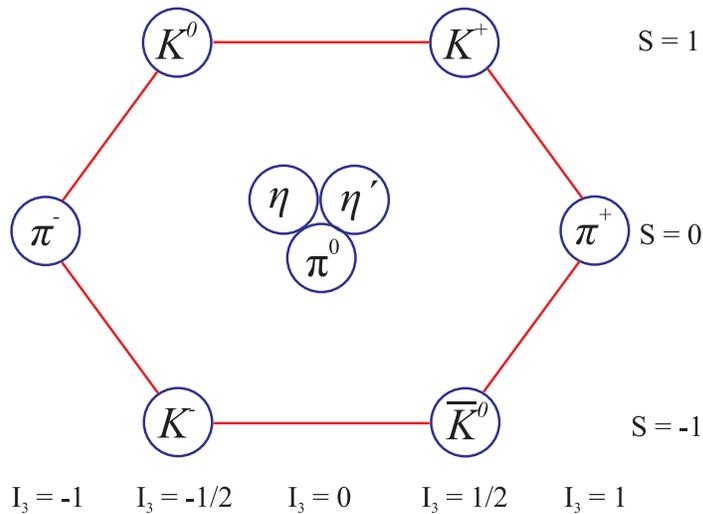}
\caption{The ground states of the pseudoscalar mesons ($J^{P} = 0^{-}$). 
       The strangeness $S$ of the mesons is plotted versus the third component of the isospin $I_{3}$.}
\label{fig:mesons}
\end{figure}  
  \end{center}

From the three lighter quarks $u$, $d$ and $s$ nine possible $q\bar{q}^{\prime}$ combinations can be built. 
This allows to built a meson nonet according to the SU(3) frame including the octet and the singlet state:
\begin{equation}
3 \otimes \bar{3} = 8 \oplus 1.
\label{equ:nonet}
\end{equation}
The assumption that these $q\bar{q}^{\prime}$ systems are the ground state combinations of the quark-antiquark pairs 
with the angular momentum $L = 0$ implies, that the parity $P$ of the constructed mesons should be equal to $P = -1$. 
With that conditions ($L = 0$ and $P = -1$) mesons with the internal spin $J = 1$ are called vector mesons, whereas
mesons with the internal spin $J = 0$ are called pseudoscalar mesons. \\
The ground states of the pseudoscalar meson nonet is presented in figure~\ref{fig:mesons}. \\

The quark structure and masses of the ground states of the pseudoscalar mesons are listed in table \ref{pseudo_mesons}.
\begin{table}[H]
\begin{center}
\begin{tabular}{c c c} \hline
 Pseudoscalar meson  & Quark structure  & Mass [ MeV ]\\ \hline
  $\pi^{0}$     & $1/\sqrt{2} (u\bar{u} - d\bar{d})$  & 134.98\\
  $\pi^{+}$     & $u\bar{d}$                          & 139.57\\
  $\pi^{-}$     & $d\bar{u}$                          & 139.57\\
  $K^{+}$       & $u\bar{s}$                          & 493.68\\
  $K^{-}$       & $\bar{u}s$                          & 493.68\\
  $K^{0}$       & $d\bar{s}$                          & 497.61\\
  $\bar{K}^{0}$ & $\bar{d}s$                          & 497.61\\
  $\eta$        & $A(d\bar{d} + u\bar{u}) + B(s\bar{s})$ & 547.85\\
  $\eta^{\prime}$       & $a(d\bar{d} + u\bar{u}) + b(s\bar{s})$ & 957.66\\ \hline
\end{tabular}
\caption{List of the pseudoscalar mesons as the quark-antiquark combinations. 
Meson masses are taken from reference~\cite{pdg}.}
\label{pseudo_mesons}
\end{center}
\end{table}

  \renewcommand{\theequation}{B.\arabic{equation}}
  \renewcommand{\thefigure}{B.\arabic{figure}}
  \renewcommand{\thetable}{B.\arabic{table}}
  \setcounter{equation}{0}  
  \setcounter{figure}{0}  
  \setcounter{table}{0}  

\chapter{Parameterization of the proton-proton Final State Interaction}
\label{Parameterization of the proton-proton Final State Interaction}
\markboth{\bf Appendix B}
         {\bf Appendix B}

The total transition amplitude for the $pp \to ppX$ process can be written (in the
Watson-Migda{\l} approximation \cite{watson}) as \cite{moalem}:
\begin{equation}
|M_{pp \to ppX}|^{2} = |M_{0}|^{2} \cdot |M_{FSI}|^{2} \cdot ISI,
\label{equ:ampl}
\end{equation}
where $M_{0}$ denotes the total production amplitude, $M_{FSI}$ refers to the elastic interaction among particles
in the exit channel, and ISI stands for the reduction factor in the initial channel of the colliding protons.\\
The element $|M_{FSI}|$ is predominantly due to the $pp$-FSI ($M_{pp \to pp}$), which can be calculated as \cite{morton}:
\begin{equation}
M_{pp \to pp} = \frac{e^{\delta_{pp}(^{1}S_{0})} \cdot sin \delta_{pp}(^{1}S_{0})}{C \cdot k},
\label{equ:ampl2}
\end{equation}
where $C^{2} = \frac{2 \pi \eta_{c}}{e^{2 \pi \eta_{c}} - 1}$ is the Coulomb penetration factor \cite{bethe},
$\eta_{c}$ is the relativistic Coulomb parameter.\\
The phase shift $\delta_{pp}(^{1}S_{0})$ is calculated using the Cini-Fubini-Stanghellini formula with the
Wong-Noyes Coulomb correction \cite{noyes, naisse, noyes1}:
\begin{equation}
C^{2}~p~ctg \delta_{pp} + 2p~\eta_{c}~h(\eta_{c}) = -
\frac{1}{a_{pp}}+\frac{b_{pp}~p^{2}}{2}-\frac{P_{pp}~p^{4}}{1 + Q_{pp}~p^{2}},
\label{equ:shift}
\end{equation}
where $h(\eta_{c})=-ln(\eta_{c})-0.57721+\eta^{2}_{c} \sum_{n=1}^{\infty}\frac{1}{n(n^{2}+\eta^{2}_{c})}$ \cite{jackson}, \\
$a_{pp}~=~-7.83$~fm, \\
and $b_{pp}~=~2.8$~fm stand for scattering length and effective range, respectively \cite{naisse}. \\
The parameters $P_{pp}~=~0.73~$fm$^{3}$, and $Q_{pp}~=~3.35~$fm$^{2}$ are related to the shape of
nuclear potential derived from one-pion-exchange model \cite{naisse}.\\

  \renewcommand{\theequation}{C.\arabic{equation}}
  \renewcommand{\thefigure}{C.\arabic{figure}} 
  \renewcommand{\thetable}{C.\arabic{table}} 
  \setcounter{equation}{0}  
  \setcounter{figure}{0}  
  \setcounter{table}{0}  

\chapter{Formalism of a combined analysis of photo- and hadro-production of the $\eta^{\prime}$ meson}
\label{Formalism kanzo}
\markboth{\bf Appendix C}
         {\bf Appendix C}

\begin{center}
\begin{figure}[H]
 \includegraphics[height=.3\textheight]{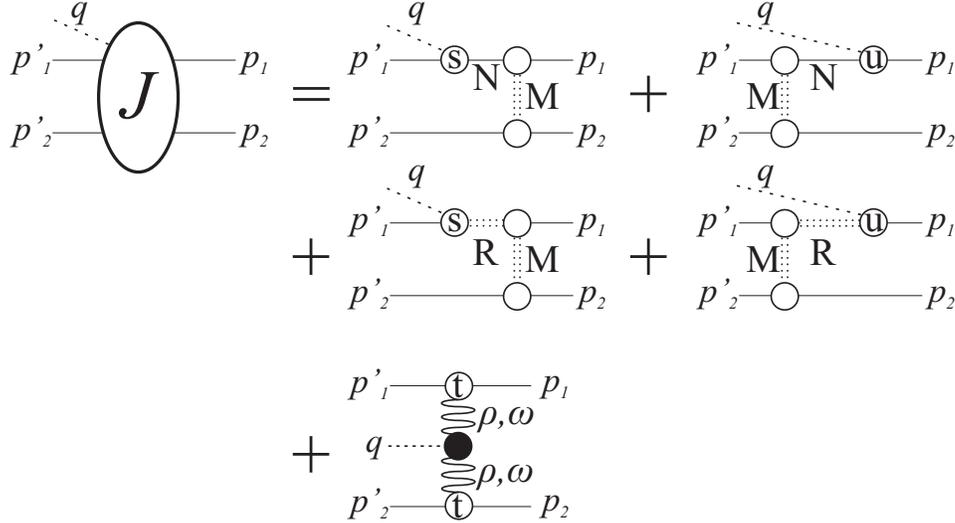}
  \caption{Basic production mechanisms for the $pp \to pp\eta^{\prime}$ reaction. 
  Time proceeds from right to left for each individual graph. $N$ and $R$ denote the intermediate nucleons and 
  resonances, respectively. $M$ stands for all possible exchanges of mesons as: 
  $\pi$, $\eta$, $\rho$, $\omega$, $\sigma$ and $a_{0}$ 
  for the nucleon graphs (the first two graphs of the sum in figure~\ref{fig:mech_kanzo}) and 
  $\pi$, $\rho$ and $\omega$ for resonance graphs (the third and the fourth graphs of the sum in figure~\ref{fig:mech_kanzo} ).}
\label{fig:mech_kanzo}
\end{figure}
\end{center}

Investigations of the $\eta$ and $\eta^{\prime}$ meson production are performed for the following reaction channels:
$$\gamma + N \to N + M,$$
$$\pi + N \to N + M,$$
\begin{equation}
N + N \to N + N + M,
\label{equ:reac_kanzo}
\end{equation}	 
based on a relativistic meson-exchange approach, where $M = \eta, \eta^{\prime}$.\\
According to the model described in~\cite{ kanzo48, kanzo43}
the transition amplitude $M$ \cite{kanzo48}\\
for the $pp \to pp\eta^{\prime}$ reaction is given by:
\begin{equation}
M = (1 + T_{f}G_{f})J(1 + T_{i}G_{i}),
\label{equ:ampl_kanzo}
\end{equation}
where $T_{i,f}$  denotes the $NN$ $T$-matrix interaction in the initial ($i$) 
or final ($f$) state, \\and $G_{i,f}$ is the corresponding two-nucleon propagator \cite{kanzo48}. \\
$J$ sums all basic $\eta^{\prime}$ production mechanisms depicted in figure \ref{fig:mech_kanzo}.\\

The free parameters of the model -- i.e. resonance parameters and the $NNM$ coupling constant 
-- are fixed to reproduce the available data for photo- and hadron-induced reactions listed in equation~\ref{equ:reac_kanzo}. \\

  \renewcommand{\theequation}{D.\arabic{equation}}
  \renewcommand{\thefigure}{D.\arabic{figure}} 
  \renewcommand{\thetable}{D.\arabic{table}} 

  \setcounter{equation}{0}  
  \setcounter{figure}{0}  
  \setcounter{table}{0}  
 
\chapter{Linear energy dependence of the production amplitude}
\label{Linear dependence}
\markboth{\bf Appendix D}
         {\bf Appendix D}

The following considerations and notation are adopted from reference~\cite{deloff}.\\

The cross section for the $pp \to pp\eta^{\prime}$ reaction can be expressed in the following manner:
\begin{equation} 
\frac{d \sigma}{dLips} = |C(k)|^{2} [a + b~ P_{2} (\hat{p} \cdot \hat{q})] +
         C_{0}~ (\eta) [d~ ReC(k) + e~ ImC(k)] P_{2} (\hat{p} \cdot \hat{k})
\label{equ:del_cross1}
\end{equation}
where in the standard notation $dLips$ stands for the invariant three body phase space element,  
$p$ denotes the initial momentum of the proton, $k$ stands for the relative momentum of the two 
protons in the exit channel, and $q$ corresponds to the relative momentum between meson 
and proton-proton pair in the final state. $C(k)$ and $C_{0} (\eta)$ denote 
enhancement factor and Coulomb penetration factor, respectively.\\
The $a$ parameter denotes the modulus of the sole squared production amplitude which survives at threshold
and is associated with the $^{3}P_{0} \to ^{1}\!\!S_{0}s$ transition. The parameter $b$ stands for the 
interference term between the latter amplitude and the $^{3}P_{0} \to ^{2}\!\!S_{0}d$ amplitude. 
Parameters $d$ and $e$ correspond to the interference with a $^{3}P_{2} \to ^{1}\!\!D_{2}s$ amplitude. 
The whole set of parameters depend on the final state momenta.\\

Looking at the threshold behaviour, one can see that the $a$ parameter turns to be constant, 
the $b$ parameter is proportional to $q^{2}$ and either $d$ or $e$ are of the order of $k^{2}$. 
In such a situation the $a$ parameter has to be expanded to the same order by setting: 
$a = a_{0} + a_{1} q^{2}$ where $a_{0}$ and $a_{1}$ are two unknown parameters. 
The $a_{0}$ parameter can be absorbed in the normalization process and the ultimate 
expression for the cross section reads as:
\begin{equation} 
\frac{d \sigma}{dLips} \propto |C(k)|^{2} \Big( 1 + \frac {q^{2}}{q^{2}_{max}}  
                [x + y~ P_{2} ( \hat{p} \cdot \hat{q})]  \Big) +
          \frac{k^{2}}{k^{2}_{max}} C_{0} (\eta) \Big[z_{r}~ ReC(k) + z_{i}~ ImC(k) \Big] P_{2} (\hat{p} \cdot \hat{k})
\label{equ:del_cross2}
\end{equation}
where $x$, $y$, $z_{r}$ and $z_{i}$ are the dimensionless parameters to be determined. The $x$ 
parameter stands for the correction of the order of $q^{2}$ to the dominant transition and 
the $y$, $z_{r}$ and $z_{i}$ parameters are the measure of the D-waves admixture.\\

   \cleardoublepage

\pagestyle{empty}
\begin{center}
{\bf Acknowledgments\\}
\end{center}

\begin{flushright}
\large{\it First of all, this dissertation would have not come into life without enormous help of Prof.~Pawe{\l}~Moskal.
I am really honoured and have a great pleasure to work with Him. I admire His
knowledge and  I would like to express my admiration for His patience, lots of
fruitful ideas and great concepts.\\

I acknowledge Prof.~Walter~Oelert for being my supervisor during the stay in J{\"u}lich.\\

I am very grateful to Dr.~Dieter~Grzonka and Prof.~Walter~Oelert for all challenging ideas and help.\\

I thank Colleagues from the COSY-11 group for nice cooperation.\\

I would like to thank Dr.~Rafa{\l}~Czy{\.z}ykiewicz for answers to all my questions. \\
I really learnt a lot from Him.\\

Especially, I thank Marcin~Zieli{\'n}ski for the whole bunch of expert's suggestions.\\

For careful reading and correcting this manuscript I am grateful to Prof.~Walter~Oelert, Dr.~Dieter~Grzonka and Prof.~Pawe{\l}~Moskal.\\

I acknowledge a big support of my family during the time of my studies.\\

Last but not least, I would like to thank my wife for being with me,\\
 sharing physics passion and daily life.\\}
\end{flushright}

   \cleardoublepage
   
   \def\bibname{References}
\newpage
\clearpage

\pagestyle{plain}
\pagestyle{myheadings}

\end{document}